\documentclass[12pt]{report}
\usepackage[english]{babel}
\usepackage[utf8]{inputenc}
\usepackage[normalem]{ulem}

\usepackage{geometry}
\usepackage{graphicx}
\graphicspath{ {images/} }
\usepackage{amsmath,amssymb}
\usepackage{caption}
\usepackage{subcaption}
\usepackage{mathtools}
\usepackage{hyperref}
\usepackage{color}
\usepackage{cleveref}
\usepackage{comment}
\usepackage{changepage}

\usepackage{fancyhdr}
\usepackage[backend=biber,style=numeric,sorting=none]{biblatex}
\newcommand{\be}{\begin{equation}}
\newcommand{\ee}{\end{equation}}
\newcommand{\bea}{\begin{eqnarray}}
\newcommand{\eea}{\end{eqnarray}}
\newcommand{\nn}{\nonumber}
\newcommand{\Tend}{T_\text{end}}
\newcommand{\Teq}{T_\text{eq}}
\newcommand{\Tosc}{T_\text{osc}}
\newcommand{\Tc}{T_\text{c}}
\newcommand{\Rend}{R_\text{end}}
\newcommand{\Req}{R_\text{eq}}
\newcommand{\Rosc}{R_\text{osc}}
\newcommand{\Rc}{R_\text{c}}

\newcommand{\rhocrit}{\rho_{c,0}}

\newcommand{\gs}{g_\star}
\newcommand{\gss}{g_{\star S}}

\newcommand{\rp}{\rho_\phi}
\newcommand{\rR}{\rho_R}

\newcommand{\TBBN}{T_\text{BBN}}
\newcommand{\LambdaQCD}{\Lambda_{\text{QCD}}}
\newcommand{\kap}{\kappa}

\newcommand{\Omegastdo}{\Omega_{\rm std}^{-3/2}}
\newcommand{\OmegastdT}{\Omega_{\rm std}^{-7/6}}

\newcommand{\Toscthree}{T_{\rm osc}^{R_3}}
\newcommand{\rRi}{\rho_{Ri}}

\usepackage{pdfpages}
\begin{document}

\includepdf[]{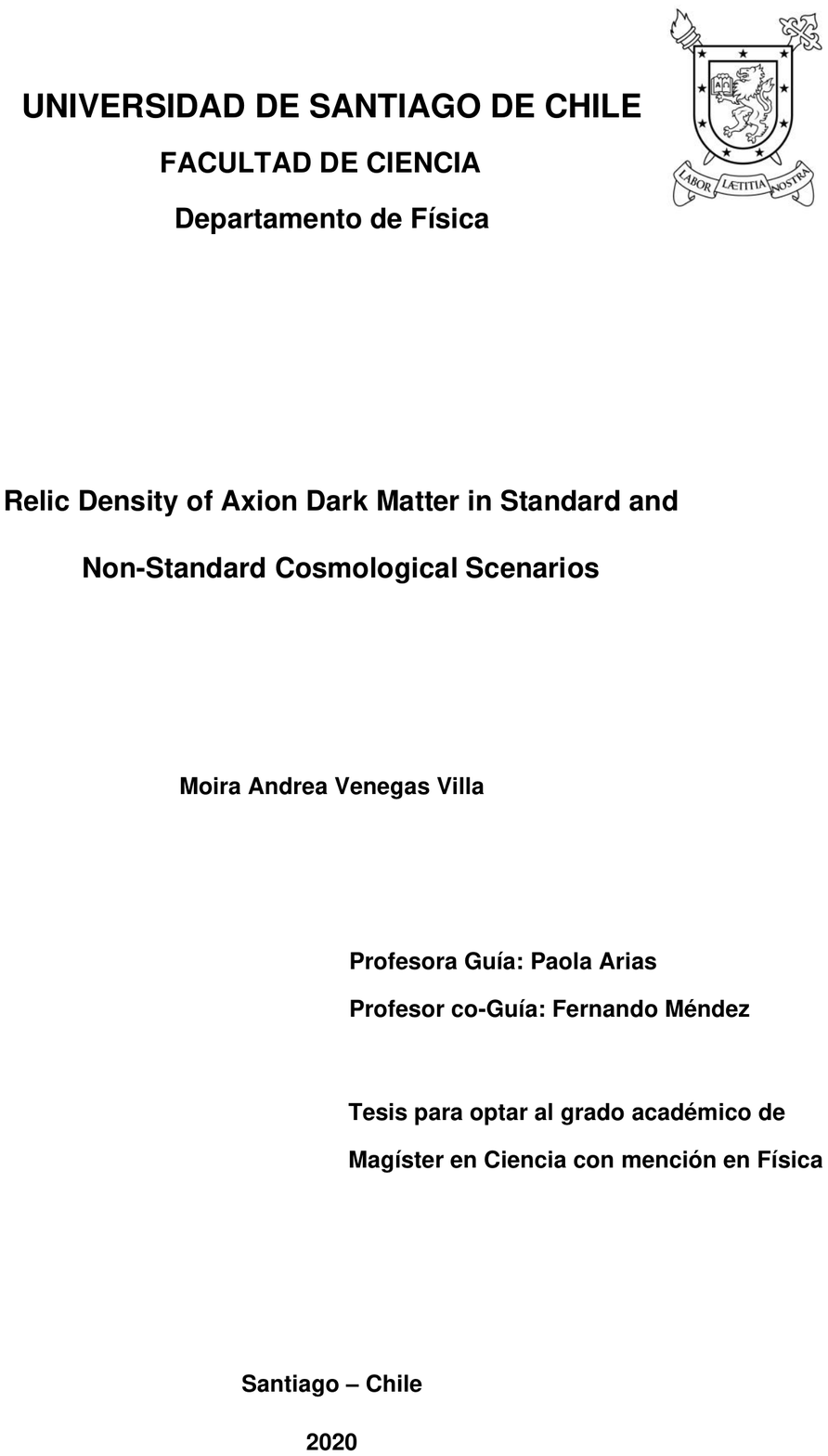}

\newpage
\pagestyle{empty}
\vspace*{\fill}

{\raggedleft\vfill{%
Moira Andrea Venegas Villa,
2021. \\ 
Reconocimiento-No Comercial 4.0 Internacional.
}\par
}


\chapter*{Abstract}
\pagenumbering{roman}
The axion is an hypothetical particle that has become a very serious candidate to explain the nature of cold dark matter. So far, no signal of axions has been found.  Nonetheless, only some experiments have had the right sensitivity to probe the axion parameter space where it could explain the whole dark matter observed today - the so called axion window - and it is expected that a whole new generation  will slowly start to dive into the window. Thus, it is expected during the following years, that either the axion is not found on that parameter space, or that a discovery is made. In the first case, it could happen that that mechanism of production of axions in the early universe is not well understood and new elements have to be considered. In the second case, a discovery will certainly teach us a lot about the first moments of the universe, allowing us to look before the epoch of nucleosynthesis. In this master's thesis we study the production of axion dark matter through the so-called misalignment mechanism by considering that during that time, the universe was dominated by a new kind of fluid, different than radiation. We perform a very detailed analysis of the oscillation temperature and the relic density today, both analytically and numerically. Our findings show that on the one hand, the oscillation temperature is strongly influenced by the non-standard cosmology, affecting the relic density, and on the other hand, the energy density of the axion gets diluted, because the new fluid eventually decays, injecting entropy into the thermal bath. We find the predicted parameter space of axion dark matter for different non-standard cosmologies and we show its impact on the coupling of axions to two photons.

The manuscript is organized as follows, first, we review the standard cosmology and the concepts that will be needed for this thesis. Then, we briefly present the axion as a solution of the strong CP problem and show the non-thermal mechanism of producing axions in the early universe. Then we introduce the non-standard cosmology to be considered in this work, and we proceed to compute the relic density of axion dark matter assuming the oscillation of the field happens in the different possible stages of the new fluid dominance. As a summary of our findings, we show the impact of different non-standard cosmologies on the axion dark matter window and we compare them with the one from standard cosmology. 
\newline
\newline
\textbf{Keywords:} Cosmology, dark matter, axion,  misalignment mechanism.



\chapter*{Resumen}
El axion es una partícula hipotética candidata a explicar la naturaleza de la materia oscura fría. Hasta ahora no se han encontrado señales del axion, no obstante, solo algunos experimentos han tenido la sensibilidad adecuada para sondear el espacio de parámetros del axion capaz de explicar el total de materia oscura observada actualmente, también llamado axion window. Se espera que una generación nueva de experimentos comience a acceder lentamente a esta ventana. Por lo tanto, durante los próximos años es posible que el axion no se encuentre en ese espacio de parámetros, o que se logre descubrir. En el primer caso, podría suceder que el mecanismo de producción de axiones en el universo temprano no ha sido bien entendido y nuevos elementos deben ser considerados. En el segundo caso, su descubrimiento nos enseñará mucho sobre los primeros momentos del universo, permitiéndonos mirar antes de la época de  nucleosíntesis. En esta tesis, estudiamos la producción de materia oscura de tipo axion mediante el misalignment mechanism en un periodo en que el universo está dominado por un nuevo tipo de fluido, diferente a radiación. Desarrollamos un detallado análisis analítico y numérico de la temperatura de oscilación y de la actual densidad reliquia. Nuestros hallazgos muestran que la temperatura de oscilación está fuertemente influenciada por la cosmología no estándar, afectando la densidad reliquia, y además, la densidad de energía del axion se diluye, debido a que el nuevo fluido eventualmente decae e inyecta entropía en el baño térmico. Encontramos que el espacio de parámetros de materia oscura de tipo axion para diferentes cosmologías no estándar y mostramos su impacto en el acoplo de axiones a dos fotones.

En el manuscrito primero revisamos la cosmología estándar. Luego presentamos al axion como una solución del problema de strong CP y mostramos el mecanismo no térmico de producción de axiones en el universo temprano. A continuación, introducimos la cosmología no estándar  y calculamos la densidad reliquia de materia oscura de tipo axion, asumiendo que las oscilaciones del campo suceden en diferentes posibles etapas del dominio del nuevo fluido. Para resumir nuestros resultados, mostramos el impacto de diferentes cosmologías no estándar en la ventana de materia oscura de tipo axion y la comparamos con la obtenida en la cosmología estándar.
\newline
\newline
\noindent\textbf{Palabras claves:} Cosmología, materia oscura, axion,  misalignmnet mechanism.

\chapter*{Acknowledgements}

I would like to express my gratitude to my supervisor, Professor Paola Arias, for her assistance at every stage of the research project, for all her enthusiasm, patience and continuous support during these years. 

I would like to thank Professor Fernando Mendez for his active role in my studies. Also, I wish to extend my special thanks to Professor Jorge Gamboa, for considering me to do my research internship and to Professor Manu Paranjape for his support during my research internship at Université de Montréal.

Finally, I would also like to thank the University of Santiago for the research support grants.

\chapter*{Agradecimientos}

Quisiera expresar mi agradecimiento a mi supervisora, la profesora Paola Arias, por su asistencia en cada etapa del proyecto de investigación, por todo su entusiasmo, paciencia y continuo apoyo durante estos años.

Me gustaria agradecer al profesor Fernando Méndez por su papel activo en mi formación, también deseo extender un especial agradecimiento al profesor Jorge Gamboa, por considerarme para realizar mi pasantía de investigación y al profesor Manu Paranjape por su apoyo durante mi pasantía de investigación en la Université de Montréal.

Agradezco a la Universidad de Santiago por las becas otorgadas de apoyo a la investigación. 

A mis amigos, Kristiansen, Francisco y Ricardo, por las tantas conversaciones que me han motivado a seguir en este camino.

Finalmente, quiero agradecer a mi familia, por el inmenso apoyo que me han brindado desde siempre.

\tableofcontents
\listoffigures 

\pagestyle{fancy}
\chapter{Introduction}
\pagenumbering{arabic}


\section{Dark Matter}

By astrophysical and cosmological observation we know that Dark Matter (DM)  currently corresponds about the 26$\%$ of the energy density of the universe \cite{2020} but its nature is still unknown, resulting in one of the biggest open questions in modern cosmology.

\begin{figure}[ht]
\centering
\includegraphics
[width=8.1 cm,height=6.81 cm]{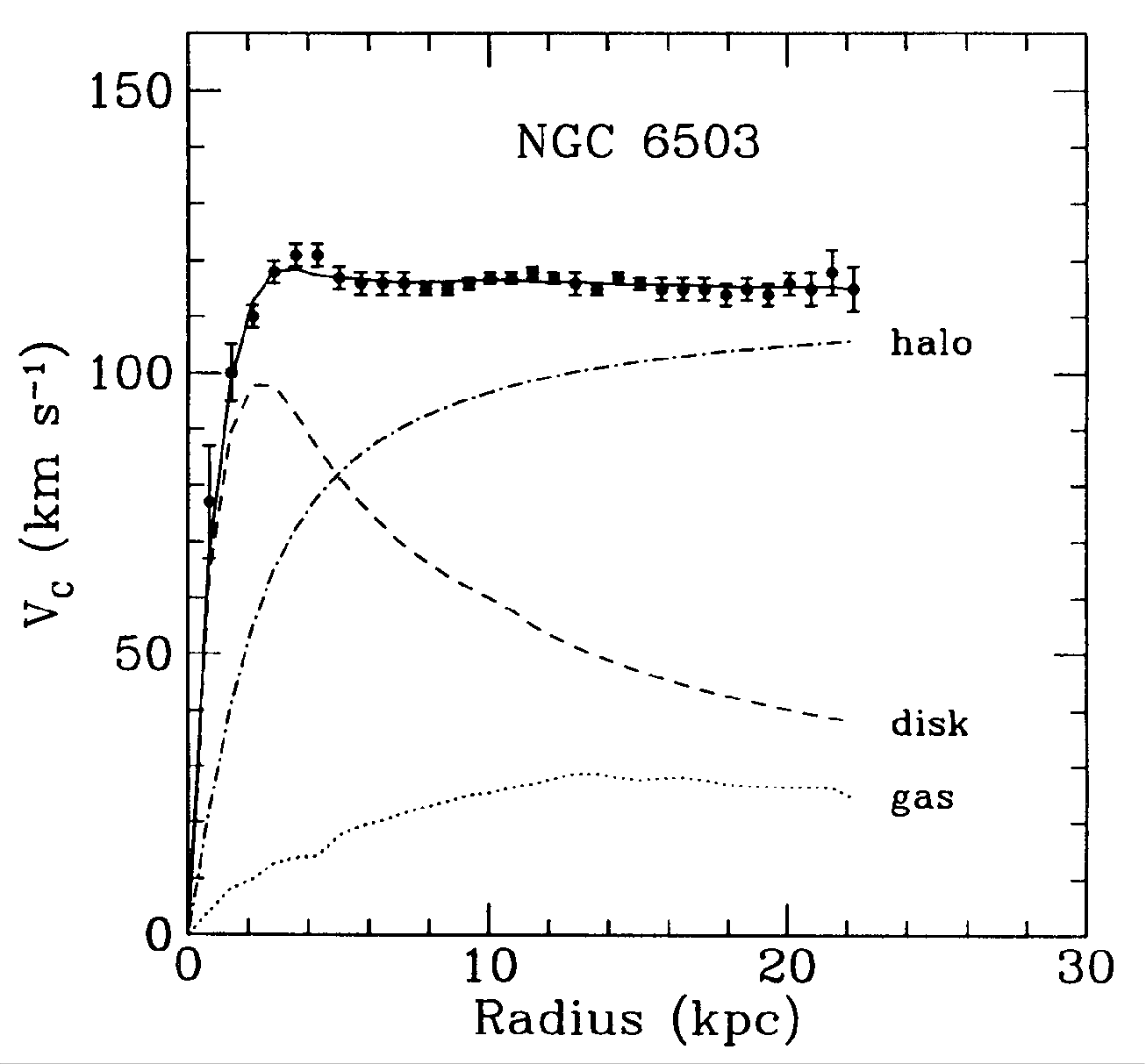}
\caption[Galactic rotation curve]{Galactic rotation curve for NGC 6503  (data from \cite{Freese:2017idy}) showing disk and gas contribution plus the dark
matter halo contribution needed to match the data.
 }
\label{fig:rotation}
\end{figure}

One of the first proofs on the DM existence came from the observation of the rotation curves of galaxies, namely the graph of velocities of stars and gas as a function of their distance from the galactic center. Data revealed that the ordinary matter, also called baryonic matter, only accounts for a very small fraction of the average mass that was expected to be measured, to explain it, galaxies must have enormous dark halos made of nonluminous matter of unknown composition as it is shown in \cref{fig:rotation}, where the velocity profile of galaxy NGC 6503 is displayed as a function of radial distance from the galactic center. The baryonic matter in the gas and disk cannot alone explain the galactic rotation curve. However, adding a DM halo allows a good fit of the data \cite{Freese:2017idy}.

\begin{figure}[ht]
\centering
\includegraphics
[width=12.5 cm,height=8.81 cm]{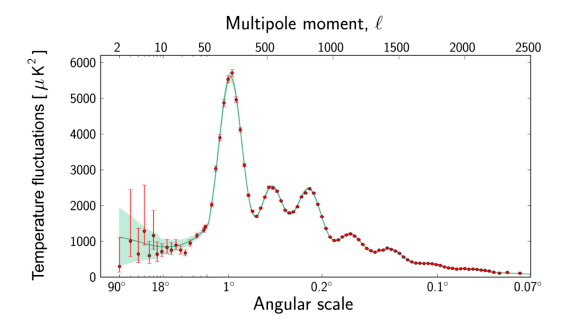}
\caption[ Planck’s power spectrum of temperature fluctuations in the CMB]{Planck’s power spectrum of temperature fluctuations in the CMB.
The fluctuations are shown at different angular scales on the sky. Red dots with error bars are the Planck data. The green curve represents the standard model of cosmology, $\Lambda$CDM. (Figure from \cite{2016}).}
\label{fig:cmb}
\end{figure}

Dark Matter is also needed to explain consistently the precise measurement from the peaks in the temperature anisotropies of the cosmic microwave background (CMB). The CMB is composed of photons which freely propagate to us since the time when atoms were formed at recombination time, therefore encoding information of early events in the universe, among them is the dynamics of the primordial plasma made up of baryons and photons. As the plasma collapses inward by gravity effects, it meets resistance from photon pressure, reversing the plasma direction,  which is called rarefaction. This cycle of compression and rarefaction triggered a distinctive pattern of peaks in the temperature anisotropies of the CMB. By measuring the height of these peaks, is possible to deduce the amount of gravitational matter that was present in the early Universe. 
The measurements \cite{2016} showed that a large extra component is required, that does not correspond to radiation pressure, but contributes to the gravitational wells and therefore further
enhances the compression peaks with respect to the rarefaction peaks. As is
illustrated in \cref{fig:cmb},  a third peak that is boosted to a height comparable to or exceeding the second peak is a sign of a
sizable dark matter component at the time of recombination \cite{Halverson_2002}.

Confirmation of the existence of dark matter has also been found in gravitational lensing, since gravitational energy density can bend light rays sufficiently enough such that a distorted image is obtained by the observer. Thus, data obtained from galaxies clusters required the existence of dark matter to explain the observed light patterns \cite{Massey_2010}.

Regarding to the nature of dark matter, the structure formation in the universe that we observe today, guide us to consider that  the majority of the dark matter observed has to be in the form of cold dark matter, which means that this component has to be non-relativistic at the time of galaxy formation. In contrast, a large amount of hot dark matter involves the fragmentation of structures, which is not in agreement with observations of large-scale structures \cite{ white1983clustering, blumenthal1984formation}. Additionally, DM has not been observed other than gravitationally, therefore, the interaction with the ordinary matter has to be very weak or inexistent. Besides, the DM has to be stable on cosmological times scales.

The fact that the Standard Model of particle physics (SM) is unable to to explain the nature of dark matter drive us to consider new physics beyond SM. Several models for DM have been proposed over the past few decades, we will briefly discuss some of them below.

Primordial black holes (PBH) became attractive DM candidates in the last years, thanks to the discovery of gravitational waves (GW) from the events of black holes with unexpected high masses \cite{Abbott_2016}. Its most popular production mechanism assumes that these compact objects are generated during the period of radiation dominance, due to the collapse of high-density perturbations formed in a stage of inflation in the very early Universe. The mass of a PBH depends
on the amplitude of the fluctuation from which it forms \cite{Niemeyer_1998}.  Through cosmological and astrophysical observations over the last few years, it has been possible to obtain a strict upper limit on PBH abundance for a wide range of PBH masses \cite{Carr_2016},  However,
the total number of detected events is quite small, therefore recent observations show that PBH can not explain the total DM abundance, and at most they can constitute a fraction of dark matter \cite{PhysRevD.96.023514}.

Another potential dark matter candidate are Weakly Interacting Massive Particles (WIMPs). They arise naturally in different theories beyond the Standard Model, some examples are the neutralino in
supersymmetry, or the lightest Kaluza-Klein particle in theories with extra spacetime dimensions  or the Heavy Photon in Little Higgs models. Most of WIMPs models place them in the range of masses around 1 to $10^{5}$ GeV and interaction cross sections from $10^{-51}$ $\rm{cm^2}$ \cite{schumann2019direct}.  Their cross sections are limited by direct DM search limits, the strongest coming from the XENON100 experiment \cite{Aprile_2018} and LUX \cite{Akerib:2016vxi}.

WIMPs are  a very well-motivated candidates for DM, since on the one hand, they are stable thanks to the fact that they have a symmetry that prevents their decay despite their high masses and on the other hand, since they are thermally produced in the early universe, along with being massive and weakly interactive, they decouple from the thermal plasma being non-relativistic and then, with the expansion of the universe, they begin to move with non-relativistic velocities, this naturally yields a CDM scenario. An interesting observation and big motivation to consider WIMPs as DM, corresponds to the “WIMP miracle”, which consist in the fact that a thermal production of WIMPs with a weak-scale cross section give us the correct relic dark matter abundance of $\Omega_{\rm{CDM}} \approx $0.26. 

Finally, the QCD axion is a solid dark matter candidate almost since the time of its proposal
\cite{Peccei:1977hh, Weinberg:1977ma, Wilczek:1977pj}, where soon enough it was realised it can be efficiently produced non-thermally in
the early universe \cite{Preskill:1982cy, Dine:1982ah}. Axions appear as a pseudo-Nambu-Goldstone bosons when the so-called
Peccei-Quinn symmetry (PQS) is spontaneously broken, at some energy $f_a$. As the universe cools down, the axion potential energy changes during the QCD
phase transition epoch, because instanton effects break explicitly the PQS, and it acquires a small mass. During this process, the value of the axion field gets realigned (the process is known as the 'misalignment mechanism'), changing from an arbitrary initial value to the true vacuum value of the field. This process is of extreme relevance, if the axion exists, since on the one hand, it solves the so-called strong CP problem of the QCD sector, and on the other hand, in the process of vacuum realignment \cite{ Dine:1982ah, Abbott:1982af, Arias:2012az, Nelson:2011sf} cold axion particles are produced during the oscillation of the field around the minimum, that contribute to the dark matter relic density of the universe.

Axion DM is the main focus of this thesis and \cref{chap: ACDM} is entirely devoted to them. But first, in the next section we will do a brief review of standard cosmology, focusing on general tools that we require to carry out this thesis.


\section{Overview of Cosmology}
\subsubsection{Thermal history of the universe}

The Friedmann-Lemaître-Robertson-Walker (FLRW) metric describing a homogeneous,
isotropic universe can be written as
\be
ds^2=dt^2-R^2(t)\left(\frac{dr^2}{1-kr^2} + r^2d\theta^2+r^2\sin ^2\theta d\phi^2 \right),
\ee
where $t$ is the physical time, $R(t)$ is the scale factor that describes the relative expansion of the universe and the parenthesis term represents the spatial line element. The parameter $k$ describes the spatial curvature of our universe.

 Solving the Einstein equations with the FRLW metric, we obtain the Friedmann equations (\ref{eq:Friedmann1}) and (\ref{eq:Friedmann2}), that describe the evolution of the universe for any given energy content. 
\begin{align}
&H^2+\frac{k}{R^2}=\frac{1}{3 M_P^2}\,\rho + \frac{\Lambda}{3}  \label{eq:Friedmann1},\\
&\frac{\Ddot{R}}{R}=-\frac{1}{6\,M_P^2}\,(\rho + 3 p),
\label{eq:Friedmann2}
\end{align}
where $M_P= 2.4\times 10^{18}\,\rm{GeV}$ is the reduced Planck mass, $\rho$ and $p$ describe the energy density and the local pressure of the fluid, respectively. The $\Lambda$ term is associated to the dark energy density. Since we are focused in the early universe, its role is negligible and hence we set $\Lambda$ =0. In \cref{eq:Friedmann1} we have introduced the Hubble rate, which is defined as
\be
H= \frac{\dot R}{R},
\label{H}
\ee
here, $\dot R$ represents the temporal derivative of $R$.

In a realistic description of the early universe, there is more than one fluid present in the theory, for that reason we
define the total energy density $\rho_{tot}$ and the total pressure $p_{tot}$ of the system as
\be
\rho_{\rm{tot}}=\sum_{i} \rho_i, \qquad p_{\rm{tot}}=\sum_{i} p_i ,
\ee
here, the subscript  i labels the i-th fluid that we are considering. There are at least three of such forms: radiation, non-relativistic matter and dark energy.

To have a full description of the system, we need a third equation in addition to \cref{eq:Friedmann1} and \cref{eq:Friedmann2}, linking the density and the pressure of each fluid, thus, we introduce the equation of state
\be
p_i= \omega_i \rho_i,
\label{eq:state}
\ee
where the parameter $\omega$ characterizes the fluid. For instance, radiation has $\omega$=1/3, while non relativistic matter is pressureless, i.e. $\omega$=0.

On the other hand, from the first Friedmann equation, we can see that the space is flat ($k = 0$) if the density of the universe equals the critical density $\rho_c$ defined as
\be
\rho_c= 3 M_P^2 H^2.
\ee
Now, we introduce the density parameter as the ratio of the total energy density for some fluid $\rho_i$ to the critical energy density
\be
\Omega_i=\frac{\rho_i}{\rho_c}, \qquad \Omega_{tot}=\frac{ \rho_{tot}}{\rho_c}=\sum_i\Omega_i.
\ee

There are important cosmological parameters characterizing the energy balance in the present Universe.
Their numerical values today are  \cite{2020}
\begin{align}
&H_0 \approx 100\, \rm{h\, km\, s^{-1} Mpc^{-1}},  &\rho_c\approx 3.44\times 10^{-47} \rm{GeV}^{-4},
\end{align}
and
\begin{align}
&\Omega_{\Lambda,0} \approx 0.68, \\ 
&\Omega_{r,0} \approx 10^{-5} ,\\
&\Omega_{m,0}
\approx 0.31,   
\end{align}
 about $\Omega_{m,0}$, we know that non-relativistic matter consists of baryons and dark matter, the current contributions of each one is 
\begin{align}
\Omega_{b,0} \approx 0.04, \quad \Omega_{DM,0} \approx 0.26,
\end{align}
here and throughout this thesis, we indicate the values of
parameters at the present time with the index 0.

The current model of the universe predicts that it is critically flat at large scales
and is verified by various simulations and experiments \cite{2020},  this leads  \cref{eq:Friedmann1} to become
\be
H^2=\frac{1}{3 M_P^2}\rho.
\label{eq:Hrad}
\ee

From the Friedmann equations, we can derive the continuity equation, which give us the evolution of energy density in an expanding universe and is written as 
\begin{align}
    \dot\rho +3 H(\rho +p )=0.
\end{align}

By plugging \cref{eq:state} and \cref{eq:Hrad} into the continuity equation we can obtain the evolution of the energy density as a function of the scale factor,
\be
\rho=\rho_0\left( \frac{R_0}{R}\right)^{3(1+\omega)}.
\label{eq:generalrho}
\ee

\begin{figure}[ht]
\centering
\includegraphics[scale=0.6]{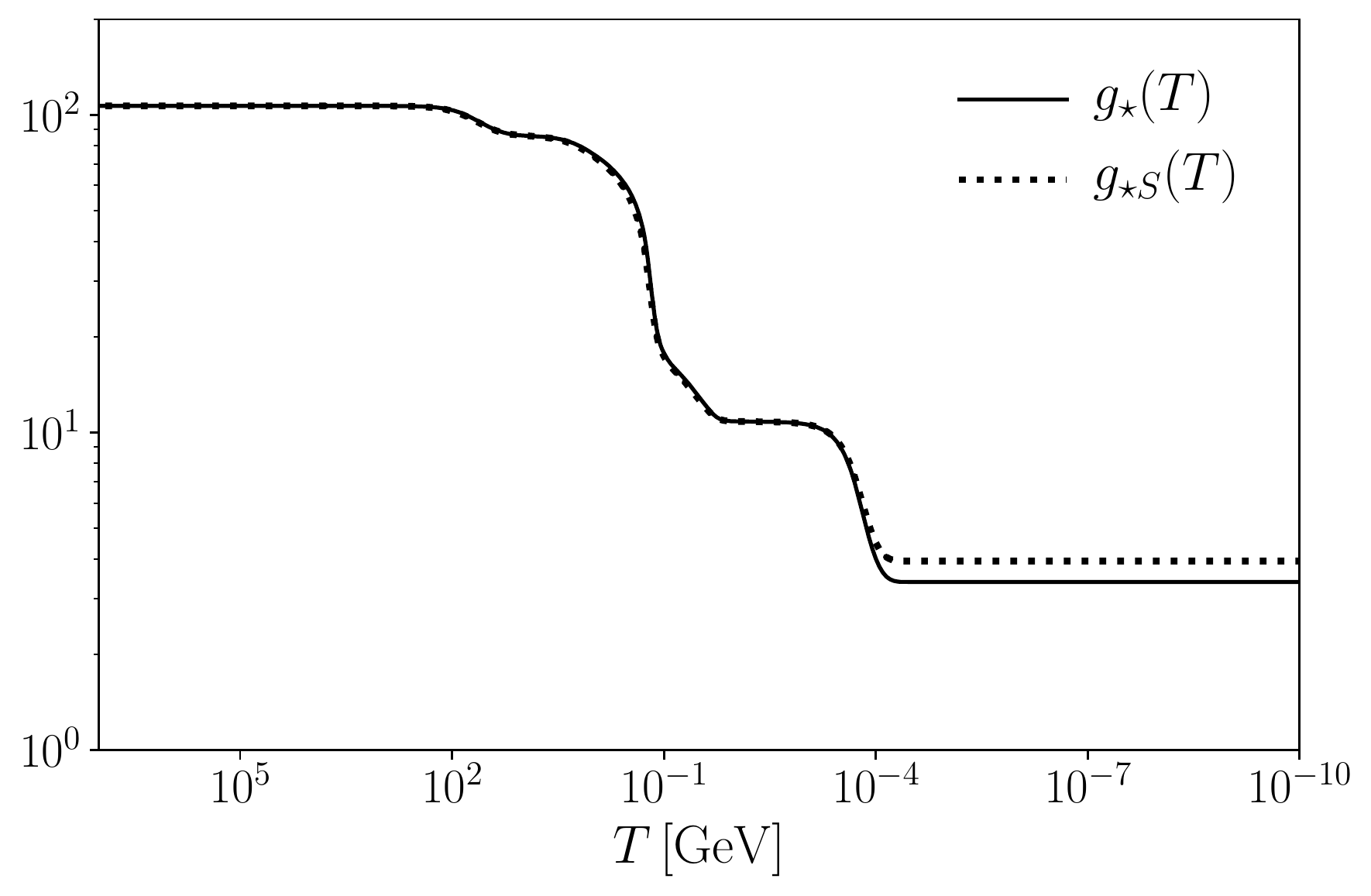}
\caption[Evolution of degrees of freedom]{Evolution of the degrees of freedom in the Standard Model with the temperature.}
\label{fig:dof}
\end{figure}

We now turn our attention to a thermal bath of particles. The energy density of relativistic particles is defined as
\be
\rho=\gs(T)\frac{\pi^2 \,T^4}{30},
\label{eq:rhora}
\ee
where $\gs(T)$  counts the number of relativistic degrees of freedom. The  number of relativistic degrees of freedom at  temperature $T$ for species $i$ that decouple from the thermal
equilibrium at temperature $T_i$ can be defined as
\be
\gs=\sum_{\rm{bosons}} g_i \left( \frac{T_i}{T}\right)^4 + \frac{7}{8} \sum_{\rm{fermions}} g_i \left( \frac{T_i}{T}\right)^4.
\ee
 Fig.\ref{fig:dof} illustrates the evolution of the relativistic degrees of freedom with the temperature. For $T>1$ TeV all the Standard Model degrees of freedom are relativistic and in equilibrium and $\gs$ = 106.75. For $T < 1$ MeV, the only
relativistic species are the photons and the three neutrinos, and $\gs$= 3.36. 

On the other hand, at early times the universe evolves adiabatically so the comoving entropy density $S$ remains constant, we can write it as
\be
d(s R^3)=0,
\label{eq:sconserv}
\ee
with $s$ the entropy density, defined by
\be
s=\frac{\rho + p}{T}=\frac{2\pi^2}{45}\gss(T)T^3,
\label{eq:entropy}
\ee
here, $g_{*S}$ counts the number of relativistic degrees of freedom that contribute to the entropy and are given by
\be
\gss=\sum_{\rm{i=bosons}} g_i \left( \frac{T_i}{T}\right)^3 + \frac{7}{8} \sum_{\rm{i=fermions}} g_i \left( \frac{T_i}{T}\right)^3
\ee

We can rewrite \cref{eq:sconserv} as \cref{eq:s2} to get the evolution of the temperature in the universe
\begin{equation}
\frac{ds}{dt}+3Hs=0.
\label{eq:s2}
\end{equation}

 The above equation can be cast in favour of the scale factor $R$ instead of time as
\begin{equation}
\frac{dT}{dR}=\left(1+\frac{T}{3\,\gss}\frac{d\gss}{dT}\right)^{-1}\left(-\frac{T}{R} \right), \label{eq:dTdR}
\end{equation}
which gives us the relation $T=T(R)$.

Finally, if we focus on a universe dominated by radiation, the expansion rate is 
\be
H(T)=\sqrt{\frac{\pi^2 \gss(T)}{30}\frac{T^4}{3M_P^2}},
\label{eq:hubble_std}\ee
then, from \cref{eq:generalrho} and \cref{eq:rhora} along with considering that the degrees of freedom remain constant, we realize that $T\propto R^{-1}$ and we get the temporal derivative of the Hubble parameter in this period as
\be
\frac{dH}{dt}= R\,H \frac{dH}{dR}= -2 H^2.
\label{eq:Hdotrad}
\ee

\newpage 
\section{Big Bang Nucleosynthesis}

One of the stringent validity test of the standard model of cosmology ($\Lambda$CDM) corresponds to  Big Bang Nucleosynthesis (BBN). This period occurs during the radiation-dominant epoch, with a typical temperature scale of $\mathcal{O}(1)$ MeV. At that scale, due to the cooling the universe generates by its expansion,  radiation is not energetic enough  to significantly break the bonds in the strong sector, such that hadrons  started to  participate in efficient fusion reactions giving way  to the creation of the first light nuclei, including deuterium (D), helium-3 ($^3$He), and helium-4 ($^4$He). This chemical element synthesis prediction is in agreement with the primordial abundances inferred from observational data acquired from, for instance, the absorption lines of ionized hydrogen region in compact blue galaxies \cite{Cooke_2018} and the spectra of metal-poor main-sequence stars \cite{Sbordone_2010}. This matching with very high-accuracy is taken as a standard cosmology model backup and therefore provides powerful constraints on possible deviations from it.
According to the above, BBN corresponds to the earliest moment of the universe that we have direct evidence, any prediction about previous history is still speculation.


\chapter{Axions}
\section{The strong CP problem}
\label{sec:SCP}
In quantum chromodynamics (QCD), the theory of strong interactions, the nontrivial topological vacuum structure generates a CP-violating term, such that we can include it in the QCD Lagrangian as 

\be
\mathcal{L}_{\rm{QCD}+\Bar{\theta}}= \mathcal{L}_{\rm{QCD}} +  \frac{g_s^2}{32\pi^2}\Bar{\theta} \,G_{\mu\nu}\tilde G^{\mu\nu},
\ee
where $G_{\mu \nu}$ corresponds to the gluon field strength tensor and $\tilde G_{\mu \nu}$ is its dual, $g_s$ is the QCD gauge coupling constant and $\Bar{\theta}$ is a combination of two parameters
\be
\Bar{\theta}=\theta+\theta_{\rm{weak}}
\label{eq:thet1},
\ee
the parameter $\theta_{\rm{weak}}$ appears when the electroweak interactions are considered and we can express it in terms of the quarks mass matrix M as
\be
\theta_{\rm{weak}}=\rm{arg}\left( \rm{Det\, M} \right),
\ee
while the term $\theta$ comes from the non trivial description of the QCD vacuum.

The $\Bar{\theta}$ term violates P and CP symmetries and gives rise to a non-vanishing neutron electric dipole moment (NEDM)\cite{RevModPhys.82.557}
\be
d_n\approx 4.5\times 10^{-15}\,\theta\,\mbox{e\,cm}, 
\label{eq:dipol1}
\ee
here e is the electron charge. By no-observation of NEDM in experiments is possible to constrain its value with a tight upper bound \cite{PhysRevLett.97.131801}
\be |d_n|< 2.9 \times 10^{-26}\, \mbox{e\,cm} \qquad \text{at 90$\%$ C.L}.
\label{eq:dipol2}
\ee
which translates into an upper bound for $\Bar{\theta}$
\be
\Bar{\theta} <0.7\times10^{-11}.
\label{eq:thet2}
\ee
From \cref{eq:thet1} and \cref{eq:thet2}, we can derive that $\theta_{\rm{weak}}$ and $\theta$ must to cancel with high precision. That is of course not forbidden, but these two quantities have a physical origin completely unrelated and there is no natural explanation for the extreme cancellation of them. Consequently, the strong CP problem it is considered as a fine-tuning problem.

\section{Axions as a solution of the strong CP problem}
Three main solutions were proposed to explain the strong CP problem: one simple alternative is to consider a massless up-quark \cite{Banks:1994yg}, another option is to spontaneously break the CP symmetry, which would allow setting $\Bar{\theta}=0$, and the third is the Peccei Quinn  mechanism, which  introduces a  new particle, the axion. About the first proposal, if the up-quark is massless (actually is the lightest quark) is possible to make the term $\mathcal{L}_{\Bar{\theta}}$ disappear by a rotation of the quark fields, such that $\Bar{\theta}$ is unobservable. But by computing the topological mass distribution with Lattice QCD, an inconsistency appears when considering a massless  up quark \cite{Alexandrou_2020}. On the other hand, in a model which spontaneously breaks the CP symmetry, one can set $\Bar{\theta}= 0$ at the Lagrangian level. However, if CP is spontaneously broken $\Bar{\theta}$ gets induced back at the loop$-$level \cite{peccei2008strong}. To get $\Bar{\theta} < 10^{-11}$  one needs, in general, to ensure that $\Bar{\theta}$ vanishes also at the 1-loop level. The major drawbacks of this solution are that their models are quite complex \cite{dine2015challenges, vecchi2014spontaneous,}, in addition to the fact that the experimental data is in excellent agreement with the CKM Model- a model where CP is explicitly broken \cite{peccei2008strong}.

The third and the most popular explanation of the smallness of $\Bar{\theta}$ corresponds to the PQ mechanism which involves an hypothetical particle called the axion \cite{Peccei:1977hh, Weinberg:1977ma, Wilczek:1977pj}. The main idea of the solution consists in promoting the $\theta$ parameter to a dynamical field and then via QCD non-perturbative effects (instantons) the new field relaxes its expectation value towards zero. 

The PQ mechanism is based on the introduction of an additional global $U(1)$ symmetry. This symmetry, called the Peccei-Quinn symmetry or $U(1)_{PQ}$, is spontaneously broken at some energy $f_a \gg \mathnormal{v}_{EW}$= 247 GeV. As a consequence of the spontaneous breaking, the axion $a(x)$ emerges as a Goldstone boson. In addition, the $U(1)_{PQ}$ is also explicitly broken by 
instanton effects, meaning that the axion field acquires the anomalous
coupling to gluons  and a residual $U(1)_{PQ}$ symmetry acts as a shift symmetry on the axion field. 

The Lagrangian of the theory has then the form
\begin{align}
\mathcal{L}_{QCD}&= \mathcal{L}_{QCD} + \mathcal{L}_{\Bar{\theta}}+\mathcal{L}_{a} \\
&= \mathcal{L}_{QCD} +  \frac{g_s^2}{32\pi^2}\Bar{\theta} \,G_{\mu\nu}\tilde G^{\mu\nu} +  \frac{g_s^2}{32\pi^2}\, \frac{a}{f_a} \,G_{\mu\nu}\tilde G^{\mu\nu} -  \frac{1}{2}\partial_\mu a \partial^\mu a + ...\,,
\label{eq:lqcd2}
\end{align}
where dots account for other possible interaction terms.

The effective potential for the axion field has a minimum at $\langle a \rangle\,=\,- f_a \Bar{\theta}$. Thus, the $\Bar{\theta}$ CP violating term can be absorbed into the axion field, defining the physical axion field 
\be
a_{{\rm{phys}}}= \frac{a}{f_a} +\Bar{\theta}. 
\ee

Therefore, the $\Bar{\theta}$ parameter has
been promoted to a dynamical field that evolves to its CP-conserving minimum $\langle a_{{\rm{phys}}}\rangle$ = 0.

For simplicity, we introduce a dimensionless field, the misalignment angle 
\be
\theta=\frac{a}{f_a},
\label{eq:misal}
\ee
here we have rewritten $a_{{\rm{phys}}}$ as $a$.

In spite of the fact that the axion appears as a Goldstone boson, it gets a non-zero mass from the QCD anomaly. The axion mass is temperature dependent and is inversely related with the decay constant $f_a$ by
\be
m_a^2(T)=\frac{\chi(T)}{f_a^2},
\ee
where $\chi(T)$ is the topological susceptibility in QCD, which has
been evaluated in the chiral limit \cite{Crewther:1977ce,DiVecchia:1980yfw}, next-to-next-to leading order in chiral perturbation
theory \cite{Gorghetto:2018ocs}, and directly via QCD lattice simulations \cite{Borsanyi:2016ksw} and they all coincide in the central
limit of $\chi$ 
\be
m_0\,=\, 5.69 \frac{10^{9}\,\rm{GeV}}{f_a}\, \rm{meV}
\ee
with a zero-temperature value of $\chi(0)\equiv \chi_0=0.0245$~fm$^{-4}$, in the symmetric isospin case estimated from Lattice QCD simulations. 
 
For our numerical calculations we will use the results of ref.\cite{Borsanyi:2016ksw} for the axion mass, but for analytical estimations, we will use an approximate expression, which has to be cut off by hand once the mass reaches the zero-temperature value, we can summarize it  that as \cite{Hertzberg_2008}
\be
m_a(T)= \begin{cases}
\displaystyle \alpha\,
m_0\left(\frac{T}{\Lambda_{QCD}}\right)^{-4}, & T\gtrsim \Lambda_{QCD} \\
\displaystyle 
\quad m_0, & T\lesssim \Lambda_{QCD} \\

\end{cases}
\label{eq:thermal_mass}
\ee
with $\Lambda_{QCD}\sim 400$~MeV, $\alpha=0.02$ and is related to the ratio of the topological susceptibility at two different temperatures.

The axion mass is a free parameter in the QCD axion theory and in \cref{fig:stdgay} we show some astrophysical and cosmological constrains
on a wide range of the Peccei-Quinn axion mass in combination with experimental searches. Some of these restrictions will be reviewed in \cref{chap: bound}.

\section{Axion couplings}
The interaction between axion and photons is described by the term
\be
\mathcal{L}_{a\gamma}=\frac{g_{a\gamma}}{4}a F^{\mu\nu} \tilde{F}_{\mu \nu}= g_{a\gamma}\,a\,\text{\textbf{E}}\cdot \text{\textbf{B}},
\ee
where $F^{\mu \nu}$ is the photon field strength, $\tilde{F}_{\mu \nu}$ is its dual field , \textbf{E} and \textbf{B} are the electric and magnetic fields respectively, $a$ is the axion field and $g_{a\gamma}$ is the axion-photon coupling, which is parametrized by \cite{Kim_1998}
\be
g_{a\gamma}=\frac{\alpha}{2\pi f_a} \left(\frac{E}{A} -1.92\right),
\label{eq:gay}
\ee
here $\alpha=\frac{e^2}{4\pi}$ is the fine structure constant,  $A$ is the color anomaly related to QCD, while $E$ is the electromagnetic anomaly. The ratio $E/A$ is a model dependent factor and the most popular models will be described in the next subsection.

The axion-photon coupling is generic to all the models and most of axion searches strategies are based on this interaction. In general, a broad
range of its values is possible as indicated by the diagonal orange band in \cref{fig:stdgay}. 

Depending on the model, axion also could have interactions with fermions, writing the interaction Lagrangian as
\be
\mathcal{L}_{af}=-i\frac{g_{af}}{2 m_f}\Bar{\Psi}_f\, \gamma_5\Psi_f\,a,
\ee
where $\Psi_f$ is the fermion field, $m_f$ the fermion mass and $g_{af}$ the axion-fermion coupling constant, given by
\be
g_{af}=\frac{C_f\,m_f}{f_a},
\label{eq:gaf}
\ee
here $C_f$ is a model-dependent
parameter that gives the PQ charge of the fermion $\Psi_f$.

\section{Axion models}
\subsection{ KSVZ model}
The KSVZ model was proposed by Kim \cite{kim1979weak} and by Shifman, Vainshtein and Zakharov \cite{shifman1980can}, in this model, the axion does not couple with SM fermions at tree level, implying $C_f$ = 0 in \cref{eq:gaf}. Axion KSVZ has a coupling at tree level to gluons and to an exotic heavy quark $Q$ being the unique particle which carries
PQ charge in the KSVZ model, while the interaction with other SM particles occurs at one loop where the gluons and the new $Q$ quark act as mediators.
Depending on the charge of $Q$, the ratio $E/N$ in \cref{eq:gay} changes, The most common value used in the model is $E/N=0$ as long as the electric charge of the new heavy quark is taken to vanish.

\subsection{ DFSZ model}
In the DSFZ model \cite{dine1981simple} the fundamental fermions
carry PQ charge and no exotic quark is needed. The DFSZ axion couples at tree level to SM photons and charged leptons, besides nucleons. This model predicts $E/N= 8/3$ and the model also establishes the value of the axion-electron coupling $C_e$  as
\be
C_e= \frac{\cos^2 \beta_H}{3},
\ee
with cot $\beta_H$ the ratio of two Higgs vacuum expectation values from this model.
\section{Astrophysical bounds and axion searches}
\label{chap: bound}

\subsection{Astrophysical bounds}
Stars and other astrophysical objects represent an optimal environment for the production of light and weakly-coupled particles such as axions. The presence of axions would add new emission channels in the well-studied astrophysical processes and therefore give us possible restrictions on the parameters of the axions. Some of these restrictions in astrophysical environments will be briefly described below.

\subsubsection{Axions from the Sun}
 In the solar plasma a photon can convert into an axion
if it interacts with an electromagnetic field generated by the charged particles in the
plasma, this is the so called Primakoff
effect. A restrictive bound on $g_{a\gamma}$ is derived from
the fact that
the emission of axions from the Sun implies an increment of the nuclear burning
and, consequently, a redistribution of the solar temperature, leading to an increase in
the $^8B$ neutrino flux. In \cite{Gondolo:2008dd} has been used
the neutrino  flux measurements of the SNO (Sudbury Neutrino Observatory) to obtain the limit $g_{a\gamma}\leq 7 \times 10^{-10}\,\rm{GeV}^{-1}$. Also, the axion-electron coupling has been
constrained to $g_{ae}\leq 2.5 \times 10^{-11}\,\rm{GeV}^{-1}$ by the Bremsstrahlung collisions.

\subsubsection{Globular clusters}
A globular cluster is a gravitationally-bound ensemble of stars which formed at about the same epoch. Two different types of stars in globular clusters are especially interesting for the axion bounds, the horizontal branch (HB) stars and the red giant branch (RGB). When axions are included in the model, helium-burning stars may consume helium faster than expected rate due to the new energy loss channel, which is negligible for RGB stars.
So the population of RGB stars should be relatively larger than the population of HB stars inside the globular clusters. By computing the observed population of HB and RGB stars \cite{Raffelt:1985nk}, the axion-photon coupling is constrained to $g_{a\gamma}< 10^{-10}\rm{GeV^{-1}}$, it translates, for instance, for the DFSZ model in $f_a \gtrsim 10^{7}$ GeV or its equivalent  $m_0 \lesssim 0.5$ eV . 

\begin{figure}[h]
\centering
\includegraphics[width=14.1 cm,height=8.81 cm]{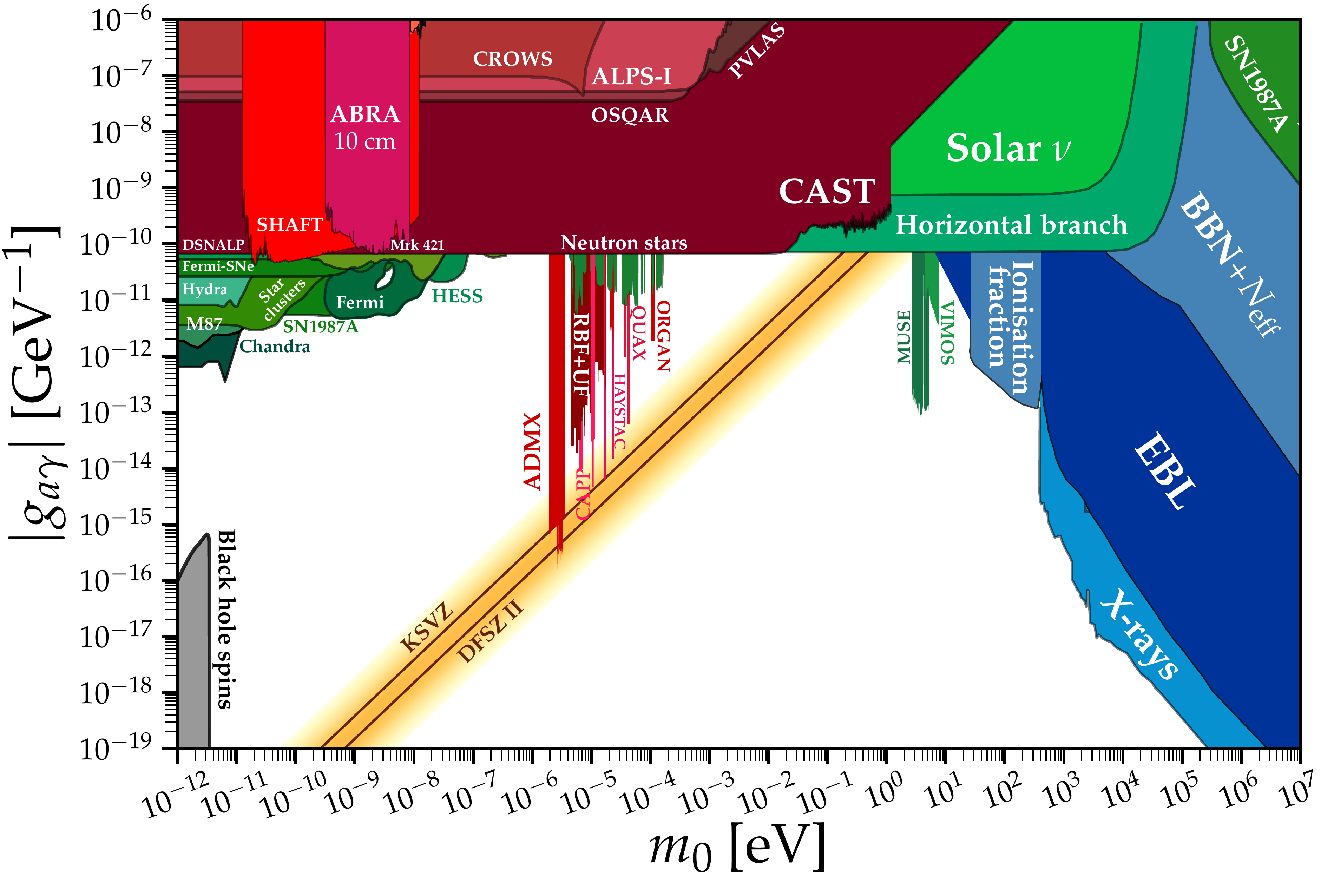}
\caption[Axions parameter space for the axion-photon coupling]{Axions parameter space for the axion-photon coupling, the most sensitive experiments are shown: laboratory experiments (ALPS), helioscopes (CAST), haloscopes (ADMX) and telescopes.
Also the most significant constrains from horizontal branch (HB) stars are drawn. The yellow band represents the most favored axion models. (Image from \url{ https://github.com/cajohare/AxionLimits}.) \label{fig:stdgay}}

\end{figure}

\subsubsection{White dwarfs }
White dwarfs (WD) are a remnant of initial low massive stars. A WD is very hot when it is
formed, but since it has no energy source it gradually cools down due to neutrino emission and later by surface photon emission. The possible emission of axions by axion-Bremsstrahlung, could imply an increase in their cooling rate. The axion emission can be constrained by comparing the observed
cooling speed with WD models. The most restrictive limit on the axion-electron coupling is $g_{ae}< 10^{-13}\,\rm{GeV^{-1}}$ \cite{Raffelt:1987yb}. Whereas the axion-photon coupling is constrained
 using the amount of linear polarization in the radiation emerging from magnetic white dwarfs. This method sets an upper limit on $g_{a\gamma}$
 that depends on the axion mass $m_0$. In ref. \cite{gill2011constraining}, they derived the bound $g_{a\gamma}< 10^{-10}\,\rm{GeV^{-1}}$ for
$m_0\leq 10^ {- 4}$ eV.

\subsection{Axion searches}

\subsubsection{Helioscopes}
 Several helioscopes, like CAST are used to detect the flux of solar axions. These helioscopes are essentially vacuum pipes, where the principle of
detection is the use of a strong magnetic field to convert solar axions back into photons. The requirement that the total luminosity of the axion be less than the known solar luminosity leads to a limit for the axion$-$photon coupling and  axion$-$electron coupling. CAST currently gives the strongest constraint for $g_{a\gamma}$ \cite{Anastassopoulos:2017ftl}, which lead to an upper limit,  $g_{a\gamma} < 0.66\times 10^{-10}\,\rm{GeV^{-1}}$
for axion mass values below $m_0 < 0.02$ eV .

\subsubsection{Haloscope}

Haloscope technique consist in the detection of non relativistic axions by a resonant cavity.  In such cavity, a strong electromagnetic field is produced, with a frequency related to the size of the cavity. There exists a narrow range
of the axion mass for which the axion would interact with the electromagnetic field and
convert into a light pulse which would be eventually detected. The most sensitive  axion haloscope currently is the Axion Dark Matter eXperiment (ADMX) \cite{Asztalos:2009yp}, where the region excluded  so far is $1.9\,{\rm{\mu eV}} < m_0 < 3.53\, {\rm{\mu eV}}$.

In fig.\ref{fig:stdgay}, we show the most sensitive experiments with the current limits on the photon coupling of axions. Also, the most relevant constrains and hints by astrophysical observations are delineate.

\chapter{Axions as Cold Dark Matter}
\label{chap: ACDM}

Previously, we wrote that the spontaneous symmetry breaking (SSB) of the PQ symmetry occurs when the temperature falls below the critical temperature, $T_{PQ} \sim f_a$. It is crucial to define whether the SSB occurs before or after inflation, because if the SSB occurs in a post-inflation era, in addition to having the misalignment mechanism (\cref{sec: misalig}), which happens at temperature $\Lambda_{QCD}<f_a$, an extra production of non-thermal axions appears.

Let us introduce the Gibbons-Hawking
temperature $T_{GH} = H_I/2\pi$ \cite{PhysRevD.15.2738}, here the expansion rate at the end of inflation $H_I$ has an upper bound by PLANCK measurements \cite{visinelli2011axions}
\be
H_I\leq 6\times 10^{14}\,\rm{GeV},
\ee
while a lower limit on $H_I$ comes from requiring the Universe to be radiation-dominated at $T\simeq 4$ MeV, so that primordial nucleosynthesis can take place \cite{hannestad2004lowest}. This temperature constraints  the smallest allowed reheating temperature by inflation, such that
\be
H_I\geq 7.2\times 10^{-24}\,\rm{GeV}.
\ee

Therefore, there are two scenarios that can be distinguished in the axion cosmological history
\begin{itemize}
\item Scenario A: SSB occurs after inflation ends ($f_a \leq \frac{H_I}{2\pi}$).
\item Scenario B: SSB occurs before/during  inflation ends ($f_a \geq \frac{H_I}{2\pi}$).
\end{itemize}

If the PQ symmetry breaks after inflation, the axion field is very inhomogeneous such that the universe consists of many patches causally disconnected. Under this condition, topological defects are present in the early universe, like axion strings, and they are an important source of cold relic axions, produced by their decay. The patches in the universe have different initial expectation values for the axion field, 
and as an initial condition it is usual to take the average contribution over a Hubble horizon, corresponding to $\theta_i=\pi/\sqrt{3}$.   The preferred mass range for axion cold DM in this scenario, is $5\times 10^{- 6}\,{\rm{eV}} < m_0 <10^{- 2}\,{\rm{eV}}$ \cite{2013production,Hertzberg:2008wr}, also referred as     “classic axion window", where astrophysical observations
give the upper bound, while the lower bound for the axion mass comes from imposing that the axion dark matter does not overclose the universe.

If the axion field is already present during inflation, it can produce isocurvature fluctuations due to quantum effects. By the restrictions in this type of fluctuations obtained from  CMB observations along with WMAP-7 + BAO + SN data \cite{wmap2011seven}, we can obtain a bound in the parameters of the axion related to the inflationary scale $H_I$. Considering that axions give the dominant contribution for cold dark matter abundance ($\Omega_{CDM} = \Omega_{a}$), the bound corresponds to
\be
H_I < 4.3\times 10^{-5} \theta_i f_a.
\ee

On the other hand, if the PQ symmetry breaks
before or during inflation (and it is not restored afterwards), due the fast expansion, the axion field becomes homogeneous, consequently, axionic strings are not present.
 In this scenario, the initial expectation
value of the field is arbitrary and gets homogenised
during inflation, such that the observable universe has just one value of $\theta_i$. In sceneario B, the axion CDM spot is bounded from below from stellar cooling and supernovae bounds, and from above by isocurvature perturbations. By assuming an initial misalignment angle of $\theta_i~1$, it is required $f_a\sim 10^{11}$~GeV, translated into $m_0\sim \mu$eV. Since $\theta_i$ is a random parameter, the value of $f_a$ can be pushed to the boundary of the isocurvature bound, but to the price of fine-tuning the $\theta_i$ angle.  In this scenario, values of $f_a\gtrsim  10^{15}~$GeV - equivalent to $m_0\sim 10^{-4}\mu$eV -  need to consider $\theta_i\lesssim 10^{-2}$ and are considered fine-tuned or anthropic.

The contribution of axions produced by decays of topological defects is so far still under discussion, where the main discrepancy comes from the rate at which the axions are emitted from them. Due to these uncertainties, in this thesis we only focus on studying the production of axions in a non-standard cosmology through the misalignment mechanism. Even so, for the sake of completeness, in \cref{sec:string}, we include a brief overview of axion production from topological string. For more details about it, see \cite{sikivie2008axion, harari1987evolution,visinelli2011axions} and in \cite{ramberg2019probing} the contribution from the decay of an axionic string network was performed in a  a scan of cosmological histories.

\section{Axions from the misalignment mechanism}
\label{sec: misalig}

After the spontaneous breaking of the PQ symmetry, the axion field acquires a residual potential with the form 
\be
V(a)= m_a^2(T)f_a^2\left(1-{\rm{cos}} \frac{a}{f_a} \right).
\ee
The axion Lagrangian density is given by

\be
\mathcal{L}=\frac{1}{2}\partial_\mu a\partial^\mu a- m_a^2(t)f_a^2\left(1-\cos\frac{a}{f_a}\right).
\ee
It is convenient to work with the dimensionless field $\theta(x) \equiv a(x)/f_a$ introduced in \cref{eq:misal} and the equation of motion considering a FRW metric corresponds to
\be
\ddot{ \theta} + 3H(t)\dot{\theta}-\frac{
1}{R^2(t)}\nabla_x^2\theta+m_a^2(T)\sin\theta=0.
\ee

Literature \cite{Sikivie_2008,2013production}  has widely studied that the axion contribution to dark matter comes from the zero momentum axion modes. For that reason, from now on we consider  $\nabla^2_{\Vec{x}}\,a =0$ and the equation of motion become

\be
\ddot{\theta} + 3H(t)\dot{\theta}+m_a^2(T)\sin\theta=0.
\label{eq:eof}
\ee

In the early universe, when the temperature is $T\gg\Lambda_{QCD}$, the axion mass is negligible compared to the friction term from the Hubble expansion, so the solution for \cref{eq:eof} is $\theta=\theta_i$, with $\theta_i$ a constant, referred as the initial misalignment angle. When the temperature drops off and become $T\sim \Lambda_{QCD}$ the axion mass term gets important, such that $3H\approx m_a$. This misaligns the axion potential and  the axion begins to rolls toward the true  minimum of the potential,  corresponding to $\langle \theta\rangle=0$. 

\begin{figure}[t]
\centering
\includegraphics[scale=0.6]{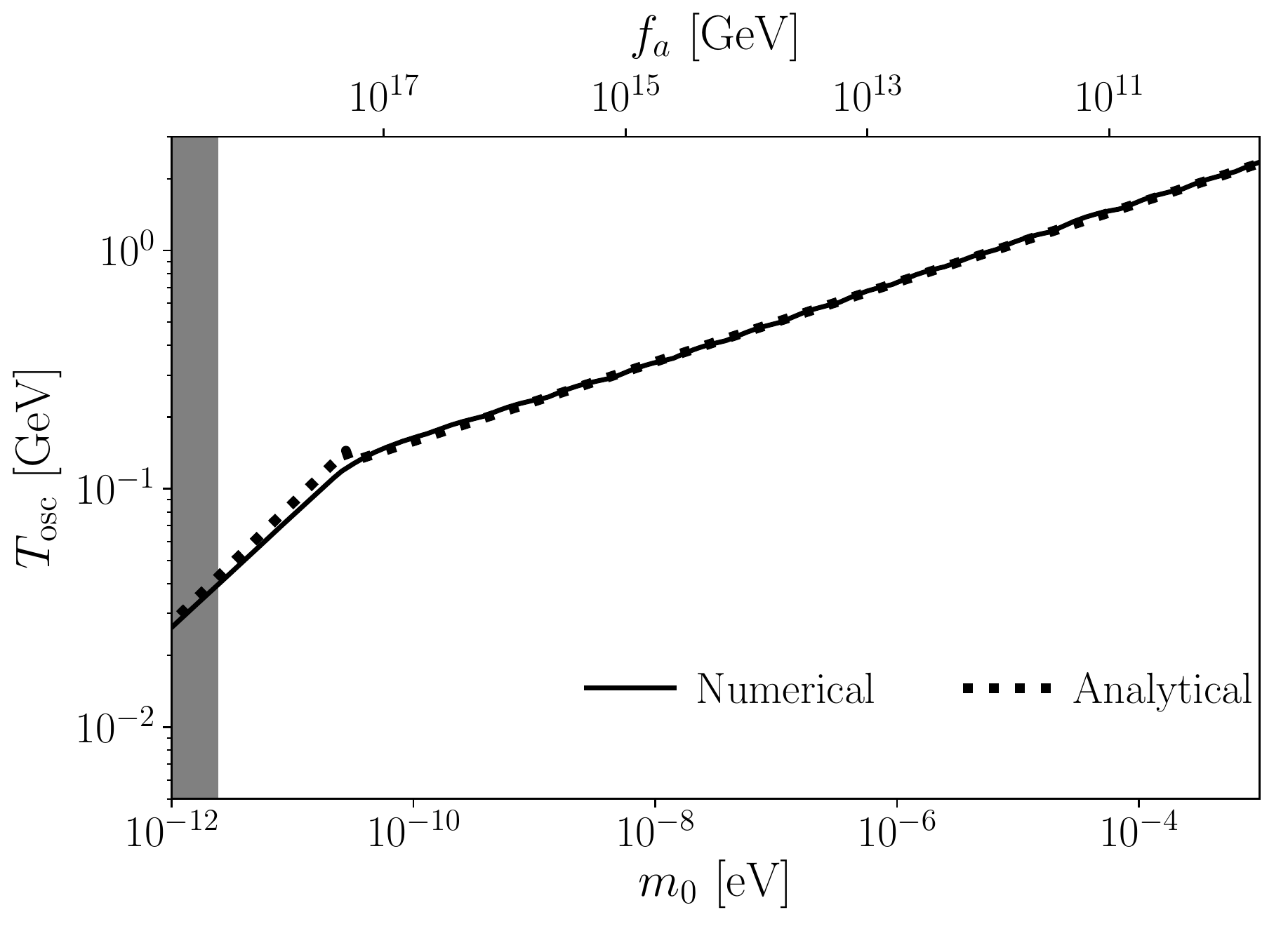}
\caption[Oscillation temperature vs axion mass]{Oscillation temperature vs axion mass. Black line is for the numerical solution and dotted black line is for analytical solution The bend around $m_0\sim10^{-5}{\rm{\mu eV}}$ corresponds to the QCD transition. The gray area corresponds to scales over the Planck scale. }
\label{fig:std_tosc}
\end{figure}

We introduce the oscillation temperature $\Tosc$, which is defined as the temperature at which the oscillations start  and is usually obtained through the relation
\be
m_a(\Tosc)=3H(\Tosc),
\label{eq:equality}
\ee
by considering the standard cosmology, the CDM axion oscillates in a radiation dominated epoch, so by inserting the Hubble parameter written in \cref{eq:hubble_std} and \cref{eq:thermal_mass} into \cref{eq:equality}, $\Tosc$ takes the form
\be
\Tosc= \sqrt{m_a(\Tosc)M_p}
(g_*(\Tosc))^{-1/4},
\ee
From the previous expression it follows that depending on whether the temperature effects are important for the mass of the axion or not, there are two expressions for $\Tosc$ given by
\be
\Tosc=\begin{cases}
\displaystyle 
\left(\frac{\alpha m_0 M_P \Lambda_{QCD}^4 }{\sqrt{g_*(\Tosc)}}\right)^{1/6} & \text{for }\quad \Tosc \gtrsim \Lambda_{QCD},\\
\displaystyle 
\left(\frac{m_0M_P}{\,\,\sqrt{ g_*(\Tosc)}}\right)^{1/2} & \text{for } \quad \Tosc\lesssim \Lambda_{QCD}.
\end{cases}
\label{eq:std_tosc}
\ee

The behaviour of $\Tosc$ is depicted in \cref{fig:std_tosc}, where the change in the slope is quite visible around the QCD transition, as expected. The mass of the axion at which this intersection takes
place can be easily computed to be
\be
m_i=\sqrt{\frac{\alpha \Lambda_{QCD}^4}{M_P^2}}\approx 10^{-5} \rm{\mu eV}.
\ee

Finally, the solution for \cref{eq:eof} is no longer trivial. We can find the evolution of $a(t)$ using WKB approximation

\be
a(t)=a_0\left(\frac{m_a(\Tosc)\,R_{osc}^3}{m_a(T)\,R^3(T)}\right)^{1/2}{\rm{cos}}(\int m_a\,dt).
\ee

\begin{figure}[h]
\centering
\includegraphics[scale=0.6]{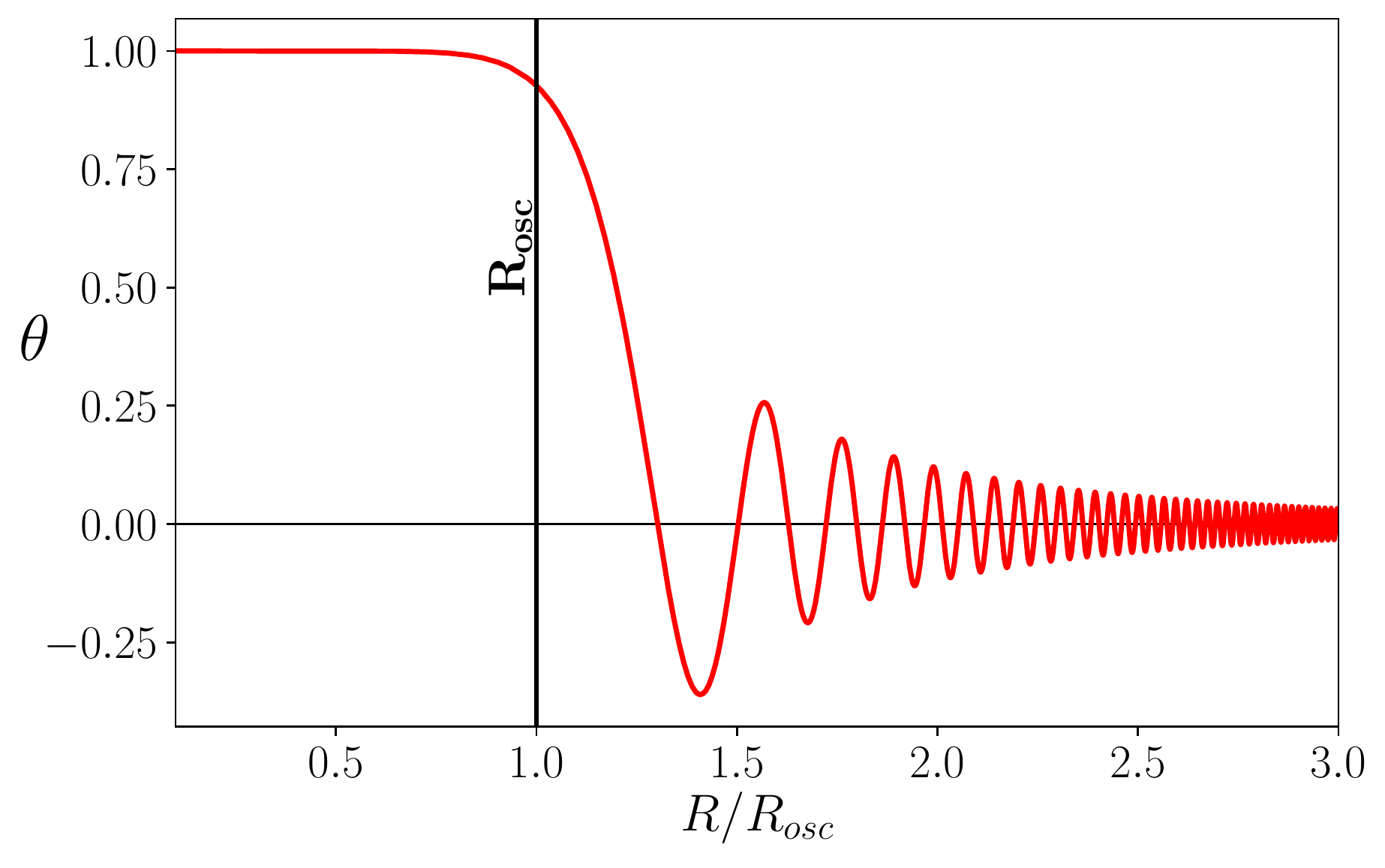}
\caption[Numerical evolution of the axion field ]{Numerical evolution of the axion field with an initial misalignment angle $\theta_i$=1 and axion mass $m_0=1\rm{\mu}$eV. The axion
starts to feel the pull of its mass at $R_{osc}$, and evolves to its minimum at $\theta = 0$, i.e. the
misalignment mechanism solves the strong CP problem.}
\label{fig:theta}
\end{figure}

For the numerical analysis, let us express the time dependence in terms of the scale factor by replacing
$\frac{d}{dt} = RH\frac{d}{dR}$. By doing this,
we obtain the following evolution equation

\be
\theta '' + \left(\frac{4}{R} + \frac{\dot H(R)}{H^2(R) \,R} \right)\theta'+\left(\frac{m_a(R)}{H(R)R}\right)^2\, \sin\theta=0,
\label{eq:Neom}
\ee
where $\theta'= d\theta/dR$ and $\dot H$ in a radiation epoch is given by \cref{eq:Hdotrad}. The numerical solution is shown in \cref{fig:theta}.

Now, from the energy tensor for a scalar field and taking the approximation of a quadratic potential, the axion energy density is given by 
\begin{align}
&\rho_a=\frac{1}{2}\dot a + V(a)=\frac{1}{2}\dot a^2(t)+\frac{1}{2}m_a^2(T)a^2(t).
\label{eq:rho_appr}
\end{align}

\begin{figure}
     \centering
         \includegraphics[scale=0.6]{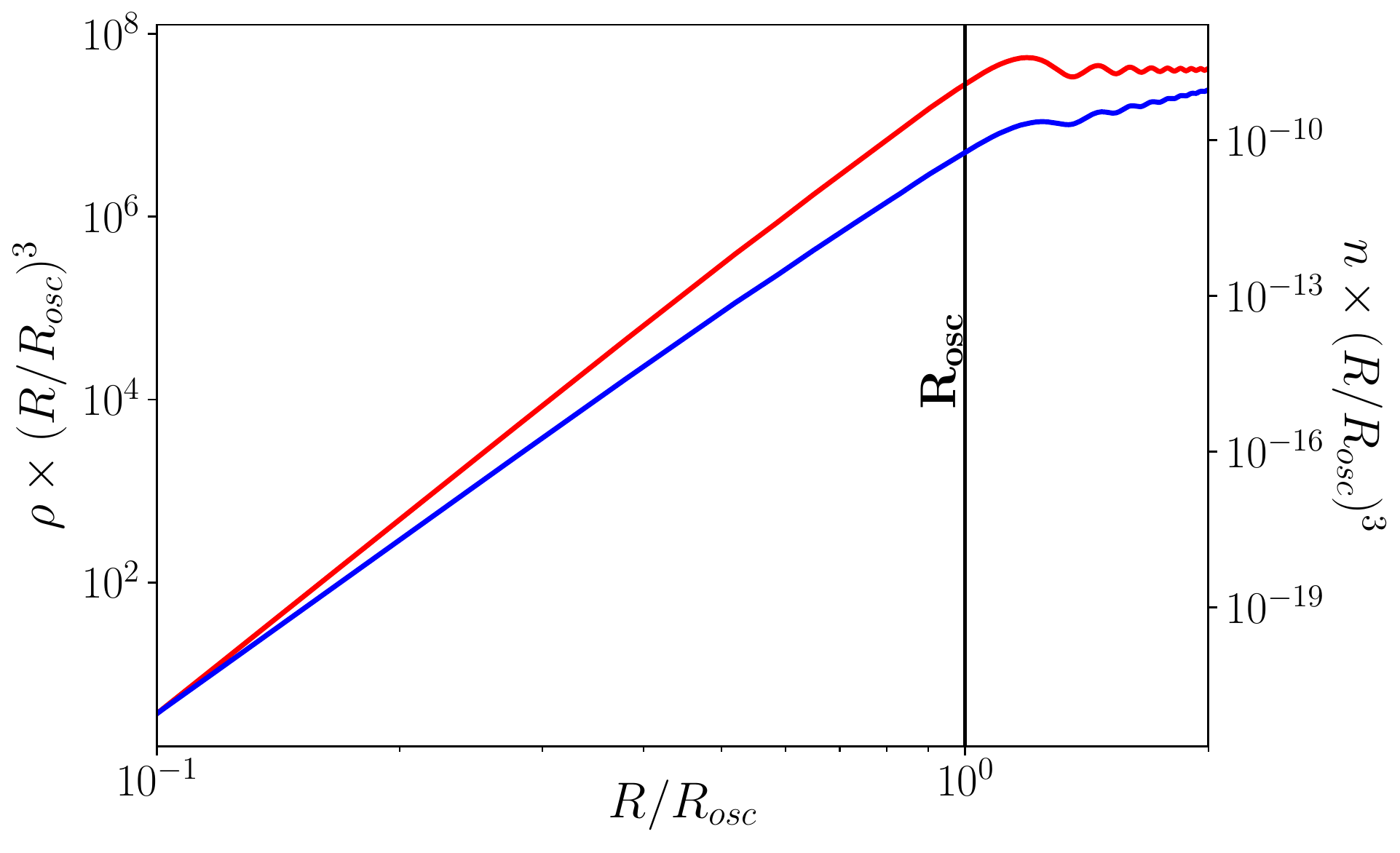}
\caption[Evolution of the energy density and axion number ]{(blue) The axion energy density keeps varying due to the temperature-dependent mass. (red) The comoving number of axions  produced via misalignment mechanism is conserved, although it is not immediately when the oscillations begin at $R_{\rm{osc}}$, this is still a good approximation. \label{fig:conserv}} 
\end{figure}

For $T<\Tosc$, we can take time average
over the period of the oscillation and obtain
$\rho\simeq \langle \dot a \rangle =\frac{1}{2} m_a^2\langle a \rangle^2$. In spite of the energy density of the axion is not conserved, since the mass varies with time (see \cref{fig:conserv}), the comoving axion number is conserved $N_a = n_aR^3$, therefore, we can find the energy of
axions today as $\rho_a(t) = m_a\,n_a(t)$, and it gives
\be
\rho_a(T_0)=\rho_a(\Tosc)\frac{m_0}{m_a(\Tosc)}\left(\frac{R_{osc}}{R_0}\right)^3,
\label{eq:std_axdensityR}
\ee
where we have used that axion number is conserved from the oscillation time until today as shown in figure \ref{fig:conserv}. From \cref{eq:std_axdensityR} we can see that energy density behaves as non-relativistic matter from  the oscillations time onwards.

Since universe evolves adiabatically, the scale factor and the temperature are related
through the condition of entropy conservation and we can write $\left(R_{osc}/R_{0}\right)^3=s(T_0)/s(\Tosc)$, consequently we have an explicit expression for the present axion energy density as a function of the temperature
\be
\rho_a(T_0)=\rho_a(\Tosc)\frac{m_0}{m_a(\Tosc)}\frac{s(T_0)}{s(\Tosc)},
\label{eq:std_axdensity}
\ee
We can insert $\Tosc$ obtained in \cref{eq:std_tosc} into \cref{eq:std_axdensity} to write down the current relic density of axions, where the general expresion is

\be
\Omega_a\approx 0.01 \left(\frac{\theta}{1}\right)^2 \sqrt{\frac{m_0}{m_{osc}}}\left(\frac{f_a}{10^{12}GeV}\right)^{3/2},
\ee
while for the two regimes we get

\be
\Omega_a \approx \begin{cases}
\displaystyle 
0.17 \left(\frac{\theta_i}{1}\right)^2 \left(\frac{m_0}{5.6 \mu \rm{eV}}\right)^{-7/6} & m_0\geq 10^{-5} \mu \rm{eV}, \\
\displaystyle 
0.006 \left(\frac{\theta_i}{1}\right)^2\left(\frac{m_0}{5.6 \mu\rm{eV}}\right)^{-3/2} & m_0\leq 10^{-5}\mu \rm{eV}.
\end{cases}
\label{eq:std_axion_abundance}
\ee

The initial misalignment angle needed at the moment of oscillation varies from $\theta_i\in [-\pi; \pi]$,
where values of $\theta_i$ below $\mathcal{O}(1)$, are considered anthropic, meaning that they seem to be unnatural. We can write down an analytical expression for the oscillation angle, $\theta_i$, 

\be
\theta_i^2 = \begin{cases}
\displaystyle 
2\, \Omega_{a} \frac{\rho_c}{\chi_0} \frac{(M_p\,m_0)^{3/2}}{T_0^3 F(T)}, & m_0\geq 10^{-5} \mu \rm{eV}, \\
\\
\displaystyle 
2\, \Omega_{a} \frac{\rho_c}{\chi_0} \frac{(M_p\,m_0)^{7/6}}{T_0^3 F(T)}  \left( \alpha\,\Lambda_{QCD}^4\right)^{1/6}, & m_0\leq 10^{-5}\mu \rm{eV},
\end{cases}
\label{eq:SC_thetai}
\ee
here $F(T)=\frac{g_{*S}(T_0)}{g_{*S}(\Tosc)}\left(\frac{\pi^2 g_*(\Tosc)}{10}\right)^{3/4}$. 

\begin{figure}[t]
\centering
\includegraphics[scale=0.6]{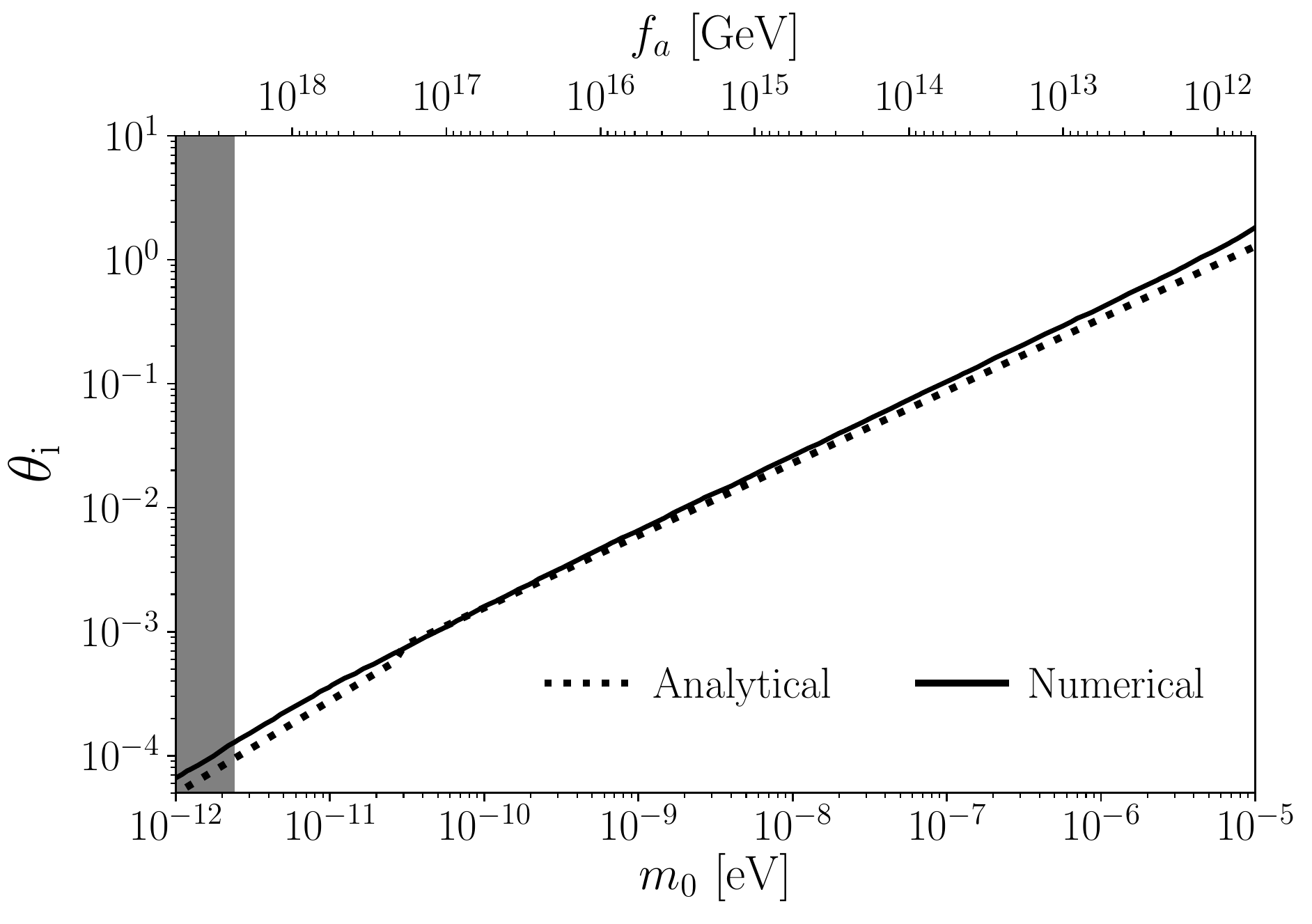}
\caption[Initial misalignment angle vs axion mass]{Initial misalignment angle vs axion mass. Black line is for the numerical solution and  dotted line is the analytical solution obtained in \cref{eq:SC_thetai}. The gray area corresponds to scales over the Planck scale.}
\label{fig:std_relic}
\end{figure}

Imposing that axions represent the total dark matter currently measured such that $\Omega_a=\Omega_{CDM} $= 0.26
, we get a relation between the parameters $m_0$ and $\theta_i$ . Fig.\ref{fig:std_relic} shows the analytic approximation of \cref{eq:SC_thetai} matches well with the full numerical result, the differences at small angles come from the approximation for the axion mass, since for our analytical computation we are taking an abrupt change so that it evolves as a constant at low temperatures whereas the full numerical data decreases slowly towards $m_0$. Moreover, for the region above the curve, we have a combination of parameters that give us an overproduction of axions, while below the curve, the axions cannot explain the total dark matter and it gives rise to consider a second component.

\subsection{Anharmonicities}
\label{anhar}
As we have seen, the full cosine-potential of the axion makes the evolution equation of $\theta$ nonlinear, we have avoided this by considering a small $\theta$  such that the potential becomes approximately harmonic, namely $V (a) =m_a^2 a^2/2$ as we wrote in \cref{eq:rho_appr}. However, for large initial displacements $\theta_i \simeq \pi$, anharmonic corrections caused by axion
self-interactions become important \cite{Marsh:2015xka} and the potential becomes flatter than the used
potential, resulting in the axion field oscillations starting in a $R_{osc}$ delayed increasing
the relic abundance relative to the harmonic approximation.

The solution to account the anharmonicities effects and thus improve the approximate computation obtained in \cref{fig:std_relic} is to introduce an anharmonic correction factor $\mathcal{F}(\theta_i)$ whose role is to parameterize the anharmonicities, i.e. $\theta_i\rightarrow \mathcal{F}(\theta_i) \theta_i$. For small values of $\theta_i$ the function converges to one and diverges for larger values. An analytic approximation to $\mathcal{F}(\theta_i)$ for the cosine potential is \cite{visinelli2009dark}
\be
\mathcal{F}(\theta_i)=\left[ \ln \left(\frac{e}{1-\theta_i^2/\pi^2} \right)\right]^{7/6}.
\label{eq:anharm}
\ee

\section{Axions from string decays}
\label{sec:string}

Essentially,  the dimensionless axion field $\theta$ can be introduced into the theory as the phase degree of freedom of a new complex scalar field, the PQ field
\be
\Phi_a=|\Phi_a|e^{i\theta}.
\label{eq:complex}
\ee
The cosmological evolution of the PQ field is given by
\be
\mathcal{L}=\frac{1}{2}|\partial_\mu\Phi|^2-\frac{\lambda}{4}\left(|\Phi|^2 -v_a^2\right)^2.
\label{eq:Lstring}
\ee
The effective potential in \cref{eq:Lstring} induces the PQ phase transition such that the scalar field gets vacuum expectation value $|\langle \Phi \rangle|^2 =v_a^2 \propto f_a^2$, and the global $U(1)_{PQ}$
symmetry is spontaneously broken, as a result, the axion, lies in the degenerate circle of minima in the  potential.
Since the symmetry breaks in different regions of space, there is a non-trivial
winding around the false vacuum, which corresponds to the formation of axion global strings.

It has been well studied that the  string networks enter the so-called scaling solution \cite{kibble1976topology},  where the large-scale
structure scales with the horizon scale and we expect about one long string per Hubble volume. During the radiation dominated regime, the energy density
of the long-string network is described by
\be
\rho_s=\frac{\mu_s}{t^2},
\ee
here $t^{-1}$ is a long-distance cutoff of the order of the horizon and $\mu_s\approx 2\pi f_a^2\ln (f_a\,d)$, the energy per unit lenght of the global string, with $d$ the characteristic
distance between strings.

When the universe expands and cosmic strings cross, they can form loops and the scaling symmetry is maintained by the
continuous emission of axions \cite{Marsh:2015xka}.
As soon as the axion mass becomes important at $\Tosc$, these loops become unstable and strings become the boundaries of so-called domain walls; as a consequence, the emission of axion particles ceases \cite{sikivie2008axion}.

In the literature there has been discussion about the shape of the energy spectrum of the radiated axions $\omega(T)$ \cite{harari1987evolution,davis1985goldstone} and numerical simulations of decay of axion strings are required in order to determine it.

The total number of axions produced by string decay in a comoving volume is given by the integral \cite{Marsh:2015xka,1990eaun} from the time of the PQ phase transition at $T=f_a$ up to $\Tosc$
\be
\frac{n_a}{s}\sim \int_{\Tosc}^{f_a}\frac{\mu_s}{\omega(T)M_p^2}\,dT.
\ee

Axions produced by string decay are dominated by the low-frequency modes, making them non-relativistic and contributing to CDM. We can write the relic energy density of axion radiated by strings compared with the misalignment production as
\be
\Omega_{\rm{a,str}}=\alpha_{str}\, \Omega_{\rm{a,mis}}.
\ee
Results in the classical literature for $\alpha_{str}$ are \cite{battye1994axion, harari1987evolution,davis1985goldstone}, with values ranging from 0.16 to 186.

\chapter{Axion CDM in a non-standard  cosmological  scenario}
There are several works that have previously studied a non-standard cosmological scenario in the context of axion physics. Besides the pioneer papers of \cite{Steinhardt:1983ia, Kawasaki:1995vt, Giudice:2000ex,Lazarides:1990xp} that considered an early matter dominance period, in \cite{Visinelli:2009kt} the authors studied the cases of low temperature reheating (LTRH) and kination, including the anarmonicities in the axion potential. In \cite{Grin:2007yg} it was consider thermally produced axions in LTRH and kination cosmologies. In \cite{Blinov:2019jqc} it was considered the misalignment production of axion-like particles in early matter domination and kination cosmologies and in \cite{Ramberg:2019dgi} a full scan of cosmological histories was performed, together with including the contribution from the decay of an axionic string network. In refs.~\cite{Nelson:2018via, Visinelli:2018wza} the impact of non-standard cosmologies on the formation of axion miniclusters was considered.

Our approach is to consider firstly a very detailed analysis of the DM production during the NSC scenario. In particular, we aim to contrast the axion relic density obtained  numerically with analytical results, such as to have a deep understanding of the relationship of the relic density on the different NSC parameters. Moreover, we would like to perform a scan of different cosmological histories that do not need to assume fine tuned initial conditions (anthropic reasons) and that can account for the whole DM of the universe. In that sense, we will share the spirit of \cite{Blinov:2019jqc} and try to keep the assumptions as minimal as possible. An important distinction between our study to others, is that we will keep track of the initial conditions of the NSC. 

\section{Non-standard cosmology }{\label{sec:nsc}}
Prior to BBN, it is assumed in standard cosmology, that the universe transitioned from the inflationary epoch, with an equation of state of $\omega=-1$, to a radiation dominated period, with $\omega=1/3$, around the time of neutrino decoupling. The details of such transition are currently not known, and in that context, a discovery of DM and its properties could be the first signal to learn how that period actually was. Non-standard cosmologies have been widely studied in many contexts, spawning an equation of state from $-1\leq \omega \leq 1$\footnote{Has been argued that cosmologies with $\omega>1$ could feature superluminal propagation, although in \cite{DEramo:2017gpl} they show that is not the case.}. Extensively studied NSC are the Early Matter Domination (EMD), firstly explored in the context of supersymmetry and string theory (e.g.\cite{Moroi:1999zb,Vilenkin:1982wt,Coughlan:1983ci}), with $\omega=0$, and the kinetic energy domination, known as Kination Dominance (KD), with $\omega=1$ \cite{Barrow:1982ei, Ford:1986sy,Spokoiny:1993kt}.  For a detailed review on NSC, see \cite{Allahverdi:2020bys}, and references therein. In this work, we will consider that the NSC is generated through the existence of one or a set of heavy meta-stable particles that potentially come to dominate the energy density of the early universe, with a general equation of state $p_\phi=\omega_\phi \rho_\phi$, such that their (combined) energy density scales as $\rho_\phi\propto R^{3(1+\omega_\phi)}$ \cite{PhysRevD.28.1243}.\footnote{We will denote the new field as $\phi$, but bearing in mind that could be a set of $n$ metastable fields whose energy density can be written as $\rho_\phi=\sum_{i=1}^n \rho_{\phi_i}$.} In order to ensure the successful period of nucleosynthesis, the field shall decay before the temperature $\Tend\sim T_{\rm BBN}\sim~1$~MeV is reached, with a a decay rate of $\Gamma_\phi$. Our assumption will be that the decay is entirely to the light degrees of freedom of the Standard Model, and that they are in local thermal equilibrium among themselves. The equations that govern the evolution of the energy densities of the new field and radiation are \cite{chung1999production} 
\begin{align}
\frac{d\rp}{dt}+3(1+\omega_\phi)\,H\,\rp&=-\Gamma_\phi\,\rp\,,\label{eq:cosmo2} \\
\frac{ds}{dt}+3\,H\,s&=\frac{\Gamma_\phi}T\,\rp,\label{eq:cosmo3}
\end{align}
Using \cref{eq:entropy} the latter equation above can be recast to find the relationship between temperature and scale factor as
\begin{equation}\label{eq:cosmo3b}
    \frac{dT}{dR}=\left(1+\frac{T}{3\,\gss}\frac{d\gss}{dT}\right)^{-1}\left[-\frac{T}{R}+\frac{\Gamma_\phi\,\rp}{3\,H\,s\,R}\right].
\end{equation}

Recall also that now the Hubble parameter is given by
\be
H=\sqrt{\frac{\rho_\phi+\rho_R+\rho_a}{3M_P^2}},
\label{eq:Hubble}\ee
Nonetheless, the axion contribution is always subdominant, so we will ignore it. Then, the evolution of the $\phi$-radiation system is decoupled form the DM evolution and can be solved numerically.

We will assume there is certain initial condition at temperature $T=T_i$, between the radiation energy content $\rho_R$ and the one of the new field $\rho_\phi$, such that
\be
\kappa\equiv \frac{\rho_{\phi}}{\rho_{Ri}}.
\ee
That is, for $\kap<1$ there is always initially a period of radiation domination, which could be undertaken by the dominance of the new field, if $\omega<1/3$\footnote{Depending on the value of $\kap$, even for $\omega\gtrsim 0.3$ can happen that $\phi$ is not able to dominate over radiation}. 

It is customary to define the temperature at which the field has mostly decayed away, $\Tend$, as \cite{chung1999production,Giudice_2001}
\begin{equation}\label{eq:Tend}
    \Tend^4\equiv\frac{90}{\pi^2\,\gs(\Tend)}\,M_P^2\,\Gamma_\phi^2.
\end{equation}

\begin{figure}[t]
\center
\includegraphics[scale=0.6]{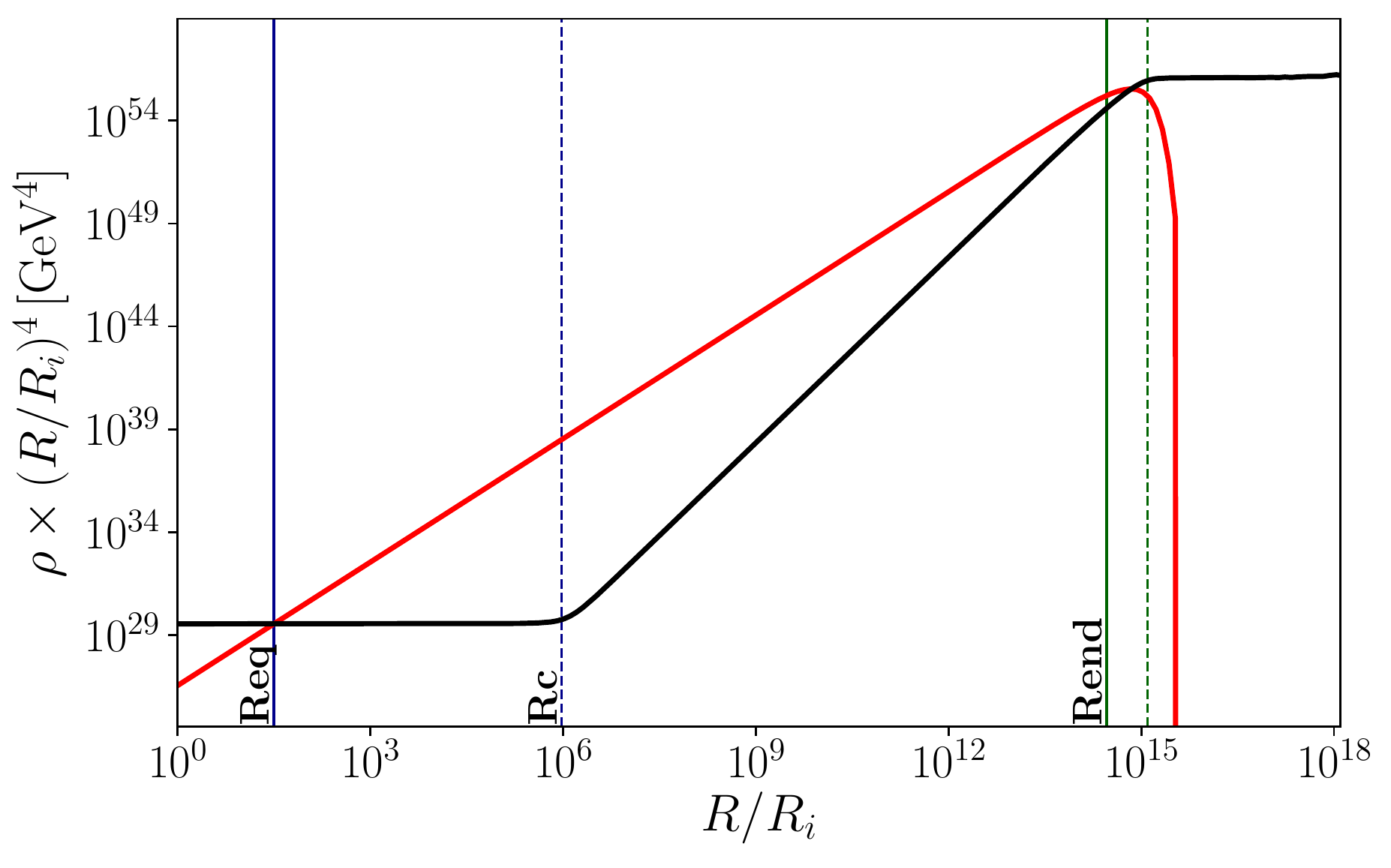}
\caption[Evolution of the energy densities for radiation and  $\phi$ field ]{Evolution of the energy densities for radiation (black) and  $\phi$ field (red) as a function of the scale factor $R$, for $\omega_\phi=0$, $T_{end}=4\,{\rm{GeV}}$, $T_{i}=10^{12}\,{\rm{GeV}}$ and $\kappa=10^{-3}$. The vertical lines are $R_{\rm{eq}}/R_{i}$ (solid blue), $R_{\rm{c}}/R_{i}$ (blue dotted), $R_{\rm{end}}/R_{i}$ (solid green). }
\label{fig:energies_nsc}
\end{figure}

As long as $\Gamma_\phi\ll H(T)$, the entropy is separately conserved in the SM and in the hidden sector of $\phi$, and thus, the temperature in the SM follows the usual relation for a radiation dominated universe $T\propto R^{-1}$. Once the dominance of $\phi$ is established, the universe will expand accordingly with the equation of state, such that $H\propto T^{\frac{3}2(1+\omega_\phi)}$. During their decay, entropy is no longer conserved and is transferred to the SM, therefore, the expansion of the universe changes as $H\propto T^4$, no matter the equation of state \cite{Maldonado_2019} . The 
 evolution of the energy densities $\rho_\phi$ and $\rho_R$, as well as the evolution of the temperature as a function of the scale factor $R$ are shown in fig.(\ref{fig:energies_nsc}){\footnote{The green dotted line corresponds to the numerical scale factor at which the entropy injection to radiation has ceased and the evolution goes back to be adiabatic. In our analytical work we consider a spontaneous decay, such the standard cosmology is recovered at $R_{\rm{end}}$, we will see throughout this work, that is a good approximation.}}. Different relevant stages of the evolution are: a) $\Req$: the moment where both energy densities from fluid and radiation become equal (this region is only present if $\kap<1$). b) $\Rc$: scale factor when the decays of $\phi$ start to affect the temperature of the SM, and c) $\Rend$: the scale factor at $T=\Tend$. We can find analytical expressions for these parameters by solving the evolution of the energy densities eqs.~(\ref{eq:cosmo2}) and (\ref{eq:cosmo3}). A more detailed derivation of these equation in the different regimes of dominance is outlined in the appendix \ref{app:phi-r},  in here we just highlight the most important results.

Before the decays of $\phi$ can affect the evolution of the temperature, meaning $\Gamma_\phi\ll H(T)$, can be easily seen from eqs.~(\ref{eq:cosmo2}) and (\ref{eq:cosmo3}), that the energy densities evolve as
\be
\rho_R=\rho_{Ri}\left(\frac{R_i}{R}\right)^4, \,\,\,\,\,\, \rho_\phi=\rho_{\phi i}\left(\frac{R_i}{R}\right)^{\beta}
\label{eq:energies_free}
\ee
Where $R_i$ is an arbitrary initial scale factor associated to the temperature $T_i$ and we have defined $\beta=3(\omega_\phi+1)$. 
The point  where both energy densities equate is what we define as  $R=\Req$, which can be written as
\be
\Req=R_i\, \kappa^{\frac{1}{\beta-4}},
\ee
 For $\beta=4$ the relation is not well defined, since this value is not compatible with $\kap<1$. For $\kap>1$ the two energies do not equate at this stage, but later, when $\phi$ field decays.
In this period the relation $T\propto R^{-1}$, still holds, so we have
{{
\be
\Teq=T_i\, \kappa^{\frac{1}{4-\beta}}.
\label{eq:Teq}
\ee}}
We assume there will be a period of  $\phi$ dominance, that in the case of $\kappa<1$ starts from $\Req$ on. At some moment, the decay of $\phi$ has an impact on the evolution of the SM temperature, this happens around the scale factor we have denoted as $R_c$.  Thus, we solve analytically eqs.~(\ref{eq:cosmo2}) and (\ref{eq:cosmo3}), with $H(T)\approx \sqrt{\frac{\rho_\phi(T)}{3M_P^2}}$. To first order in $\Gamma_\phi/H_i$, where $H_i\equiv H(R_i)$,  we find{\footnote{These expressions are only valid for $\beta>0$. The solution for the case $\beta=0$ can be found in the \cref{app:phi-r}}}
\bea
\rho_\phi(R)&\eqsim& \rho_{Ri} \kappa \left[\left(\frac{R_i}{R}\right)^\beta-\kappa^{-1/2}\frac{2\Gamma_\phi}{\beta\, H_i}\left( \frac{R_i}{R}\right)^{\beta/2}\right], \label{eq:phi_density}\\
\rR(R)&\eqsim&\rho_{Ri}\left[\left(\frac{R_i}{R}\right)^4+\kappa^{1/2}\frac{2\Gamma_\phi}{(8-\beta)H_i}\left(\frac{R_i}{R}\right)^{\beta/2}\right]. \label{eq:rad_density}
\eea
For $R\ll R_c$ the first terms of the rhs of eqs.~(\ref{eq:phi_density}) and (\ref{eq:rad_density}) are the dominant ones. Eventually, as the decays of $\phi$ start to be important,  the second terms of the rhs can not be neglected. We find, by equating the first and second term of the rhs of eq.~(\ref{eq:rad_density}) that,
\be
\Rc\eqsim R_i\left(\frac{(8-\beta)}{\sqrt{\kappa}}\frac{H_i}{2\Gamma_\phi}\right)^\frac{1}{4-\beta/2}.
\ee
And by requiring the second and third term of the rhs of eq.~(\ref{eq:phi_density}) are comparable, we find an expression for when the decays start to be important, meaning $R=\Rend$, to be
\be
\Rend\eqsim R_i\left(\frac{\beta\, \sqrt{\kappa}\, H_i}{2\Gamma_\phi}\right)^{2/\beta}.
\label{eq:rend}
\ee
And the corresponding temperatures can be obtained from
\bea
\rho_{Ri}\left(\frac{R_i}{\Rc}\right)^{4}\approx \rR(\Tc),\quad\mbox{and}\quad {{\frac{\Gamma_\phi}{H_i}=\frac{\Tend^2}{T_i^2},}}
\eea
to find
{{
\bea
\Tc&\eqsim&T_i \left(\frac{2\Gamma_\phi\, \sqrt{\kappa}}{H_i(8-\beta)}\right)^{\frac{2}{8-\beta}}=
T_i\left(\frac{4\kappa}{(8-\beta)^2} \frac{\Tend^4}{T_i^4}\right)^\frac{1}{8-\beta}.
\label{eq:Tc}
\eea}}
From this analysis it can be also extracted that deep during the $\phi$ domination, the relation between temperature and scale factor is
\be
T\eqsim T_i\left[\frac{2\sqrt{\kappa}}{8-\beta}\frac{\Tend^2}{T_i^2} \right]^{1/4} \left(\frac{R_i}R\right)^{\beta/8}.
\label{eq:T_NSC}
\ee
We emphasize the equation above, tells us that for smaller $\beta$, the temperature decreases more slowly with the universe expansion and therefore it implies that the universe is bigger when reaching the temperature $\Tend$. Let us stress that the range of equations (\ref{eq:phi_density})-(\ref{eq:T_NSC}) are only valid if the fluid dominates strongly over radiation. There can be certain cosmologies where this is not the case - and still have an impact on the axion relic density - and thus, the above referred expressions have to be handle carefully.

We can differentiated 3 regions in a NSC (see \cref{fig:energies_nsc}), these regions can be characterised by the scale factor (or temperature) and they are :

\begin{itemize}
\item[i)] Region 1, $R\ll\Req$ : dominance of radiation much before $\phi$ decay. This period will be present only when $\kappa<1$ and $\Req$ sets the moment when the energy of radiation and $\phi$ equate.
\item[ii)] Region 2, $\Req\ll R\ll \Rc$: dominance of $\phi$, prior its decay becomes significant. Again, $\Req$ is only present for $\kappa<1$. For $\kappa>1$ the lower bound of the region has the be replaced by the initial condition $R_i$.
\item[iii)] Region 3, $\Rc\ll  R\ll \Rend$: dominance of $\phi$, which significantly decays into radiation.
\end{itemize}

As  we  will  see  in  \cref{sec:oscillation}, axion oscillations can take place within these three period and lead to a different relic density today than in the standard cosmology.

\subsection{Effects of the NSC on the expansion of the universe}

Different NSC scenarios will result in different cosmological histories for the Universe. The expansion of the universe, parametrised by the Hubble parameter, $H$ is affected according to eq.~\eqref{eq:Hubble}.
	\begin{figure}[ht!]
		\centering 
		\begin{subfigure}[b]{0.5\textwidth}
			\centering\includegraphics[width=1\textwidth ]{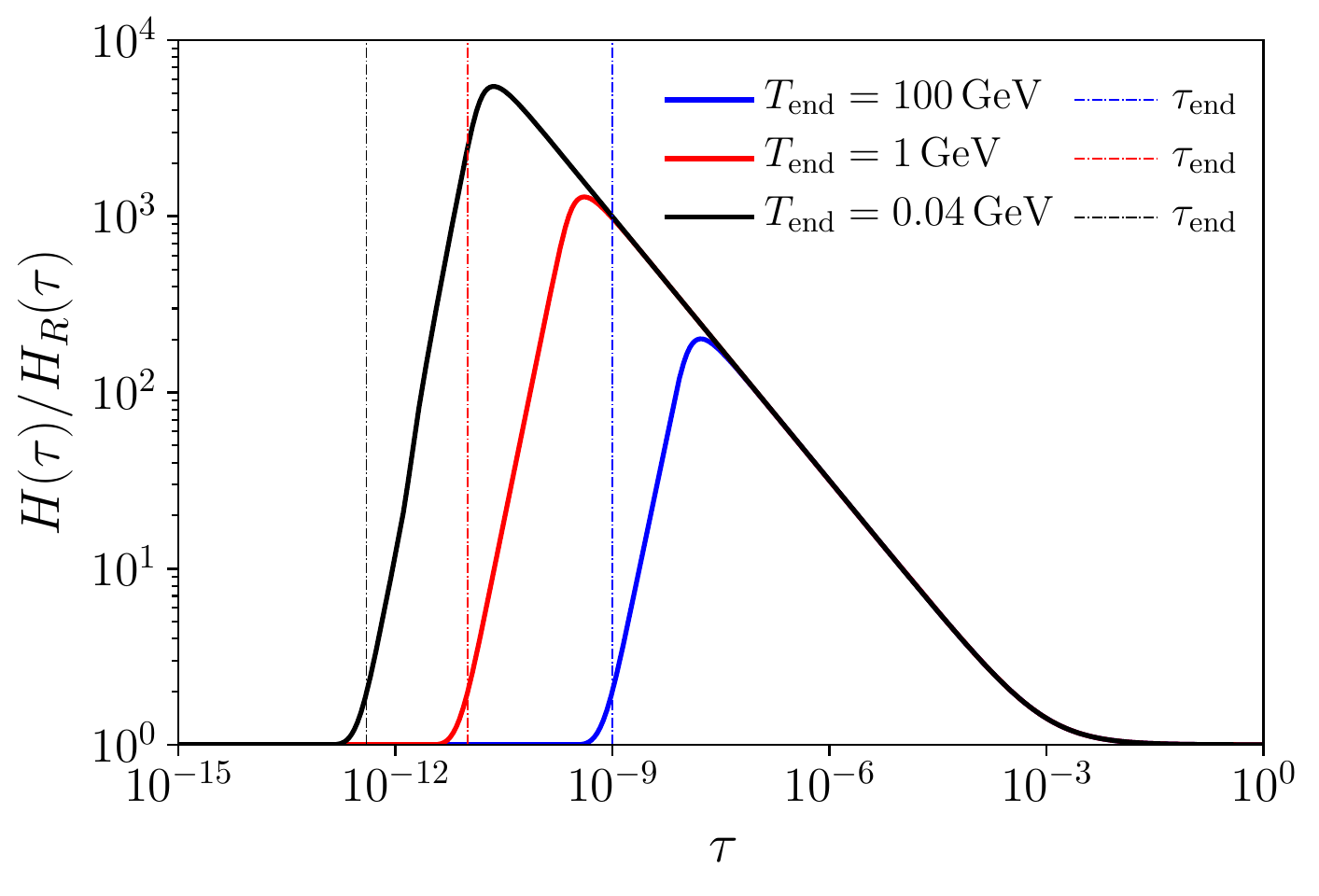}
	\caption{\label{fig:NSC_H_TEND}}
		\end{subfigure}%
		\begin{subfigure}[b]{0.5\textwidth}
			\centering\includegraphics[width=1\textwidth]{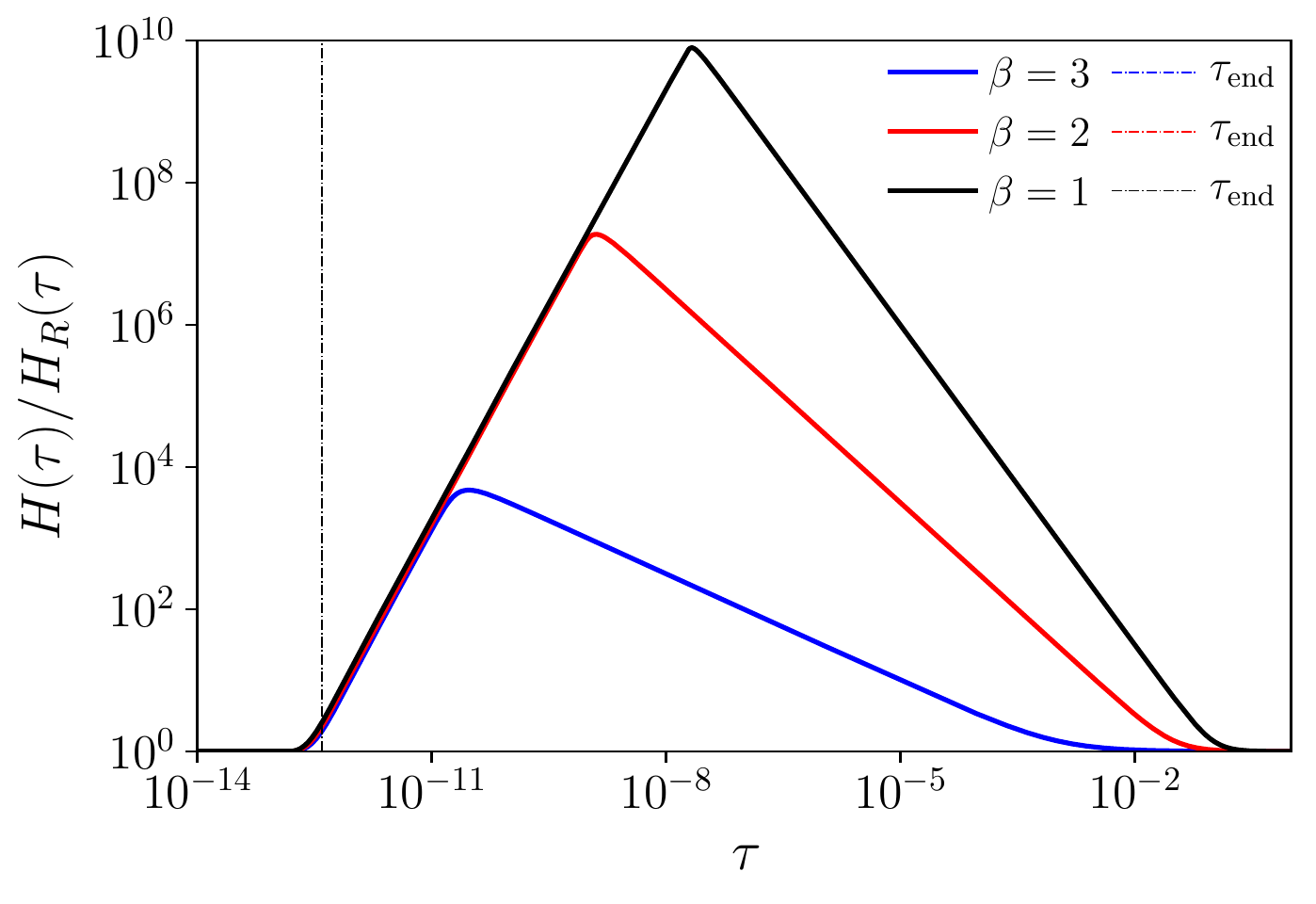}
	\caption{\label{fig:NSC_H_beta}}
		\end{subfigure}
	\begin{subfigure}[b]{0.49\textwidth}
	\centering\includegraphics[width=1\textwidth]{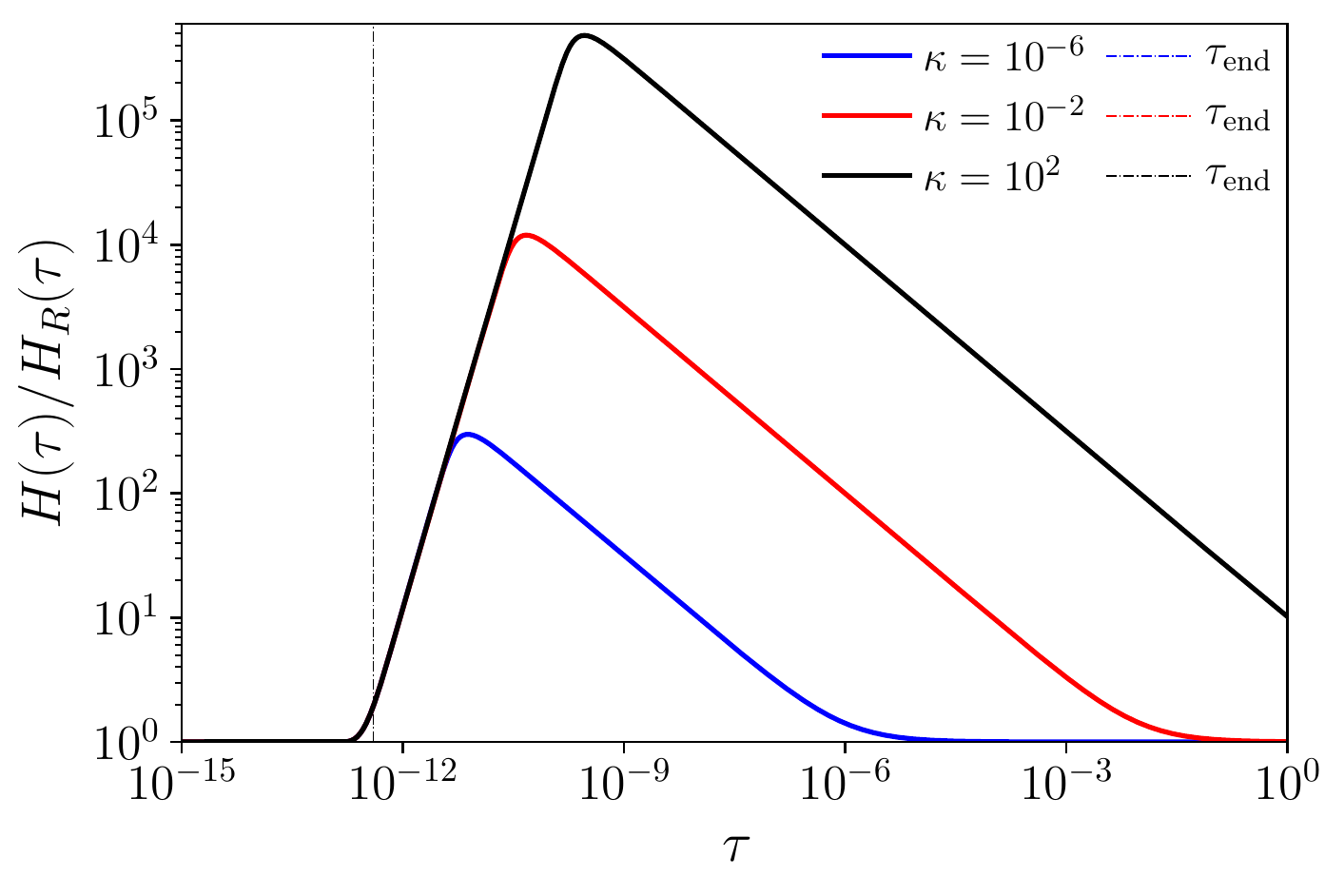}	\caption{\label{fig:NSC_H_kappa}}
		\end{subfigure}
	\begin{subfigure}[b]{0.49\textwidth}
	\centering\includegraphics[width=1\textwidth]{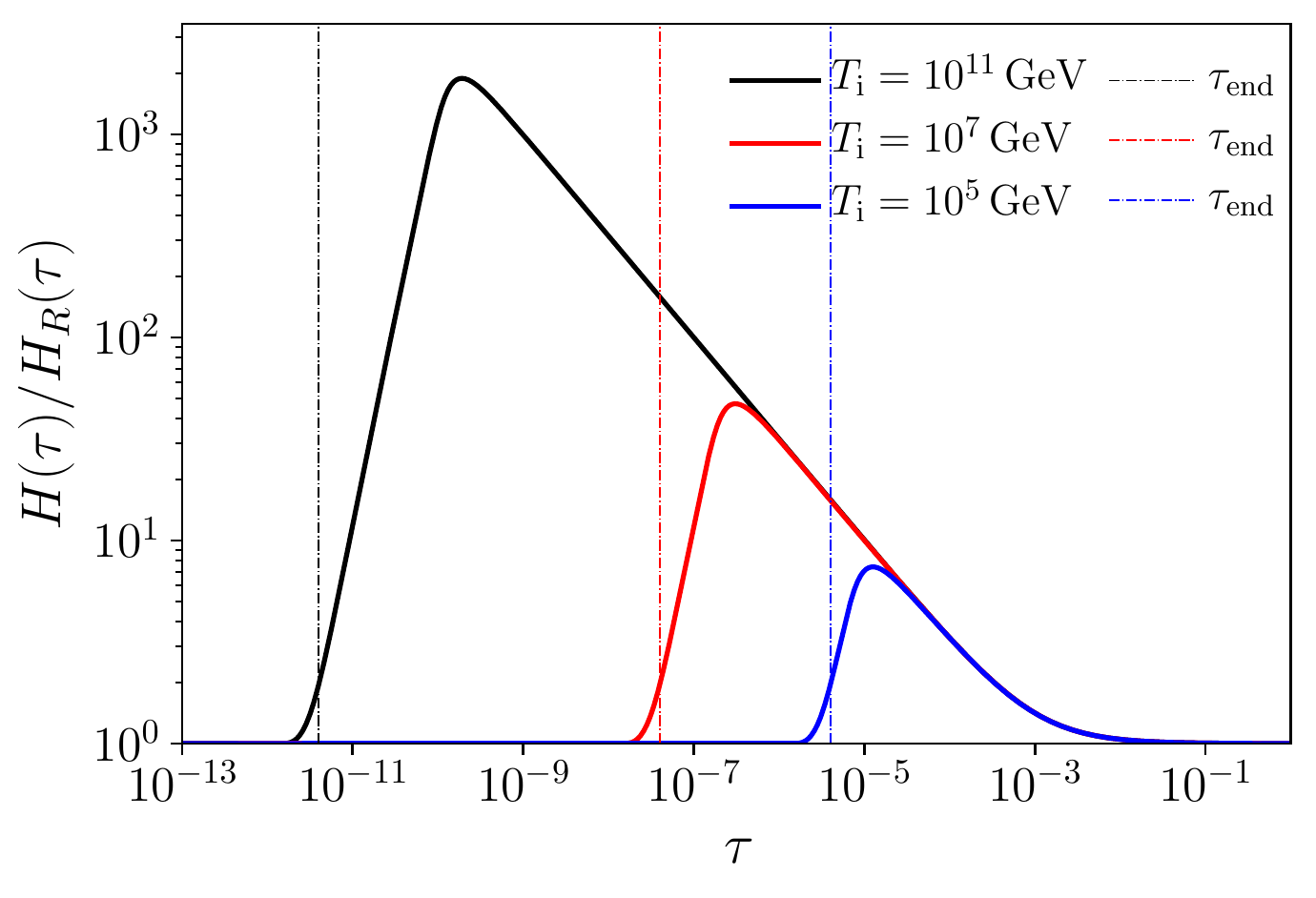}
	\caption{\label{fig:NSC_H_Ti}}
		\end{subfigure}
		\label{fig:NSC_H}
		\caption[Evolution of the Hubble parameter in a NSC]{The evolution of the Hubble parameter as a function of temperature for different NSC. In (a) we vary $\Tend$, in (b) we vary $\beta$ , in (c) we change $\kappa$ and finally, in (d) we vary $T_i$. For the rest of parameters we are keeping  $\beta=3$ , $T_i=10^{12}$ GeV, $\kap=10^{-3}$ and $\Tend=0.4$~GeV.}
		\label{fig:rel_H}
	\end{figure}
In fig.~(\ref{fig:rel_H}) we show the evolution of $H/H_{R}$ - where $H_{R}$ is the Hubble parameter in a radiation dominated universe -  as a function of the temperature\footnote{In this case we have chosen the temperature as a variable, because it allow us to 
place easily the relevant temperatures $\Teq, \Tc,\Tend$, than using the scale factor as a variable.}, where we have defined
\be
\tau\equiv T/T_i.
\ee

From the \cref{fig:rel_H} we can visualize Region 1 as the radiation domination period from $\tau$=1 until the curves begin to rise at the temperature $\tau_{\rm{eq}}$. Then,  Region 2 begins, here the field $\phi$ dominates over radiation, so we have an increase in the expansion rate and the region ends when the curve reaches the peak at $\tau_c$. From there it starts Region 3 until the temperature $\tau_{end}$ (vertical line). Finally,  we come back to the period dominated by radiation, recovering the standard cosmology.
We know that for the same $\beta$, a lower $\Tend$ implies a bigger universe leaving an increase in the expansion rate during the domination of $\phi$ as shows \cref{fig:NSC_H_TEND}. While with a higher $\Tend$, the field $\phi$ has a shorter period to dominate and decay, which results in Region 2 and 3 being shortened. 

On the other hand, fields with a smaller parameter $\omega_\phi$ in their equation of state, suffer less dilution by the expansion of the universe if we compare them with radiation. So from \cref{fig:NSC_H_beta} can be seen that a smaller $\beta$ implies an early domination of $\phi$, leaving a shortening of the Region 1. Since the temperature evolves slowly with small $\beta$, Region 2 finishes at higher temperatures and given that the universe is bigger, a faster expansion of the universe is required to achieve $\Tend$.

By changing $\kappa$ in \cref{fig:NSC_H_kappa} for the same $\beta$, a smaller $\kappa$ means a less initial energy density of the field $\phi$, so it takes longer to start to dominate the total energy density of the universe, its delay leads to a short NSC period and a lower expansion rate. For $\kappa>1$, the field $\phi$ always starts dominating the energy density of the universe. Finally, from \cref{fig:NSC_H_Ti} it can be seen that a higher initial temperature $T_i$ makes the dominance of $\phi$ longer, thus, the NSC period is extended.  We note that similar effects can be obtained either changing $\Tend$ or $T_i$.

\section{Axion relic density in a NSC}{\label{sec:oscillation}}
In this new scenario, entropy is not generally conserved, since the decay of $\phi$ gives an entropy injection that it is not shared by the axion, since the latter is thermally decoupled from the bath.  Therefore, we expect a dilution of the dark matter relic density if the DM abundance is set during the period of $\phi$ dominance. In order to compute the relic density today, we follow the same lines as the standard case, except that we use axion number conservation from the oscillation time until the scale factor corresponding to $T=\Tend$, which we call $\Rend$. From that period on, entropy is again conserved\footnote{ Numerically, we consider that the evolution goes back to be adiabatic at $T_{\rm ad}$, the temperature corresponding to the scale factor $R_{\rm ad}$. }, until today. Therefore, we have
\bea
\rho_a(T_{0})&=& \rho_a(\Tosc) \frac{m_0}{m_a(\Tosc)}\left(\frac{\Rosc}{R_0}\right)^3,\\
&=&\rho_a(\Tosc) \frac{m_0}{m_a(\Tosc)}\left(\frac{\Rosc}{\Rend}\right)^3 \frac{s(T_0)}{s(\Tend)}.
\label{rho_today}
\eea
Expression eq.~(\ref{rho_today}) can be recast by writing the ratio $\frac{s(T_0)}{s(\Tend)}$ as
\be
\frac{s(T_0)}{s(\Tend)}= \frac{s(T_0)}{s(\Tosc)} \times \frac{s(\Tosc)}{s(\Tend)}.
\ee
Plugging back into (\ref{rho_today}), we find
\be
\rho_a(T_0)=\rho_a(\Tosc)\frac{m_0}{m_a(\Tosc)}\frac{s(T_0)}{s(\Tosc)} \times \frac{S_{\rm osc}}{S_{\rm end}}.
\label{eq:NSC_axion_density}
\ee

Comparing with \cref{eq:std_axdensity}, we can see that the first part looks exactly like the one found in the standard scenario, with the important difference that the oscillation temperature will be in general different to the standard case, and there is also a factor that takes into account the change in the total entropy from the time of axion oscillation until the decay of $\phi$. This entropy dilution factor can be expressed accordingly, in terms of the parameter of the NSC, depending on which stage of the $\phi$ dominance the axion starts to oscillate. 

As we will see in the next subsection, we have found that the oscillation temperature, $\Tosc$ and the relic abundance are different than in the standard cosmology. Thus, it is expected that the axion prefered DM parameter space will have different boundaries than in the SC, depending on the details of the NSC and its characteristic parameters.

\subsection{Axion oscillation}
To start with, let us recognize what are the potential interesting scenarios for axion cosmology. In order to have some impact on the relic abundance, the $\phi$ field needs to at least inject entropy to the SM bath, otherwise, its existence will be ignored by the dark matter relic density. For $\kappa<1$, it is quite clear that only for $\omega_\phi<1/3$ this is possible, so we will stick to consider cosmologies with those equations of state. 

\begin{figure}[t]
\center
\includegraphics[scale=0.6]{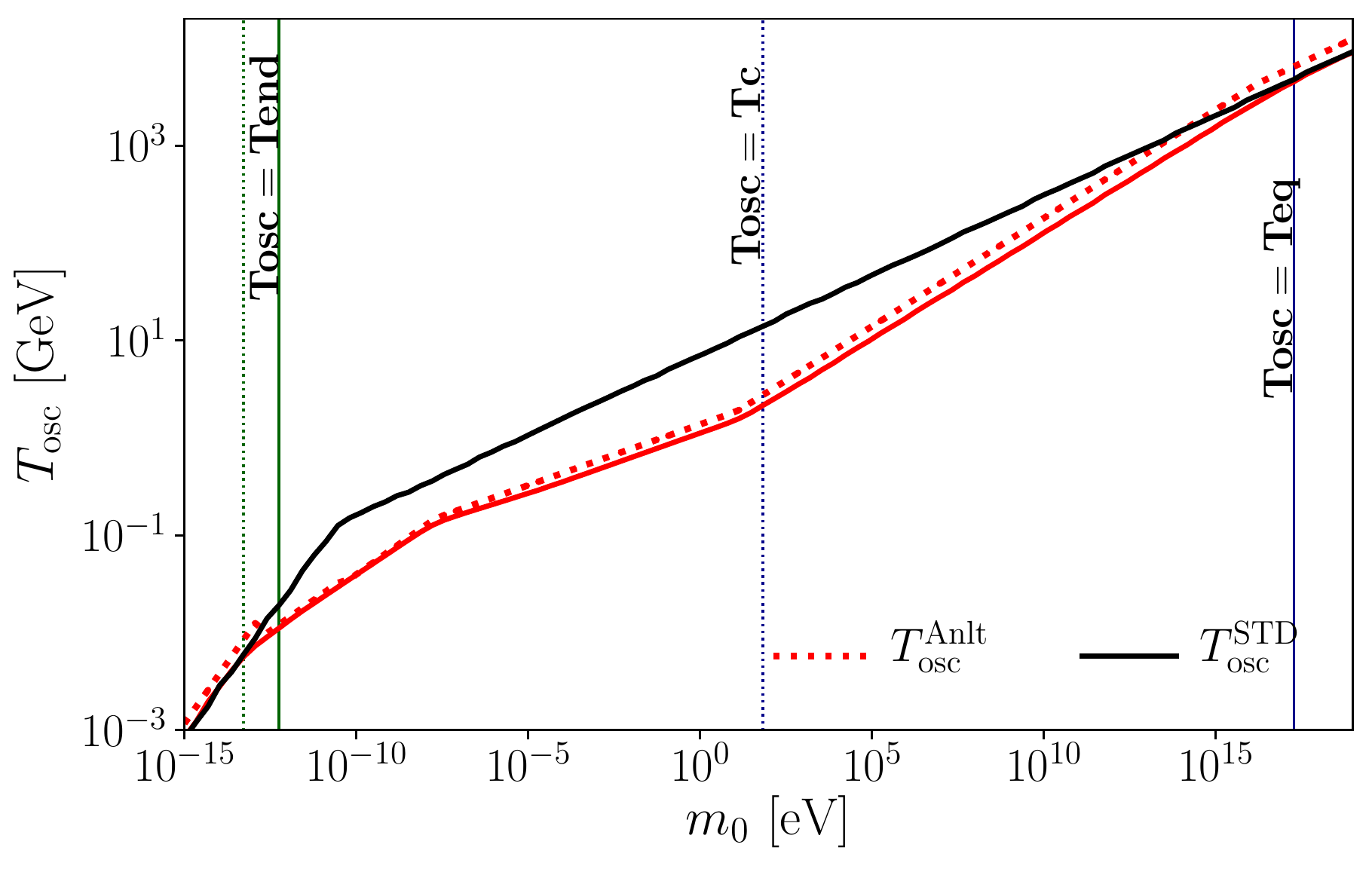}
\caption[The oscillation temperature as a function of the axion mass for a non-standard and standard cosmology.]{The oscillation temperature as a function of the axion mass for a non-standard (red) and standard (black) cosmology. For the non-standard cosmology we have used $\beta=1$, $\kappa=10^{-4}$, $\Tend= 10$ MeV, $T_i=10^{5}$ GeV. The vertical lines divide the regions mentioned above and dotted line is for the analytical $\Tosc$.}
\label{fig:Tosc_nsc}
\end{figure}

Following this line of thought, the oscillations of the axion field can take place in the 3 periods mentioned above. First, is the high temperature period of Region 1, where $\Tosc$ is the same as in the SC. Therefore, if it is possible to get the right abundance for those range of (high) masses, the relic density has to be well diluted. Then, the oscillation temperature decreases, \footnote{Note that always the oscillation temperature in the NSC will be lower than in the SC for a given mass, because the expansion rate  $H(T)$ in a non-standard cosmology is greater than in the standard cosmology, such that equality with the axion mass occurs at lower temperatures. } and according to \cref{eq:NSC_axion_density}, a smaller $\Tosc$ will imply  a higher energy density, but, oscillations in this region will also have the same dilution as in the previous region. Thus, in here there will be the competing of both effects. If the oscillation happens in Region 3, the temperature is lower than in the previous region, but also the dilution from the decay of $\phi$. In \cref{fig:Tosc_nsc} we show the oscillation temperature $\Tosc$ vs the axion mass $m_0$ in a NSC scenario with $\beta=1$ and we have highlighted the boundaries of the 4 periods. Thus, it can be seen that for every region the oscillation temperature has a change in its slope, which in addition to the entropy injection to the thermal bath, could potentially open the axion CDM window. In the next section we will focus on developing a comprehensive analysis of how the evolution of the oscillation temperature and the axion dilution in every regions lead to a different relic density today than in the standard cosmology.

\subsection{Region 1:  \texorpdfstring{$\Rosc\ll\Req$}{TEXT}}
In this region, the oscillations of the axion field occur at $\Rosc\ll \Req$, thus, in a period of radiation domination,  before the influence of $\phi$. Such period is only present in cosmologies where $\kap<1$ and it only differs from the standard cosmological scenario in that there is entropy transfer to the SM, not to axions, causing a dilution of the dark matter density.
The oscillation temperature is the same as the standard cosmological scenario, eq.~(\ref{eq:std_tosc}).
Since $\Rosc\ll\Rc$ in this region, the entropy is conserved between these two scale factors, such that,  $S_{\rm osc}/S_{\rm c}=1$, where $S_{\rm c}$ is the total entropy at $R=\Rc$. Therefore, we can find an expression for  the entropy dilution factor, valid when the fluid gets to dominate the expansion of the universe\footnote{Meaning valid only when the chosen parameters $\beta$, $\kap$, $T_i$ and $\Tend$ lead to a solid fluid domination. }, as 
{{\be
\gamma_{R_1}=\frac{S_{\rm osc}}{S_{\rm end}}=\left[\left( \frac{\kappa\beta^2}{4}\right) \left(\frac{T_i}{\Tend}\right)^{4-\beta} \right]^{-3/\beta}.
\label{eq:gamma_R1}
\ee}}
So, in this regime,
\be
\Omega_a=\Omega_a^{\rm std} \gamma_{R_1}.\label{eq:axion_densityR1}
\ee
By inspecting $\gamma_{R_1}$ it can be found that it gets smaller as $\beta$ decreases, meaning that the dilution it is much more significant for those cosmologies. Thus, from the above expression, it is quite clear that for axion masses such that in the standard cosmology lead to an overproduction of  DM, they will get much diluted if the oscillation of the field takes place in region 1. And that - in principle, but we will discuss it below - a lower $\beta$ will widen the axion window much more dramatically. 

There is, however, a condition for the oscillation of the axion field to  occur during this period, and is given by $\Rosc\ll \Req$ (or, equivalently, $\Teq\ll \Tosc)$
\be
T_i\,\kappa^\frac{1}{4-\beta} \ll {\Tosc}. 
\label{eq:region_1_window}
\ee

We can rewrite the requirement \eqref{eq:region_1_window} in terms of the axion mass, since the oscillation temperature is defined as $3H(\Tosc)=m_a(\Tosc)$.  Thus replacing the expressions for $\Tosc$, eqs.~(\ref{eq:std_tosc}), we find for $\beta=3$
\bea
4\times 10^{-14}\,\mbox{eV}\, \left(\frac{T_i}{10^{3}\,\mbox{GeV}}\, \frac{\kap}{10^{-5}}\right)^2\ll m_0,\,\,\,\,\,m_0\lesssim m_{R_1},\label{eq:r1_w1}\\
8\times 10^{-7}\,\mbox{eV}\, \left(\frac{T_i}{10^{5}\,\mbox{GeV}}\, \frac{\kap}{10^{-5}}\right)^6\ll m_0,\,\,\,\,\,m_0\gtrsim m_{R_1},\label{eq:r1_w2}
\eea
and considering $\beta=2$
\bea
0.8 \,\mbox{MeV}\, \left(\frac{T_i}{10^{4}\,\mbox{GeV}}\, \frac{\kap}{10^{-4}}\right)^2\ll m_0,\,\,\,\,\,m_0\gtrsim m_{R_1},\label{eq:r1_w1b2}
\eea
Where $m_{\rm{R_1}}$ is the mass at which $\Tosc\sim \LambdaQCD$ and  marks  the separation between temperature effects on the axion mass. We see that for $\beta=3$, the first window -  for masses satisfying $m_0\lesssim 10^{-5}\mu\mbox{eV}$, such that temperature effects are not important on the axion mass - it is  required a rather low $T_i$, which is potentially dangerous, since even by choosing $\Tend \sim \TBBN$, $\phi$ might not get to dominate over radiation in such short-lived NSC. On the other hand, the second window for $\beta=3$, where temperature effects on the mass are important,  could be achieved more naturally, although  a portion in the upper bound is already ruled out by astrophysics, since by observations of SN1987A, the axions are required to be lighter than  $m_0 \leq 10^{-3}$ eV \cite{peccei2008strong,visinelli2011axions}. On the other hand, for $\beta=2$ -  or smaller -  we have found that it is quite hard to make the oscillation happening when the mass does not get temperature corrections, the reason is that for cosmologies with $\beta\lesssim 2$, the field $\phi$ starts to dominate earlier such that the Region 1 ends earlier (high $\Teq$). Thus, in order for the oscillation to happen in region 1, we require high oscillation temperatures (see also fig.~(\ref{fig:Tosc_nsc})), which in turn implies high axion masses. That is the reason why in eq.~(\ref{eq:r1_w1b2}) we only show one condition on the axion mass - where temperature effects are important - since for the other case, that requires $m_0\lesssim m_{R_1}$, there is no cosmology that could make an axion of such small mass to oscillate in that period. We then conclude that for $\beta \leq 2$ is quite unlikely for an axion to oscillate in this period. 

Moving now to analyze the relic density of axion dark matter, we have seen that in this region, the latter can be written as $\Omega_a=\Omega_{\rm std} \gamma_{R_1}$, then, defining $\tau\equiv T/T_i$, it can be expressed as 
\bea
\Omega^{R_1}_a =
\begin{dcases}
0.21 \left(\frac{m_0}{10^{-13}\,\rm{eV}}\right)^{-3/2} \left(\frac{\tau_{\rm end}}{3 \times 10^{-12
}}\frac{10^{-2}}{\kappa}\right) {\theta_i}^2, & m_0\lesssim m_{R_1},\\
0.15\left(\frac{m_0}{10^{-7}\,\rm{eV}}\right)^{-7/6} \left(\frac{\tau_{\rm end}}{10^{-6}}\,\frac{10^{-4}}{\kappa}\right) {\theta_i}^2, & m_0\gtrsim m_{R_1}.
\end{dcases}
\label{eq:relic_R1values}
\eea
Where, inspired by the discussion above we only write the case $\beta=3$, corresponding to an early matter domination. Since the relic density depends only on $\tau_{\rm end}$, and the mass windows for Region 1, given in eqs.~\eqref{eq:r1_w1} and \eqref{eq:r1_w2}, do not depend on $\Tend$, it is quite easy to fulfill the required $\tau_{\rm end}$ as a combination of $T_i$. For instance, a $\tau_{\rm end}=10^{-12}$ can be obtained for $T_i \gtrsim 10^{9}~$Gev, and $\tau_{\rm end}=10^{-6}$ can be achieved by $T_i\gtrsim 10^3$~GeV (by requiring  $\Tend \gtrsim \TBBN $).

So, from eq.(\ref{eq:relic_R1values}), for the same mass than in a standard cosmology, in Region 1 we will have an underproduction of axions, due to they are highly diluted in the universe. By this reason we expect that the axion band lie to lower masses than the standard.

\subsection{Region 2:\texorpdfstring{ $\Req\ll \Rosc\ll \Rc$}{TEXT}}

In this regime, the axion oscillation happens during the domination of the new field $\phi$, but their decays still do not affect the evolution of the temperature in the SM sector, so $T\propto R^{-1}$. 
The Hubble parameter - assuming the field completely dominates the expansion - in this case is given by
\begin{equation}
H(R)\approx\sqrt{\frac{\rho_\phi(R)}{3M_P^2}},
\end{equation}
where the expression for $\rho_\phi$ to be considered is as in eq.~(\ref{eq:energies_free}). By equating $3H(\Tosc)=m_a(\Tosc)$  we find the temperature at the moment of oscillation, which again can be split in two, depending on whether temperature effects are important or not, meaning $\Tosc \gtrsim \LambdaQCD$ and $\Tosc \lesssim \LambdaQCD$, respectively. Now the expression for the oscillation temperatures are more involved, so the full analytical expressions are given in \eqref{app:eq:ToscR2} in the Appendix. For the sake of clarity, let us write this temperature for some benchmark NSC scenarios:
\begin{align}
\beta=2;\quad \Tosc^{R_2}=
0.26~\mbox{GeV} \left(\frac{m_0}{1\,\mbox{meV}}\, \frac{10^7\,\mbox{GeV}}{T_i}\right)^{1/5} \left(\frac{10^{-3}}{\kap}\right)^{1/10},\quad& m_0\gtrsim m_{R_2}.
\label{eq:R2_Toscbeta2}
\end{align}

\begin{align}
\beta=3;\,\,\,\Tosc^{R_2}=
\begin{dcases}
0.01~\mbox{GeV} \left(\frac{m_0}{10^{-4}\,\mu\mbox{eV}}\right)^{2/3} \left(\frac{10^7\,\mbox{GeV}}{T_i}\,\frac{10^{-4}}{\kap}\right)^{1/3},& m_0\lesssim
m_{R_2}, \\
1.5~\mbox{GeV} \left(\frac{m_0}{1\,\mbox{meV}}\right)^{2/11}\left( \frac{10^7\,\mbox{GeV}}{T_i}\frac{10^{-4}}{\kap}\right)^{1/11},& m_0\gtrsim m_{R_2}, 
\end{dcases}
\label{eq:R2_Toscbeta3}
\end{align}
where $m_{\rm{R_2}}$ sets  the  axions  mass  above  which  the  temperature  effects  start  to  be important and their analytical expression can be found in (\ref{app:eq:mR2}). We write their values for the cosmologies shown, below.
Let us note that for the same reason as the one invoked in region 1,  we write only one condition, eq.~(\ref{eq:R2_Toscbeta2}) for $\beta=2$, corresponding to $\Tosc\gtrsim \LambdaQCD$. Namely, there are no cosmologies for $\beta\leq2$, such an axion of mass $m_0\lesssim m_{R_2}$ will oscillate in that range of (high) temperatures. 
The intersection mass $m_{R_2}$, marks the transition from temperatures higher than $\LambdaQCD$ ($m_0\lesssim m_{R_2}$) and lower than $\LambdaQCD$ ($m_0\gtrsim m_{R_2}$),  and for the above mentioned cosmologies takes the values 
\be
m_{R_2}=\begin{dcases}
58\,\mu\mbox{eV}\, \left(\frac{T_i}{10^7\,\mbox{GeV}}\right)\left(\frac{\kap}{10^{-3}}\right)^{1/2}, & \mbox{for}\,\,\,\,\beta=2,\\
10^{-3}\,\mu\mbox{eV} \left(\frac{T_i}{10^7\,\mbox{GeV}}\frac{\kap}{10^{-4}}\right)^{1/2},& \mbox{for}\,\,\,\,\beta=3.
\end{dcases}
\label{eq:NSC_R2_mass}
\ee
As we can see, for  cosmologies with equations of state smaller than radiation, the QCD transition could be shifted to happen at higher axion masses (thus, higher temperatures) than the standard cosmological scenario, where it happens at $\sim 10^{-5}\,\mu$eV. 

The condition to assure that the oscillation of the axion field happens in this period is $\Req\ll \Rosc\ll \Rc$. Thus, in the same way we did for the previous region, imposing their consistency leads to ranges for the axion mass, such that for them, the oscillation can take place at Region 2.  The analytical expressions are in the Appendix \ref{app:reg2} eqs.~(\ref{app:eq:NSC_R2_Tbound}) and (\ref{app:eq:NSC_R2_0bound}).
We will write here the mass ranges  for $\beta=2$ and $\beta=3$, respectively 
\begin{equation}
17{\rm{meV}}\! \left[\!\left(\frac{\kappa}{10^{-3}}\right)^{2} \left(\frac{\Tend}{\rm{MeV}}\!\right)^{5}\! \left(\!\frac{T_i}{10^{7}\rm{GeV}}\!\right)^4\right]^{2/3}\!\ll m_0\ll
10^{18}{\rm{GeV}} \left(\frac{\kappa}{10^{-3}}\right)^{3} \left(\!\frac{T_i}{10^{7}\rm{GeV}}\!\right)^{6}\!,
\label{eq:R2_range2} 
\end{equation}

\begin{equation}
\begin{aligned}
10^{-5}\,\mu{\rm{eV}}\left(\frac{\kappa}{10^{-4}} \frac{T_i}{10^7\rm{GeV}}\right)^{4/5}\left(\frac{\Tend}{\rm{ MeV}}\right)^{6/5} \ll m_0\ll  10^{3}\,{\rm{GeV}}\left(\frac{\kappa}{10^{-4}} \frac{T_i}{10^{7}\rm{GeV}}\right)^{6}. \label{eq:R2_range3}
\end{aligned}
\end{equation}

Let us immediately note that for $\beta=2$, the whole mass range is above the intersection mass $m_{R_2}$, shown in eq.~\cref{eq:NSC_R2_mass}, meaning that for axions in that mass range oscillate at temperatures such that we must always take into account thermal effects on the mass. The range could be lowered to smaller masses, but at the price of choosing smaller (rather fine-tuned) values  of $T_i$ and $\kappa$, and paying special attention that it might be that they will lead to cosmologies where $\phi$ never gets to dominate over radiation, such that the above expressions are not longer valid.   The situation is better for $\beta=3$, where the mass range for the oscillation to happen in this region includes both, temperatures such that the mass receives thermal corrections and where it does not. We also notice that for both $\beta=2$ and 3, the upper part of the mass range can be in conflict with the bound from SN1987a and the HB stars such $f_a>10^9$~GeV, which translates to $m_0\sim $~meV.
\begin{figure}[t]
	\centering 
		\includegraphics[scale=0.6]
		{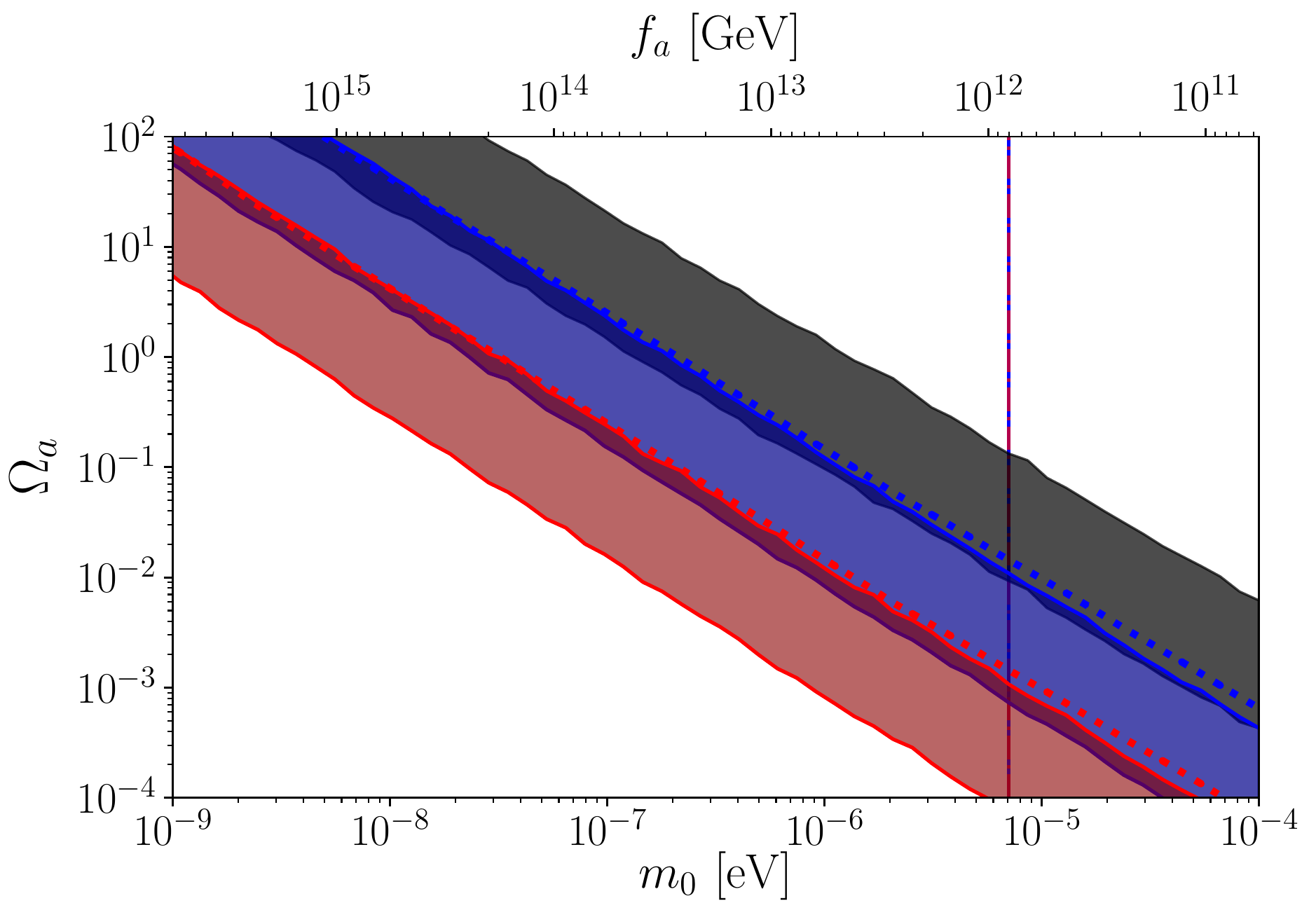}

\caption[ The axion abundance as a function of the axion mass for a non standard cosmology with $\beta=3$]{The axion abundance as a function of the axion mass for a non standard with $\beta=3$ and a standard cosmology (black band). The parameters for the NSC are: $\Tend=4$ MeV (red band) and $\Tend=40$ MeV (blue band), the rest of parameter correspond to $\kappa=10^{-5}$ and $T_i=10^{5}$ GeV. The vertical lines are the mass at which $\Tosc=T_{\rm{eq}}$ for each case, since the upper limit in Region 2 does not depend on $\Tend$, they are overlapping. The bandwidth is obtained by considering $\theta_i \in [0.25,1]$. The dotted line corresponds to the analytical result for the relic obtained in appendix eq.~(\ref{eq:NSC_R2_relic0}) and eq.~(\ref{eq:NSC_R2_relicT}) with an initial angle $\theta_i$= 1 which describes upper edge of the each NSC band.\label{fig:OmegR2}}
\end{figure}

Now we tackle the most important point: is it possible that the right DM abundance is produced from the oscillation of the axion field inside Region 2? To find out, in Appendix (\ref{app:reg2}) we detail the whole calculation that leads to the expression of the relic density, here we will write it down the main result:
\bea
\Omega_{R_2}=
\begin{dcases}
\Omegastdo\left(\frac{\beta^2}{4}\right)^{-3/\beta} \left(\frac{\Tend^2}{m_0 M_P}\right)^{\frac{6}\beta-\frac{3}2} & \mbox{for}\,\, m_0\lesssim m_{R_2},\\
\OmegastdT  \gamma_{R_2} \left( \frac{\kappa^{7/(4-\beta)}\, T_i^7}{\left(\alpha \LambdaQCD^4 \, m_0 M_P\right)^{7/6}}\right)^{\frac{4-\beta}{\beta+8}} &
\mbox{for}\,\, m_0\gtrsim m_{R_2}.
\end{dcases}
\eea
Where $\gamma_{R_2}$ is the entropy dilution factor, and is the same as the one for region 1, eq.~(\ref{eq:gamma_R1}), which, as we have previously discussed, is  a coefficient always smaller than 1.  From the above expression, we note that for masses $m_0\lesssim m_{R_2}$, or temperatures $\lesssim \LambdaQCD$, the relic density only depends on $\beta $ and $\Tend$. Also, from the same expression, seems clear that for a given axion mass, we could have cosmologies where the relic density is lower, higher or the same as the SC scenario. Nonetheless, we have argued before that if the oscillation happens during this region of $\phi$ dominance, it will be at rather high temperatures, making more likely to be in the case where the axion mass receives thermal corrections.  On the other hand, for the latter case, the expression for the relic density is more involved, and depends on $\Tend$ and $\beta$ - as it was expected -  and also on both  $T_i$ and $\kappa$. After a quick inspection of the coefficient that multiplies at $\OmegastdT$, it can be seen that it gets smaller as $\beta$ decreases. Thus, making more likely to open up the classical mass window to smaller masses for those cosmologies. 
For a better insight, we have evaluated the dark matter relic density using some benchmark values of the NSC, for a matter dominated period ($\beta=3$), and we get

\begin{align}
\Omega^{R_2}_a =
\begin{dcases}
0.4 \left(\frac{10^{-3}\,\mu\rm{eV}}{m_0} \right)^{2}  \frac{\Tend}{
 \rm{MeV}} {\theta_i}^2, & m_0\lesssim m_{R_2}, \nn\\
0.01 \left[ \left(\frac{ \rm{meV}}{m_0}\right)^{14} \left(\frac{10^{7} \rm{GeV}} {T_i}\frac{10^{-4}}{\kappa}\right)^{4} \right]^{\frac{1}{11}} \frac{\Tend}{ 10^{2}\,\rm{GeV}}\,{\theta_i}^2 & m_0\gtrsim m_{R_2}.
\end{dcases}  \nn
\end{align}

Let us note that for both regimes, the dependence on $m_0$ it is stronger in the NSC than in the SC. Also, the dependence of the relic density on $\Tend$ comes from the entropy dilution from the decay of $\phi$. Thus, as expected, for smaller $\Tend$ temperatures, the relic density gets more diluted. The latter is supported
by figure \cref{fig:OmegR2}, where we have chosen a low $\Tend$ equal to 4 MeV (red band) to ensure a high dilution rate compared to $\Tend$ equal to 40 MeV (blue band).

\subsection{Region 3: \texorpdfstring{$\Rc\ll \Rosc\ll \Rend$ }{TEXT} }
For this regime, oscillations of the axion field happen during the domination of  $\phi$, and its decays  affect the temperature in the SM.  The Hubble parameter is taken to be $H\sim \sqrt{\frac{\rho_\phi}{3 M_P}}$ and $T$ and $R$ are related now by $T\propto R^{-\beta/8}$, see  \cref{eq:T_NSC}.
Let us also remind that during this period, due to the entropy injection, the expansion of the universe follows the behaviour $H\propto T^4$.
Following the analysis presented in the previous section, we start by finding the expressions for the oscillation temperature by using $3H(\Tosc)=m_a(\Tosc)$. The analytical expressions can be found in \cref{eq:Tosc_R3}. We notice that the dependence on $\beta$ is very mild, thus, we can consider it as valid for any equation of state:
\begin{align}
\Tosc^{R_3}=
\begin{dcases}
0.04~\mbox{GeV} \left(\frac{m_0}{10^{-4}\,\mu\mbox{eV}}\right)^{1/4} \left(\frac{\Tend}{10 \mbox{MeV}}\right)^{1/2} & m_0\lesssim m_{R_3}
 \\
1~\mbox{GeV} \left(\frac{m_0}{1\,\mu \mbox{eV}} \right)^{1/8} \left( \frac{\Tend} {10\,\mbox{GeV}}\right)^{1/4} & m_0\gtrsim m_{R_3}
\end{dcases}
\label{eq:R3_Tosc}
\end{align}
\begin{figure}[t]
	\centering 
\includegraphics[scale=0.6]{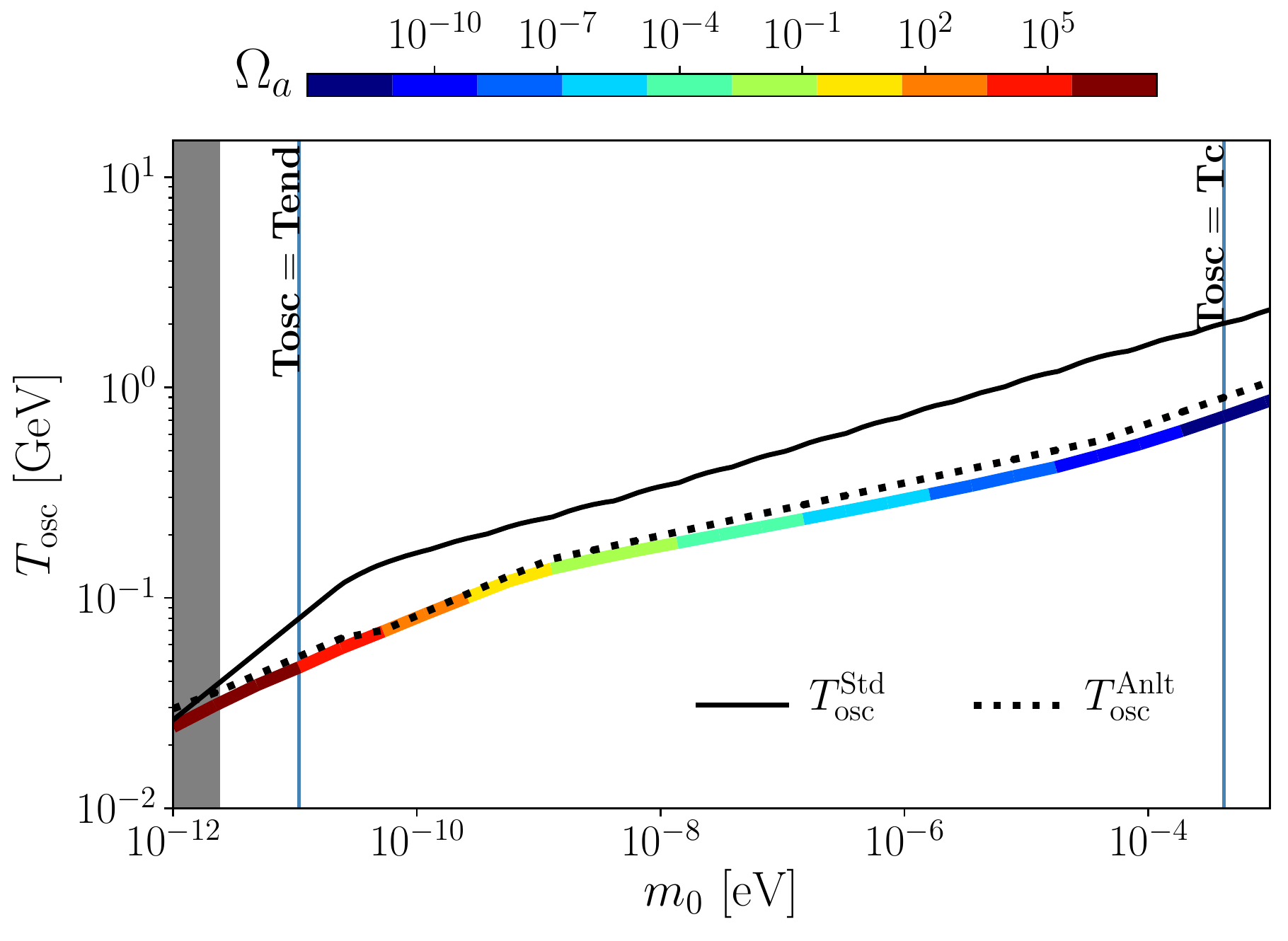}
\caption[The oscillation temperature as a function of the axion mass for a non-standard cosmology with $\beta=2$]{The oscillation temperature as a function of the axion mass for a non-standard and standard (black) cosmology. The dotted line corresponds to \cref{eq:Tosc_R3}. For the non-stardad cosmology we have used $\beta=2$, $\kappa=10^{-5}$, $\Tend= 40$ MeV, $T_i=10^{5}$ GeV. The intersection of mass happens at $m_0=10^{-9}$ eV and the correct abundance for axion CDM is achieves with a mass $1.1\times 10^{-9}$ eV. For bigger masses there is an overproduction of axions. The gray area is the bound for scales over the Planck scale. \label{fig:Tosc_R3}}
\end{figure}
The oscillation temperature in Region 3 is the lowest of the three regions analyzed here. As can be seen from \cref{fig:Tosc_R3}, after $T=\Tc$, the oscillation temperature departs the most from the one of the SC. Let us notice that the dependence on $m_0$ is - in both cases -  softer than in the SC, thus, varying more smoothly. We can also add that the most important parameter in this region is $\Tend$, and considering high values will make $\Tosc$ to go up, as expected since the NSC gets shorter. The intersection of the two regimes sets the moment where the axion mass can be considered as temperature independent, which for Region 3 happens at  eq.~(\ref{eq:int_mass_R3}) 
\be
m_{R_3}\approx 10^{-3}\, \mu{\rm{eV}}\,(8-\beta) \left(\frac{10^{-2}\,{\mbox{GeV}}}{\Tend}\right)^2.
\label{eq:mR3}
\ee
Thus, below $m_{R_3}$ the temperature effects can be safely ignored, while above, they have to be considered. In here the dependence on $\Tend$ is inversely proportional, meaning that cosmologies with higher $\Tend$ move the intersection mass to smaller values, therefore, for those cosmologies is very likely that thermal effects on the mass have to be considered.

In order to find the mass range that sets the axion oscillation during this period, we again require the consistency of the relation $\Req\ll \Rosc\ll \Rend$
see appendix~(\ref{app:reg3}) for details.  For $\beta=2$ and $\beta=3$, respectively, we get
\begin{equation}
\begin{aligned}
10^{-7}\,\mu \mbox{eV}\left(\frac{\Tend}{10^{-2}\,\rm{GeV}}\right)^2\ll m_0 \ll  2\, \mbox{MeV}\left[\left(\frac{\Tend}{10^{-2}\,\mbox{GeV}}\right)^{10}\left(\frac{T_i}{10^{10}\,\mbox{GeV}}\right)^8 \left(\frac{\kappa}{10^{-5}}\right)^4\right]^{1/3},
\end{aligned}
\end{equation}

\begin{equation}
\begin{aligned}
10^{-8}\,\mu\mbox{eV}\left(\frac{\Tend}{10^{-2}\,\rm{GeV}}\right)^2\ll  m_0\ll 10^{-5}\,\mbox{eV}\left[\left(\frac{\Tend}{10^{-2}\,\mbox{GeV}}\right)^{22} \left(\frac{T_i}{10^{11}\mbox{GeV}}\right)^8 \left(\frac{\kappa}{10^{-4}}\right)^8\right]^{1/5}, 
\end{aligned}
\end{equation}
here, considering a low $\Tend$ and a high $T_i$ - thus, a long NSC - the upper limit remains high, while the lower limit can reach very small masses, allowing Region 3 to include a wide range of masses. In this case, for both $\beta=2$ and $\beta=3$, the mass range where oscillations of the axion happen in Region 3 can accommodate both regimes: temperatures higher than $\LambdaQCD$ and smaller, but only if rather small $\Tend$ temperatures are considered. Otherwise, the temperature of oscillation is too high, and always the thermal effects on the mass have to be considered. 

Our next step is to find the axion  abundance that different cosmologies predict in this region. Since the oscillation of the axion field happens already when decays are important, it is not possible to use the dilution factor found in the previous regions to compute the relic density of DM. But, it is still straightforward to find it, since it is given by $S_{\rm{osc}}/S_{\rm{end}}$, and with the aid of the expressions for $\Rosc$ and $\Rend$ in \cref{eq:rend},  we obtain
\be
\gamma_{R_3}=\left(\frac{4}{\beta(8-\beta)}\right)^{6/\beta} \left(\frac{\Tend}{\Tosc}\right)^{24/\beta-3},
\label{eq:gamma_R3}
\ee
here, for smaller $\beta$, as expected, the dilution is more important. On the other hand, since $\Tend/\Tosc$ is bigger than the ratio $\Tend/T_i$ from the \cref{eq:gamma_R1}, we obtain the relation $\gamma_3 > \gamma_1$. We interpret it as, since in Region 3 the field $\phi$ is already decaying, the entropy injection into the thermal bath is smaller than in the previous regions. 
Then, the axion abundance is given by the following expression:
\bea
\Omega_{R_3}=\begin{dcases}
 \Omegastdo\left(\frac{2}{\beta}\right)^{6/\beta} \left(\frac{\Tend^2}{m_0 M_P}\right)^{\frac{3}\beta (4-\beta)} & \mbox{for}\,\, m_0\lesssim m_{R_3},\\
\OmegastdT\, (8-\beta)^{\frac{\beta+6}{2\beta}} \left(\frac{\Tend^6}{\alpha \LambdaQCD^4 m_0 M_P}\right)^{3/\beta-2/3} & \mbox{for}\,\, m_0\gtrsim m_{R_3}.
\end{dcases}
\label{eq:RelicR3}
\eea

\begin{figure}[t]
    \centering
    \includegraphics[scale=0.6]
    {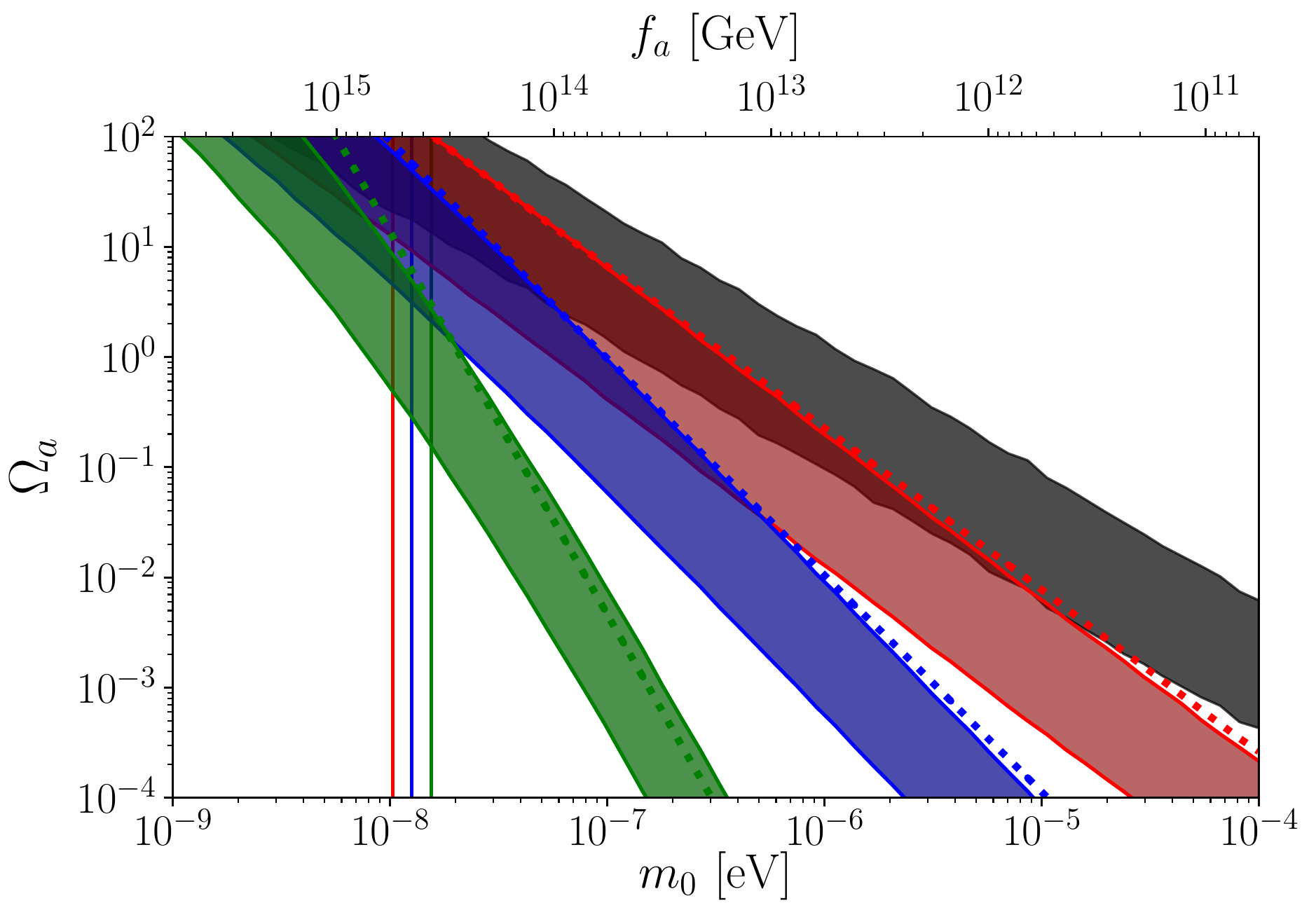}
\caption[The axion abundance as a function of the axion mass for three non standard  cosmologies]{The axion abundance as a function of the axion mass for three non standard  cosmologies. Black band is for standard cosmology, red band is for $\beta=3$, blue band for $\beta=2$ and the green band is for $\beta=1$.  The rest of parameters are:  $\Tend=0.3$ GeV, $\kappa=10^{-3}$ and $T_i=10^{6}$ GeV. The bandwidth is given by $\theta_i \in [0.25,1]$. The dotted line corresponds to analytical result for the relic obtained in \cref{eq:R3_relic} with an initial angle $\theta_i=1$, which describes the upper edge of the each NSC band. The vertical lines corresponds to the mass at which $\Tosc=\Tend$ for each $\beta$.}
    \label{fig:relic_R3}
\end{figure}

 From eq.~(\ref{eq:RelicR3}) we can see that again it is possible to write the relic density as the standard cosmology one, times a factor related to the NSC, which in this case corresponds only to $\Tend$ and $\beta$.  For the expression $m_0\lesssim m_{R_3}$, where the axion mass can be considered as temperature independent, we can first see that in order for the intersection mass $m_{R_3}$ not to get extremely small (recall that it is inversely proportional to the scale $f_a$, so it translates into a very high PQ scale), $\Tend$ shall be considered to be $\lesssim$~ GeV. For those values of $\Tend$ and axion masses, the coefficient in front of $\Omegastdo$ is always smaller than 1, thus, in order to get the right abundance, we have to consider masses smaller than we would  have to in the SC.  
On the other hand, in the mass range where temperature effects have to be considered, $m_0\gtrsim m_{R_3}$, $\Tend$ should also remain far from high, since in that case, again $m_{R_3}$ goes up, producing a subproduction of axion dark matter.

By way of estimation, we compute the axion abundance for $\beta=2$, as in the previous regions, the abundance was computed by  \cref{eq:NSC_axion_density}  for masses temperature independent and dependent.
\be
\Omega^{R_3}_a =
\begin{dcases}
0.4\left(\frac{10^{-4}\,\mu \rm{eV}}{m_0}\right)^3
\left(\frac{\Tend}{ 10^{-2} \,\rm{GeV}}\right)^{3} {\theta_i}^2 &\mbox{for}\,  m_0\leq m_{R_3}. \\
0.23 \left(\frac{10^{2}\,\mu\rm{eV}}{m_0} \right)^{2} \left( \frac{\Tend}{2 \,\rm{GeV}} \right)^5 {\theta_i}^2, &\mbox{for}\, m_0\geq m_{R_3}.
\end{dcases}
\label{eq:relicR3}
\ee

In \cref{fig:relic_R3} we show the axion relic abundance when the axion oscillates in Region 3 for $\beta=3$ (red), $\beta=2$ (blue) and $\beta=1$ (green).  We have offset the high mass range by choosing a lower initial temperature $T_i=10^{6}$ GeV and a long NSC by $\Tend=0.3$ GeV. Here, the 3 cosmologies belong to the regimen when the temperature effects are important on the mass.  We can see that after the mass at which $\Tosc=\Tend$ (vertical line), each cosmology begins to converges to the standard cosmology (black band). During Region 3, it is clear that the difference in slope between the 3 non-standard cosmologies is attributed to dilution effects.

Another effect of the non-standard evolution of the temperature is that the relic bandwidth is reduced because in Region 3, for small $\beta$, the relic abundance has a stronger dependence on the axion mass, which means that a  variation in axion mass generates a large change in its abundance, moving away from $\Omega_{DM,0}=0.26$.

The mismatch in \cref{fig:relic_R3} between the analytic result regarding to the upper edge in the numerical band is due to the fact that we considered the degrees of freedom as a constant value.

\subsection{Axion coupling to two photons}
\begin{figure}[h!]
	\centering 
	\begin{subfigure}[b]{1\textwidth}
	\centering
		\includegraphics[scale=0.3]{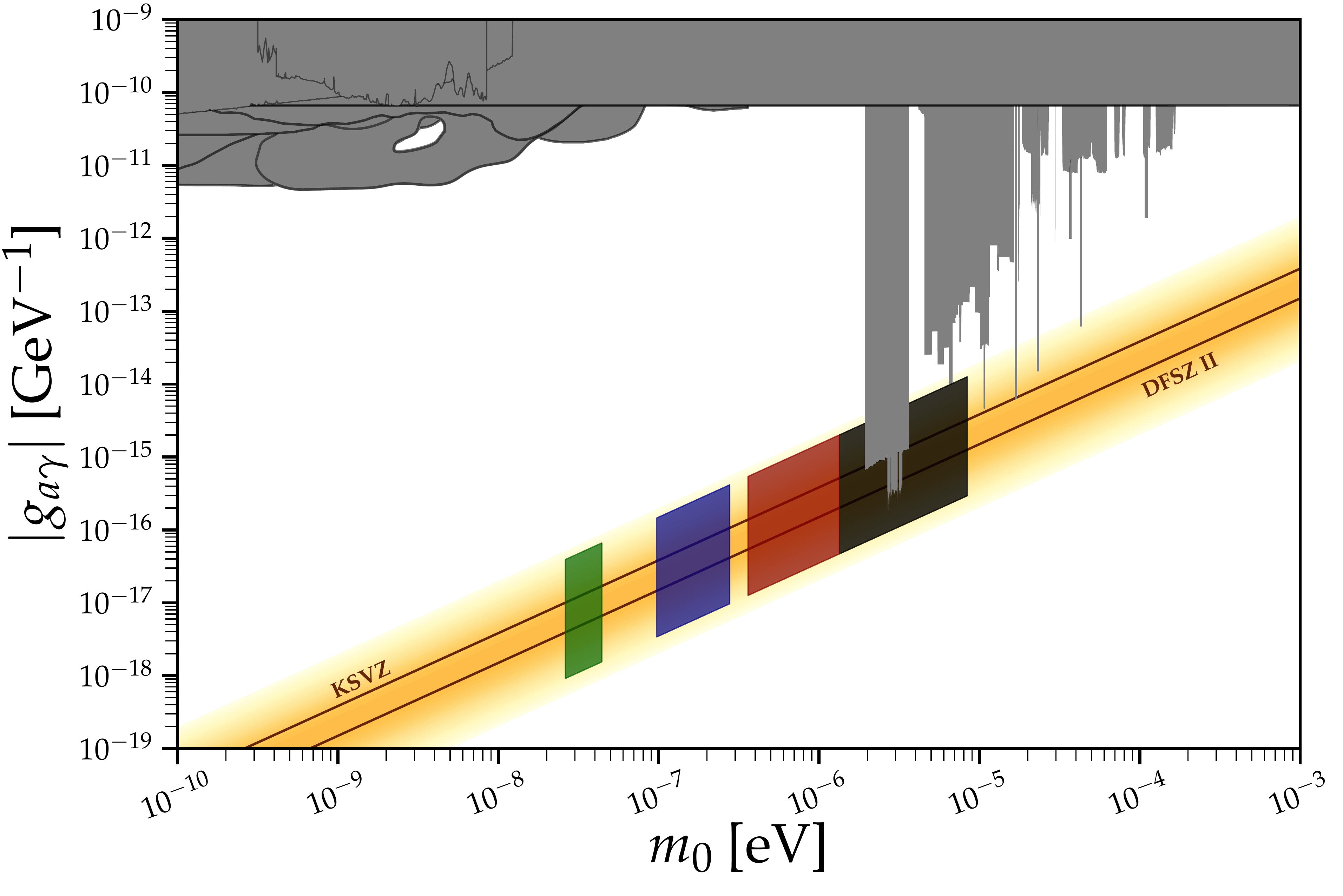}
    \caption{ \label{fig:ga_NSC}}
    \end{subfigure} %
    \par
    \begin{subfigure}[b]{1\textwidth}
	\centering 
		\includegraphics[scale=0.3]{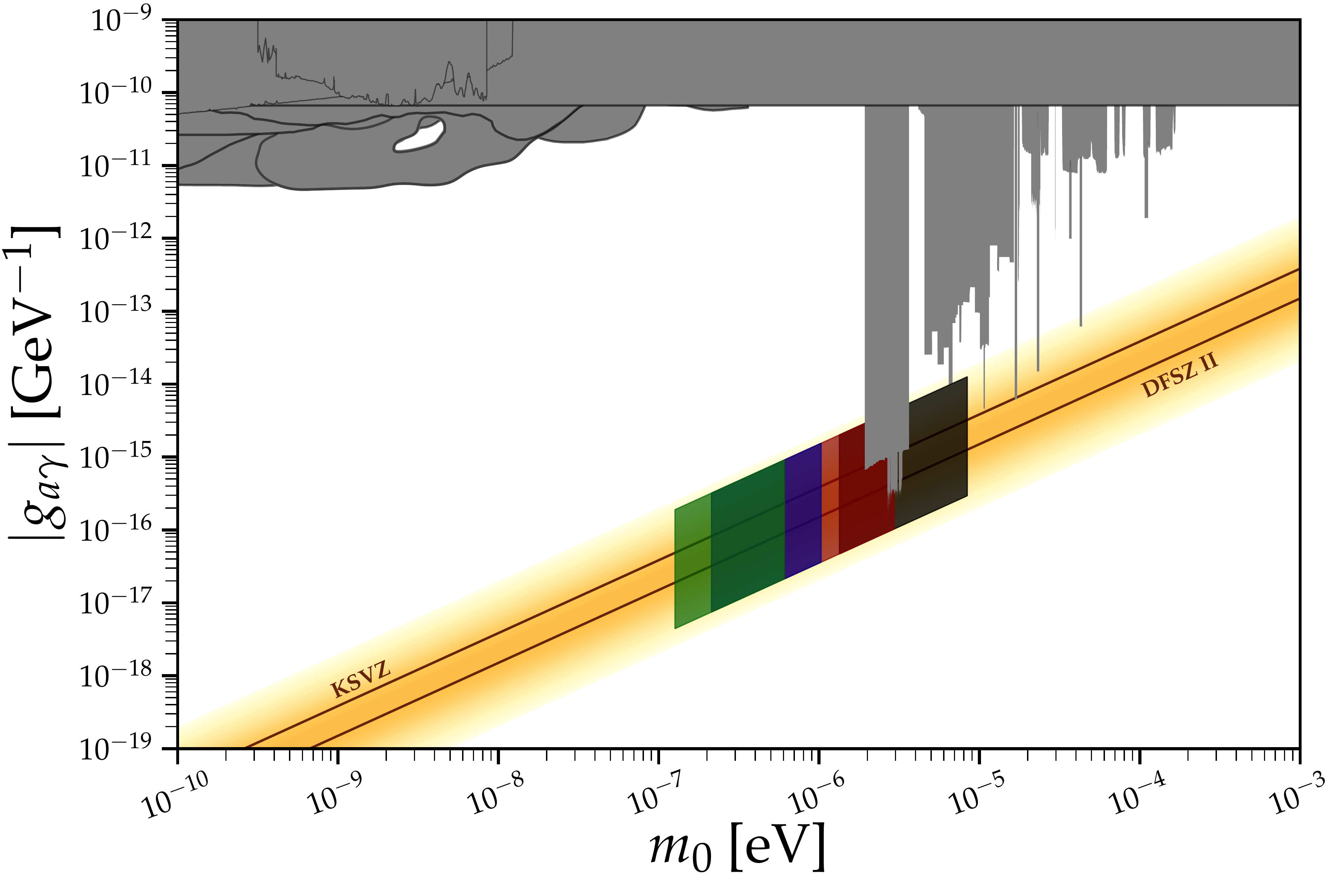}
    \caption{ \label{fig:ga_NSC_2}}
    \end{subfigure}
\caption[The axion parameter spaces for the axion-photon coupling in standard and non-standard cosmologies]{The axion parameter spaces for the axion-photon coupling in standard and non-standard cosmologies with the current experimental constraints. The colored areas show the parameters where we can obtain the correct axion CDM abundance for standard cosmology (black) and cosmologies with $\beta=3$ (red area), $\beta=2$ (blue area) and $\beta=1$ (gren area). The bandwidth is obtained by considering the initial angle $\theta_i \in [0.5,1.8]$ such to obtain the correct abundance $\Omega_{DM,0}=0.26$ (a) For the upper plot we take the rest of parameter as $\Tend=0.3$ GeV, $T_i=10^{6}$ GeV and $\kappa= 10^{-3}$. (b) We keep $\Tend$ as 0.3 GeV but we consider a different $T_i$ and $\kappa$ for every cosmology, such that for $\beta = 3$ : $T_i=10^{7}$ GeV and $\kappa=10{-7}$, for $\beta=2$: $T_i=10^{3}$ GeV and $\kappa=10^{-6}$ and for $\beta=1$: $T_i=10$ GeV and $\kappa=10^{-4}$. }
\end{figure}

Now we would like to take what has been learned from the previous analysis into the axion
phenomenology. As we have mentioned in the introductory chapter, the most exploited coupling to search for axions is the one to two photons. Thus, it would be very useful to see how different NSC can change the scenario for axion dark matter. In \cref{fig:ga_NSC} and \cref{fig:ga_NSC_2} we present the plot $g_{a\gamma}$ vs. the axion mass, the corresponding axion band for QCD axions and the dark matter windows for cosmologies $\beta=1$ (green, $\omega_\phi=-2/3$), $\beta=2$ (blue, $\omega_\phi=-1/3$) and $\beta=3$ (red, $\omega_\phi=0$) and the black band corresponds to the axion CDM window in the standard cosmology\footnote{As a standard window for axion cold dark matter we have considered masses such $1.5\times10^{-6}\, {\rm{eV}} \lesssim m_0 \lesssim 7 \times10^{-6}$~eV, such that we include masses from the 2 scenarios described in \cref{chap: ACDM}.}. We have considered a range of initial angles $\theta_i \in \left[0.5, 1.8\right]$ for all of them, to include both the inflationary and postinflationary scenarios. For \cref{fig:ga_NSC} we have considered the same parameters for all three non-standard cosmologies, with $\Tend=0.3$~GeV. As we have seen in our previous analyses, smaller $\beta$ dilutes the relic density much more, thus, allowing to reach smaller masses (higer $f_a$).
Our findings support the conclusions found in \cite{Blinov:2019jqc} that cosmologies with higher $\omega_\phi$ can be tested experimentally first than cosmologies with smaller equations of state. We also notice that the bandwidth of the dark matter region gets slimmer as $\omega_\phi$ decreases. This effect is because for those cosmologies, the relic density  depends much more strongly on the axion mass, that is, the range of masses where we can satisfy the correct relic abundance for the selected window of misalignment angles is smaller.  


One of the goals and novelties of this thesis is to keep a careful track of the initial condition for the energy densities of $\phi$ and radiation, parametrized by $\kappa$, and the initial temperature at that moment, parametrized by $T_i$. From the previous analysis performed for the different regions, we have learned that if the oscillation of the axion happens during the Region 3, there is not dependence of the relic density on $\kappa $ and $T_i$. We have seen that Region 3 is able to spawn a wide range in masses, thus, if we would like $\kappa$ and $T_i$ to have an impact on the relic density, the oscillation of the field should happen during what we have called Region 1 or 2. Nonetheless, these two regions feature rather high temperatures, thus, they can be reached by (high) axion masses that are in principle ruled out by astrophysics. Therefore, one way of lowering the oscillation temperature - and therefore the axion mass - is considering a low initial temperature and delaying the dominance of $\phi$.  That is, consider a low $T_i$ and a small $\kappa$.  With that aim in mind, we have produced  \cref{fig:ga_NSC_2} and following the line of thought argued above, we kept $\Tend=0.3$ GeV, for all three cosmologies, but  we considered for $\beta=3$, $T_i=10^{7}$~GeV and $\kappa=10^{-7}$, while for $\beta=2$, we take $T_i=10^{3}$ GeV and $\kappa=10^{-6}$ and for $\beta=1$ the parameters are $T_i=10$ GeV and $\kappa=10^{-4}$. With these values, the oscillation for $\beta=3$ occurs in Region 2 as well as for $\beta=2$, while for $\beta=1$ we move on to Region 1.
 
To achieve an overlap of the non-standard cosmologies with the axion window from the standard cosmology (black area), it is difficult, because it is needed to move to the right in the mass, which means high oscillation temperatures. We  have seen that one way of doing that is to lower $T_i$ and decreasing $\kappa$. The problem is that for the range of masses of the standard axion window and above, the axion is sub-produced in NSC, thus, the axion could only be a component of the dark matter, but not the whole observed in the universe.  Also, by lowering $T_i$ too much, we shorten the period of $\phi$ dominance, making  hard to find a NSC cosmology where $\phi$ gets to dominate the expansion of the universe after all.   

On the other hand, it is possible to move the dark matter window further to smaller masses by lowering the temperature of oscillation. In order to do that, we  can consider a low initial temperature at which the non-standard cosmology develops or to lengthen the period of $\phi$ domination, \textit{i.e} choosing a small  $\Tend$.

\chapter*{Conclusions}
\addcontentsline{toc}{chapter}{Conclusions} 

In this thesis we have studied the axion dark matter production due to the misalignment mechanism, considering that prior to BBN, there was an alternative period to radiation, set by the dominance of a new fluid, $\phi$. We have firstly studied the effects  of a non-standard cosmology on the Hubble parameter and then we have solved analytically the equations for the evolution of the energy densities, assuming a dominance of the field $\phi$. Then, we have studied the impact on the axion relic density, by keeping a careful track of the initial conditions. We have divided the NSC into 3 periods: Region 1, Region 2 and Region 3 and we have studied the axion relic abundance depending on which of those the oscillation took place. 

Next, we have studied the impact of non-standard cosmology on the universe and 
concerning to our results, we have found that during the period when $\phi$ dominates the energy density with a small $\omega_\phi$, the  size of the universe is bigger as well as the expansion rate is faster. Then, when the thermal bath undergoes the entropy injection, the temperature begins to drop more slowly than in a standard cosmology. 

Then, we have moved to our main aim, to study how is affected the axion production in anon-standard cosmology,  we found that is possible obtain two effects that potentially move the axion window. The first effect, is to change the temperature at which the axion starts to oscillate. This effect can be obtained in Region 2 and Region 3. In the former, it is because to the domination of $\phi$, the expansion rate changes its dependence on the temperature, while in the later is due to the change in the relation between the temperature and the scale factor. As we expected, we obtained that at a lower decrease in temperature, the axion relic abundance is suppressed due to high masses. The second effect is the dilution of the axions energy density due to the entropy injection. We found that for Region 1 and 2, the dilution is higher than in Region 3, being increased for a smaller $\beta$, implying that the axion window moves to lower masses.

Finally, we have presented the resulting parameter space in the coupling of axions to two photons considering three different non-standard cosmologies.  The non-standard cosmologies shift to different extent, the axion CDM window to lower axion masses. We have obtained that cosmologies with a higher $\omega_\phi$ can be tested experimentally first than cosmologies with smaller equations of state.

\addcontentsline{toc}{chapter}{Bibliography}
\printbibliography

@article{Borsanyi:2016ksw,
    author = "Borsanyi, Sz. and others",
    title = "{Calculation of the axion mass based on high-temperature lattice quantum chromodynamics}",
    eprint = "1606.07494",
    archivePrefix = "arXiv",
    primaryClass = "hep-lat",
    reportNumber = "DESY-16-105",
    doi = "10.1038/nature20115",
    journal = "Nature",
    volume = "539",
    number = "7627",
    pages = "69--71",
    year = "2016"
}

@article{Maldonado_2019,
   title={Establishing the dark matter relic density in an era of particle decays},
   volume={2019},
   ISSN={1475-7516},
   url={http://dx.doi.org/10.1088/1475-7516/2019/06/037},
   DOI={10.1088/1475-7516/2019/06/037},
   number={06},
   journal={Journal of Cosmology and Astroparticle Physics},
   publisher={IOP Publishing},
   author={Maldonado, Carlos and Unwin, James},
   year={2019},
   month={Jun},
   pages={037–037}
}

@article{Allahverdi:2020bys,
    author = "Allahverdi, Rouzbeh and others",
    title = "{The First Three Seconds: a Review of Possible Expansion Histories of the Early Universe}",
    eprint = "2006.16182",
    archivePrefix = "arXiv",
    primaryClass = "astro-ph.CO",
    reportNumber = "FERMILAB-PUB-20-242-A, KCL-PH-TH/2020-33, KEK-Cosmo-257,
  KEK-TH-2231, IPMU20-0070, PI/UAN-2020-674FT, RUP-20-22",
    doi = "10.21105/astro.2006.16182",
    month = "6",
    year = "2020"
}

@article{PhysRevD.28.1243,
  title = {Coherent scalar-field oscillations in an expanding universe},
  author = {Turner, Michael S.},
  journal = {Phys. Rev. D},
  volume = {28},
  issue = {6},
  pages = {1243--1247},
  numpages = {0},
  year = {1983},
  month = {Sep},
  publisher = {American Physical Society},
  doi = {10.1103/PhysRevD.28.1243},
  url = {https://link.aps.org/doi/10.1103/PhysRevD.28.1243}
}

@article{Giudice:2000ex,
    author = "Giudice, Gian Francesco and Kolb, Edward W. and Riotto, Antonio",
    title = "{Largest temperature of the radiation era and its cosmological implications}",
    eprint = "hep-ph/0005123",
    archivePrefix = "arXiv",
    reportNumber = "SNS-PH-00-05, FERMILAB-PUB-00-075-A, CERN-TH-2000-107",
    doi = "10.1103/PhysRevD.64.023508",
    journal = "Phys. Rev. D",
    volume = "64",
    pages = "023508",
    year = "2001"
}

@article{Lazarides:1990xp,
    author = "Lazarides, George and Schaefer, Robert K. and Seckel, D. and Shafi, Q.",
    title = "{Dilution of Cosmological Axions by Entropy Production}",
    reportNumber = "BA-90-10",
    doi = "10.1016/0550-3213(90)90244-8",
    journal = "Nucl. Phys. B",
    volume = "346",
    pages = "193--212",
    year = "1990"
}

@article{Hertzberg:2008wr,
    author = "Hertzberg, Mark P and Tegmark, Max and Wilczek, Frank",
    title = "{Axion Cosmology and the Energy Scale of Inflation}",
    eprint = "0807.1726",
    archivePrefix = "arXiv",
    primaryClass = "astro-ph",
    reportNumber = "MIT-CTP-3950",
    doi = "10.1103/PhysRevD.78.083507",
    journal = "Phys. Rev. D",
    volume = "78",
    pages = "083507",
    year = "2008"
}

@article{Abbott:1982af,
    author = "Abbott, L.F. and Sikivie, P.",
    doi = "10.1016/0370-2693(83)90638-X",
    journal = "Phys.\ Lett.\ B",
    pages = "133--136",
    reportNumber = "PRINT-82-0695 (BRANDEIS)",
    title = "{A Cosmological Bound on the Invisible Axion}",
    volume = "120",
    year = "1983"
}

@article{Preskill:1982cy,
    author = "Preskill, John and Wise, Mark B. and Wilczek, Frank",
    doi = "10.1016/0370-2693(83)90637-8",
    journal = "Phys.\ Lett.\ B",
    pages = "127--132",
    reportNumber = "HUTP-82-A048, NSF-ITP-82-103",
    title = "{Cosmology of the Invisible Axion}",
    volume = "120",
    year = "1983"
}

@article{Nelson:2011sf,
    author = "Nelson, Ann E. and Scholtz, Jakub",
    archivePrefix = "arXiv",
    doi = "10.1103/PhysRevD.84.103501",
    eprint = "1105.2812",
    journal = "Phys.\ Rev.\ D",
    pages = "103501",
    primaryClass = "hep-ph",
    title = "{Dark Light, Dark Matter and the Misalignment Mechanism}",
    volume = "84",
    year = "2011"
}

@article{Dine:1982ah,
    author = "Dine, Michael and Fischler, Willy",
    doi = "10.1016/0370-2693(83)90639-1",
    journal = "Phys.\ Lett.\ B",
    pages = "137--141",
    reportNumber = "UPR-0201T",
    title = "{The Not So Harmless Axion}",
    volume = "120",
    year = "1983"
}

@article{Arias:2012az,
    author = "Arias, Paola and Cadamuro, Davide and Goodsell, Mark and Jaeckel, Joerg and Redondo, Javier and Ringwald, Andreas",
    archivePrefix = "arXiv",
    doi = "10.1088/1475-7516/2012/06/013",
    eprint = "1201.5902",
    journal = "JCAP",
    pages = "013",
    primaryClass = "hep-ph",
    reportNumber = "DESY-11-226, MPP-2011-140, CERN-PH-TH-2011-323, IPPP-11-80, DCPT-11-160",
    title = "{WISPy Cold Dark Matter}",
    volume = "06",
    year = "2012"
}

@article{Marsh:2015xka,
    author = "Marsh, David J. E.",
    title = "{Axion Cosmology}",
    eprint = "1510.07633",
    archivePrefix = "arXiv",
    primaryClass = "astro-ph.CO",
    reportNumber = "KCL-PH-TH-2015-50",
    doi = "10.1016/j.physrep.2016.06.005",
    journal = "Phys. Rept.",
    volume = "643",
    pages = "1--79",
    year = "2016"
}

@article{Raffelt:1987yb,
    author = "Raffelt, Georg G. and Dearborn, David S.P.",
    title = "{Bounds on Weakly Interacting Particles From Observational Lifetimes of Helium Burning Stars}",
    reportNumber = "UCRL-96457",
    doi = "10.1103/PhysRevD.37.549",
    journal = "Phys. Rev. D",
    volume = "37",
    pages = "549--551",
    year = "1988"
}

@article{Visinelli:2009kt,
    author = "Visinelli, Luca and Gondolo, Paolo",
    title = "{Axion cold dark matter in non-standard cosmologies}",
    eprint = "0912.0015",
    archivePrefix = "arXiv",
    primaryClass = "astro-ph.CO",
    doi = "10.1103/PhysRevD.81.063508",
    journal = "Phys. Rev. D",
    volume = "81",
    pages = "063508",
    year = "2010"
}

@article{Ramberg:2019dgi,
    author = "Ramberg, Nicklas and Visinelli, Luca",
    title = "{Probing the Early Universe with Axion Physics and Gravitational Waves}",
    eprint = "1904.05707",
    archivePrefix = "arXiv",
    primaryClass = "astro-ph.CO",
    doi = "10.1103/PhysRevD.99.123513",
    journal = "Phys. Rev. D",
    volume = "99",
    number = "12",
    pages = "123513",
    year = "2019"
}

@article{Crewther:1977ce,
    author = "Crewther, R.J.",
    title = "{Chirality Selection Rules and the U(1) Problem}",
    reportNumber = "CERN-TH-2350",
    doi = "10.1016/0370-2693(77)90675-X",
    journal = "Phys. Lett. B",
    volume = "70",
    pages = "349--354",
    year = "1977"
}

@article{DiVecchia:1980yfw,
    author = "Di Vecchia, P. and Veneziano, G.",
    title = "{Chiral Dynamics in the Large n Limit}",
    reportNumber = "CERN-TH-2814",
    doi = "10.1016/0550-3213(80)90370-3",
    journal = "Nucl. Phys. B",
    volume = "171",
    pages = "253--272",
    year = "1980"
}

@article{Gorghetto:2018ocs,
    author = "Gorghetto, Marco and Villadoro, Giovanni",
    title = "{Topological Susceptibility and QCD Axion Mass: QED and NNLO corrections}",
    eprint = "1812.01008",
    archivePrefix = "arXiv",
    primaryClass = "hep-ph",
    doi = "10.1007/JHEP03(2019)033",
    journal = "JHEP",
    volume = "03",
    pages = "033",
    year = "2019"
}

@article{Grin:2007yg,
    author = "Grin, Daniel and Smith, Tristan L. and Kamionkowski, Marc",
    title = "{Axion constraints in non-standard thermal histories}",
    eprint = "0711.1352",
    archivePrefix = "arXiv",
    primaryClass = "astro-ph",
    doi = "10.1103/PhysRevD.77.085020",
    journal = "Phys. Rev. D",
    volume = "77",
    pages = "085020",
    year = "2008"
}

@article{Steinhardt:1983ia,
    author = "Steinhardt, Paul J. and Turner, Michael S.",
    title = "{Saving the Invisible Axion}",
    reportNumber = "EFI-83-28-CHICAGO",
    doi = "10.1016/0370-2693(83)90727-X",
    journal = "Phys. Lett. B",
    volume = "129",
    pages = "51",
    year = "1983"
}

@article{Kawasaki:1995vt,
    author = "Kawasaki, M. and Moroi, T. and Yanagida, T.",
    title = "{Can decaying particles raise the upper bound on the Peccei-Quinn scale?}",
    eprint = "hep-ph/9510461",
    archivePrefix = "arXiv",
    reportNumber = "ICRR-345-95-11, UT-730, LBL-37943",
    doi = "10.1016/0370-2693(96)00743-5",
    journal = "Phys. Lett. B",
    volume = "383",
    pages = "313--316",
    year = "1996"
}

@article{Nelson:2018via,
    author = "Nelson, Ann E. and Xiao, Huangyu",
    title = "{Axion Cosmology with Early Matter Domination}",
    eprint = "1807.07176",
    archivePrefix = "arXiv",
    primaryClass = "astro-ph.CO",
    doi = "10.1103/PhysRevD.98.063516",
    journal = "Phys. Rev. D",
    volume = "98",
    number = "6",
    pages = "063516",
    year = "2018"
}

@article{Blinov:2019jqc,
    author = "Blinov, Nikita and Dolan, Matthew J and Draper, Patrick",
    title = "{Imprints of the Early Universe on Axion Dark Matter Substructure}",
    eprint = "1911.07853",
    archivePrefix = "arXiv",
    %primaryClass = "astro-ph.CO",
    reportNumber = "FERMILAB-PUB-19-560-A-T",
    doi = "10.1103/PhysRevD.101.035002",
    journal = "Phys. Rev. D",
    volume = "101",
    number = "3",
    pages = "035002",
    year = "2020"
}

@article{Visinelli:2018wza,
    author = "Visinelli, Luca and Redondo, Javier",
    title = "{Axion Miniclusters in Modified Cosmological Histories}",
    eprint = "1808.01879",
    archivePrefix = "arXiv",
    primaryClass = "astro-ph.CO",
    reportNumber = "NORDITA-2018-063; MPP-2018-237, NORDITA-2018-063",
    doi = "10.1103/PhysRevD.101.023008",
    journal = "Phys. Rev. D",
    volume = "101",
    number = "2",
    pages = "023008",
    year = "2020"
}

@article{Peccei:1977hh,
    author = "Peccei, R.D. and Quinn, Helen R.",
    title = "{CP Conservation in the Presence of Instantons}",
    reportNumber = "ITP-568-STANFORD",
    doi = "10.1103/PhysRevLett.38.1440",
    journal = "Phys. Rev. Lett.",
    volume = "38",
    pages = "1440--1443",
    year = "1977"
}

@article{Weinberg:1977ma,
    author = "Weinberg, Steven",
    title = "{A New Light Boson?}",
    reportNumber = "HUTP-77/A074",
    doi = "10.1103/PhysRevLett.40.223",
    journal = "Phys. Rev. Lett.",
    volume = "40",
    pages = "223--226",
    year = "1978"
}

@article{Wilczek:1977pj,
    author = "Wilczek, Frank",
    title = "{Problem of Strong  $P$  and  $T$  Invariance in the Presence of Instantons}",
    reportNumber = "Print-77-0939 (COLUMBIA)",
    doi = "10.1103/PhysRevLett.40.279",
    journal = "Phys. Rev. Lett.",
    volume = "40",
    pages = "279--282",
    year = "1978"
}

@article{PhysRevD.96.023514,
  title = {Primordial black hole constraints for extended mass functions},
  author = {Carr, Bernard and Raidal, Martti and Tenkanen, Tommi and Vaskonen, Ville and Veerm\"ae, Hardi},
  journal = {Phys. Rev. D},
  volume = {96},
  issue = {2},
  pages = {023514},
  numpages = {10},
  year = {2017},
  month = {Jul},
  publisher = {American Physical Society},
  doi = {10.1103/PhysRevD.96.023514},
  url = {https://link.aps.org/doi/10.1103/PhysRevD.96.023514}
}

@article{Niemeyer_1998,
   title={Near-Critical Gravitational Collapse and the Initial Mass Function of Primordial Black Holes},
   volume={80},
   ISSN={1079-7114},
   url={http://dx.doi.org/10.1103/PhysRevLett.80.5481},
   DOI={10.1103/physrevlett.80.5481},
   number={25},
   journal={Physical Review Letters},
   publisher={American Physical Society (APS)},
   author={Niemeyer, J. C. and Jedamzik, K.},
   year={1998},
   month={Jun},
   pages={5481–5484}
   }

@article{Carr_2016,
   title={Primordial black holes as dark matter},
   volume={94},
   ISSN={2470-0029},
   url={http://dx.doi.org/10.1103/PhysRevD.94.083504},
   DOI={10.1103/physrevd.94.083504},
   number={8},
   journal={Physical Review D},
   publisher={American Physical Society (APS)},
   author={Carr, Bernard and Kühnel, Florian and Sandstad, Marit},
   year={2016},
   month={Oct}
}

@article{Abbott_2016,
   title={Observation of Gravitational Waves from a Binary Black Hole Merger},
   volume={116},
   ISSN={1079-7114},
   url={http://dx.doi.org/10.1103/PhysRevLett.116.061102},
   DOI={10.1103/physrevlett.116.061102},
   number={6},
   journal={Physical Review Letters},
   publisher={American Physical Society (APS)},
   author={Abbott, B. P. and Abbott, R. and Abbott, T. D. and Abernathy, M. R. and Acernese, F. and Ackley, K. and Adams, C. and Adams, T. and Addesso, P. and Adhikari, R. X. and et al.},
   year={2016},
   month={Feb}
}

@article{Freese:2017idy,
    author = "Freese, Katherine",
    editor = "Bianchi, Massimo and Jantzen, Robert T. and Ruffini, Remo",
    title = "{Status of Dark Matter in the Universe}",
    eprint = "1701.01840",
    archivePrefix = "arXiv",
    primaryClass = "astro-ph.CO",
    doi = "10.1142/S0218271817300129",
    journal = "Int. J. Mod. Phys.",
    volume = "1",
    number = "06",
    pages = "325--355",
    year = "2017"
}

@article{Alexandrou_2020,
   title={Ruling Out the Massless Up-Quark Solution to the Strong 
CP
 Problem by Computing the Topological Mass Contribution with Lattice QCD},
   volume={125},
   ISSN={1079-7114},
   url={http://dx.doi.org/10.1103/PhysRevLett.125.232001},
   DOI={10.1103/physrevlett.125.232001},
   number={23},
   journal={Physical Review Letters},
   publisher={American Physical Society (APS)},
   author={Alexandrou, Constantia and Finkenrath, Jacob and Funcke, Lena and Jansen, Karl and Kostrzewa, Bartosz and Pittler, Ferenc and Urbach, Carsten},
   year={2020},
   month={Dec}
}

@article{Anastassopoulos:2017ftl,
    author = "Anastassopoulos, V. and others",
    collaboration = "CAST",
    title = "{New CAST Limit on the Axion-Photon Interaction}",
    eprint = "1705.02290",
    archivePrefix = "arXiv",
    primaryClass = "hep-ex",
    doi = "10.1038/nphys4109",
    journal = "Nature Phys.",
    volume = "13",
    pages = "584--590",
    year = "2017"
}

@article{Gondolo:2008dd,
    author = "Gondolo, Paolo and Raffelt, Georg G.",
    title = "{Solar neutrino limit on axions and keV-mass bosons}",
    eprint = "0807.2926",
    archivePrefix = "arXiv",
    primaryClass = "astro-ph",
    reportNumber = "MPP-2008-69",
    doi = "10.1103/PhysRevD.79.107301",
    journal = "Phys. Rev. D",
    volume = "79",
    pages = "107301",
    year = "2009"
}

@article{Raffelt:1985nk,
    author = "Raffelt, Georg G.",
    title = "{ASTROPHYSICAL AXION BOUNDS DIMINISHED BY SCREENING EFFECTS}",
    reportNumber = "MPI-PAE-PTH-51-85",
    doi = "10.1103/PhysRevD.33.897",
    journal = "Phys. Rev. D",
    volume = "33",
    pages = "897",
    year = "1986"
}

@article{Asztalos:2009yp,
    author = "Asztalos, S. J. and others",
    collaboration = "ADMX",
    title = "{A SQUID-based microwave cavity search for dark-matter axions}",
    eprint = "0910.5914",
    archivePrefix = "arXiv",
    primaryClass = "astro-ph.CO",
    doi = "10.1103/PhysRevLett.104.041301",
    journal = "Phys. Rev. Lett.",
    volume = "104",
    pages = "041301",
    year = "2010"
}

@article{Halverson_2002,
   title={Degree Angular Scale Interferometer First Results: A Measurement of the Cosmic Microwave Background Angular Power Spectrum},
   volume={568},
   ISSN={1538-4357},
   url={http://dx.doi.org/10.1086/338879},
   DOI={10.1086/338879},
   number={1},
   journal={The Astrophysical Journal},
   publisher={American Astronomical Society},
   author={Halverson, N. W. and Leitch, E. M. and Pryke, C. and Kovac, J. and Carlstrom, J. E. and Holzapfel, W. L. and Dragovan, M. and Cartwright, J. K. and Mason, B. S. and Padin, S. and et al.},
   year={2002},
   month={Mar},
   pages={38–45}
}

@article{2016,
   title={Planck2015 results},
   volume={594},
   ISSN={1432-0746},
   url={http://dx.doi.org/10.1051/0004-6361/201526926},
   DOI={10.1051/0004-6361/201526926},
   journal={Astronomy \& Astrophysics},
   publisher={EDP Sciences},
   author={Aghanim, N. and Arnaud, M. and Ashdown, M. and Aumont, J. and Baccigalupi, C. and Banday, A. J. and Barreiro, R. B. and Bartlett, J. G. and Bartolo, N. and et al.},
   year={2016},
   month={Sep},
   pages={A11}
}

@article{Massey_2010,
   title={The dark matter of gravitational lensing},
   volume={73},
   ISSN={1361-6633},
   url={http://dx.doi.org/10.1088/0034-4885/73/8/086901},
   DOI={10.1088/0034-4885/73/8/086901},
   number={8},
   journal={Reports on Progress in Physics},
   publisher={IOP Publishing},
   author={Massey, Richard and Kitching, Thomas and Richard, Johan},
   year={2010},
   month={Jul},
   pages={086901}
}

@article{blumenthal1984formation,
  title={Formation of galaxies and large-scale structure with cold dark matter},
  author={Blumenthal, George R and Faber, SM and Primack, Joel R and Rees, Martin J},
  journal={Nature},
  volume={311},
  number={5986},
  pages={517--525},
  year={1984},
  publisher={Nature Publishing Group}
}

@article{white1983clustering,
  title={Clustering in a neutrino-dominated universe},
  author={White, Simon DM and Frenk, Carlos S and Davis, Marc},
  journal={The Astrophysical Journal},
  volume={274},
  pages={L1--L5},
  year={1983}
}

@article{Aprile_2018,
   title={Dark Matter Search Results from a One Ton-Year Exposure of XENON1T},
   volume={121},
   ISSN={1079-7114},
   url={http://dx.doi.org/10.1103/PhysRevLett.121.111302},
   DOI={10.1103/physrevlett.121.111302},
   number={11},
   journal={Physical Review Letters},
   publisher={American Physical Society (APS)},
   author={Aprile, E. and Aalbers, J. and Agostini, F. and Alfonsi, M. and Althueser, L. and Amaro, F. D. and Anthony, M. and Arneodo, F. and Baudis, L. and Bauermeister, B. and et al.},
   year={2018},
   month={Sep}
}

@article{Akerib:2016vxi,
    author = "Akerib, D. S. and others",
    collaboration = "LUX",
    title = "{Results from a search for dark matter in the complete LUX exposure}",
    eprint = "1608.07648",
    archivePrefix = "arXiv",
    primaryClass = "astro-ph.CO",
    doi = "10.1103/PhysRevLett.118.021303",
    journal = "Phys. Rev. Lett.",
    volume = "118",
    number = "2",
    pages = "021303",
    year = "2017"
}

@article{2020,
   title={Planck 2018 results},
   volume={641},
   ISSN={1432-0746},
   url={http://dx.doi.org/10.1051/0004-6361/201833910},
   DOI={10.1051/0004-6361/201833910},
   journal={Astronomy \& Astrophysics},
   publisher={EDP Sciences},
   author={Aghanim, N. and Akrami, Y. and Ashdown, M. and Aumont, J. and Baccigalupi, C. and Ballardini, M. and Banday, A. J. and Barreiro, R. B. and Bartolo, N. and et al.},
   year={2020},
   month="10",
   pages={A6}
}

@article{Cooke_2018,
   title={One Percent Determination of the Primordial Deuterium Abundance},
   volume={855},
   ISSN={1538-4357},
   url={http://dx.doi.org/10.3847/1538-4357/aaab53},
   DOI={10.3847/1538-4357/aaab53},
   number={2},
   journal={The Astrophysical Journal},
   publisher={American Astronomical Society},
   author={Cooke, Ryan J. and Pettini, Max and Steidel, Charles C.},
   year={2018},
   month={Mar},
   pages={102}
}

@article{Sbordone_2010,
   title={The metal-poor end of the Spite plateau},
   volume={522},
   ISSN={1432-0746},
   url={http://dx.doi.org/10.1051/0004-6361/200913282},
   DOI={10.1051/0004-6361/200913282},
   journal={Astronomy \& Astrophysics},
   publisher={EDP Sciences},
   author={Sbordone, L. and Bonifacio, P. and Caffau, E. and Ludwig, H.-G. and Behara, N. T. and González Hernández, J. I. and Steffen, M. and Cayrel, R. and Freytag, B. and Van’t Veer, C. and et al.},
   year={2010},
   month={Oct},
   pages={A26}
}

@article{RevModPhys.82.557,
  title = {Axions and the strong $CP$ problem},
  author = {Kim, Jihn E. and Carosi, Gianpaolo},
  journal = {Rev. Mod. Phys.},
  volume = {82},
  issue = {1},
  pages = {557--601},
  numpages = {0},
  year = {2010},
  month = {Mar},
  publisher = {American Physical Society},
  doi = {10.1103/RevModPhys.82.557},
  url = {https://link.aps.org/doi/10.1103/RevModPhys.82.557}
}

@article{PhysRevLett.97.131801,
  title = {Improved Experimental Limit on the Electric Dipole Moment of the Neutron},
  author = {Baker, C. A. and Doyle, D. D. and Geltenbort, P. and Green, K. and van der Grinten, M. G. D. and Harris, P. G. and Iaydjiev, P. and Ivanov, S. N. and May, D. J. R. and Pendlebury, J. M. and Richardson, J. D. and Shiers, D. and Smith, K. F.},
  journal = {Phys. Rev. Lett.},
  volume = {97},
  issue = {13},
  pages = {131801},
  numpages = {4},
  year = {2006},
  month = {Sep},
  publisher = {American Physical Society},
  doi = {10.1103/PhysRevLett.97.131801},
  url = {https://link.aps.org/doi/10.1103/PhysRevLett.97.131801}
}

@inproceedings{Banks:1994yg,
    author = "Banks, Tom and Nir, Yosef and Seiberg, Nathan",
    title = "{Missing (up) mass, accidental anomalous symmetries, and the strong CP problem}",
    booktitle = "{2nd IFT Workshop on Yukawa Couplings and the Origins of Mass}",
    eprint = "hep-ph/9403203",
    archivePrefix = "arXiv",
    reportNumber = "WIS-94-14-PH, RU-94-24",
    pages = "26--41",
    month = "2",
    year = "1994"
}

@article{Sikivie_2008,
   title={Axion Cosmology},
   ISBN={9783540735182},
   ISSN={0075-8450},
   url={http://dx.doi.org/10.1007/978-3-540-73518-2_2},
   DOI={10.1007/978-3-540-73518-2_2},
   journal={Axions},
   publisher={Springer Berlin Heidelberg},
   author={Sikivie, Pierre},
   year={2008},
   pages={19–50}
}

@phdthesis{2013production,
  title={Production and evolution of axion dark matter in the early universe},
  author={K. Saikawa},
  year={2013},
  school={ University of Tokyo}
}

@article{Kim_1998,
   title={Constraints on very light axions from cavity experiments},
   volume={58},
   ISSN={1089-4918},
   url={http://dx.doi.org/10.1103/PhysRevD.58.055006},
   DOI={10.1103/physrevd.58.055006},
   number={5},
   journal={Physical Review D},
   publisher={American Physical Society (APS)},
   author={Kim, Jihn E.},
   year={1998},
   month={Jul}
}

@article{shifman1980can,
  title={Can confinement ensure natural CP invariance of strong interactions?},
  author={Shifman, Mikhail A and Vainshtein, AI and Zakharov, Valentin I},
  journal={Nuclear Physics B},
  volume={166},
  number={3},
  pages={493--506},
  year={1980},
  publisher={Elsevier}
}

@article{kim1979weak,
  title={Weak-interaction singlet and strong CP invariance},
  author={Kim, Jihn E},
  journal={Physical Review Letters},
  volume={43},
  number={2},
  pages={103},
  year={1979},
  publisher={APS}
}

@article{gill2011constraining,
  title={Constraining the photon-axion coupling constant with magnetic white dwarfs},
  author={Gill, Ramandeep and Heyl, Jeremy S},
  journal={Physical Review D},
  volume={84},
  number={8},
  pages={085001},
  year={2011},
  publisher={APS}
}

@article{dine1981simple,
  title={A simple solution to the strong CP problem with a harmless axion},
  author={Dine, Michael and Fischler, Willy and Srednicki, Mark},
  journal={Physics letters B},
  volume={104},
  number={3},
  pages={199--202},
  year={1981},
  publisher={Elsevier}
}

@article{battye1994axion,
  title={Axion string constraints},
  author={Battye, RA and Shellard, EPS},
  journal={Physical Review Letters},
  volume={73},
  number={22},
  pages={2954},
  year={1994},
  publisher={APS}
}

@article{harari1987evolution,
  title={On the evolution of global strings in the early universe},
  author={Harari, Diego and Sikivie, Po},
  journal={Physics Letters B},
  volume={195},
  number={3},
  pages={361--365},
  year={1987},
  publisher={Elsevier}
}

@article{davis1985goldstone,
  title={Goldstone bosons in string models of galaxy formation},
  author={Davis, Richard Lynn},
  journal={Physical Review D},
  volume={32},
  number={12},
  pages={3172},
  year={1985},
  publisher={APS}
}

@incollection{sikivie2008axion,
  title={Axion cosmology},
  author={Sikivie, Pierre},
  booktitle={Axions},
  pages={19--50},
  year={2008},
  publisher={Springer}
}

@BOOK{1990eaun,
       author = {{Kolb}, Edward W. and {Turner}, Michael S.},
        title = "{The early universe}",
         year = 1990,
       volume = {69},
       adsurl = {https://ui.adsabs.harvard.edu/abs/1990eaun.book.....K}
}

@article{kibble1976topology,
  title={Topology of cosmic domains and strings},
  author={Kibble, Thomas WB},
  journal={Journal of Physics A: Mathematical and General},
  volume={9},
  number={8},
  pages={1387},
  year={1976},
  publisher={IOP Publishing}
}

@article{PhysRevD.15.2738,
  title = {Cosmological event horizons, thermodynamics, and particle creation},
  author = {Gibbons, G. W. and Hawking, S. W.},
  journal = {Phys. Rev. D},
  volume = {15},
  issue = {10},
  pages = {2738--2751},
  numpages = {0},
  year = {1977},
  month = {May},
  publisher = {American Physical Society},
  doi = {10.1103/PhysRevD.15.2738},
  url = {https://link.aps.org/doi/10.1103/PhysRevD.15.2738}
}

@misc{visinelli2011axions,
      title={Axions in Cold Dark Matter and Inflation Models}, 
      author={Luca Visinelli},
      year={2011},
      eprint={1111.5281},
      archivePrefix={arXiv},
      primaryClass={astro-ph.CO}
}

@article{hannestad2004lowest,
  title={What is the lowest possible reheating temperature?},
  author={Hannestad, Steen},
  journal={Physical Review D},
  volume={70},
  number={4},
  pages={043506},
  year={2004},
  publisher={APS}
}

@article{ramberg2019probing,
  title={Probing the early Universe with axion physics and gravitational waves},
  author={Ramberg, Nicklas and Visinelli, Luca},
  journal={Physical Review D},
  volume={99},
  number={12},
  pages={123513},
  year={2019},
  publisher={APS}
}

@article{wmap2011seven,
  title={Seven-year Wilkinson microwave anisotropy probe (WMAP) observations: cosmological interpretation},
  author={WMAP collaboration and others},
  journal={Astrophys. J. Suppl},
  volume={192},
  number={18},
  pages={1001--4538},
  year={2011}
}

@article{visinelli2009dark,
  title={Dark matter axions revisited},
  author={Visinelli, Luca and Gondolo, Paolo},
  journal={Physical Review D},
  volume={80},
  number={3},
  pages={035024},
  year={2009},
  publisher={APS}
}

@incollection{peccei2008strong,
  title={The strong CP problem and axions},
  author={Peccei, Roberto D},
  booktitle={Axions},
  pages={3--17},
  year={2008},
  publisher={Springer}
}

@misc{dine2015challenges,
      title={Challenges for the Nelson-Barr Mechanism}, 
      author={Michael Dine and Patrick Draper},
      year={2015},
      eprint={1506.05433},
      archivePrefix={arXiv},
      primaryClass={hep-ph}
}

@misc{vecchi2014spontaneous,
      title={Spontaneous CP violation and the strong CP problem}, 
      author={Luca Vecchi},
      year={2014},
      eprint={1412.3805},
      archivePrefix={arXiv},
      primaryClass={hep-ph}
}

@article{Hertzberg_2008,
   title={Axion cosmology and the energy scale of inflation},
   volume={78},
   ISSN={1550-2368},
   url={http://dx.doi.org/10.1103/PhysRevD.78.083507},
   DOI={10.1103/physrevd.78.083507},
   number={8},
   journal={Physical Review D},
   publisher={American Physical Society (APS)},
   author={Hertzberg, Mark P. and Tegmark, Max and Wilczek, Frank},
   year={2008},
   month={Oct}
}

@article{chung1999production,
  title={Production of massive particles during reheating},
  author={Chung, Daniel JH and Kolb, Edward W and Riotto, Antonio},
  journal={Physical Review D},
  volume={60},
  number={6},
  pages={063504},
  year={1999},
  publisher={APS}
}

@article{Giudice_2001,
   title={Largest temperature of the radiation era and its cosmological implications},
   volume={64},
   ISSN={1089-4918},
   url={http://dx.doi.org/10.1103/PhysRevD.64.023508},
   DOI={10.1103/physrevd.64.023508},
   number={2},
   journal={Physical Review D},
   publisher={American Physical Society (APS)},
   author={Giudice, Gian Francesco and Kolb, Edward W. and Riotto, Antonio},
   year={2001},
   month={Jun}
}

@article{Vilenkin:1982wt,
    author = "Vilenkin, Alexander and Ford, L. H.",
    title = "{Gravitational Effects upon Cosmological Phase Transitions}",
    reportNumber = "TUTP-82-7",
    doi = "10.1103/PhysRevD.26.1231",
    journal = "Phys. Rev. D",
    volume = "26",
    pages = "1231",
    year = "1982"
}

@article{Moroi:1999zb,
    author = "Moroi, Takeo and Randall, Lisa",
    title = "{Wino cold dark matter from anomaly mediated SUSY breaking}",
    eprint = "hep-ph/9906527",
    archivePrefix = "arXiv",
    reportNumber = "IASSNS-HEP-99-54, PUPT-1873, MIT-CTP-2873",
    doi = "10.1016/S0550-3213(99)00748-8",
    journal = "Nucl. Phys. B",
    volume = "570",
    pages = "455--472",
    year = "2000"
}

@article{Coughlan:1983ci,
    author = "Coughlan, G. D. and Fischler, W. and Kolb, Edward W. and Raby, S. and Ross, Graham G.",
    title = "{Cosmological Problems for the Polonyi Potential}",
    reportNumber = "LA-UR-83-1423",
    doi = "10.1016/0370-2693(83)91091-2",
    journal = "Phys. Lett. B",
    volume = "131",
    pages = "59--64",
    year = "1983"
}

@article{Barrow:1982ei,
    author = "Barrow, John D.",
    title = "{MASSIVE PARTICLES AS A PROBE OF THE EARLY UNIVERSE}",
    doi = "10.1016/0550-3213(82)90233-4",
    journal = "Nucl. Phys. B",
    volume = "208",
    pages = "501--508",
    year = "1982"
}

@article{Ford:1986sy,
    author = "Ford, L. H.",
    title = "{Gravitational Particle Creation and Inflation}",
    reportNumber = "TUTP-86-8",
    doi = "10.1103/PhysRevD.35.2955",
    journal = "Phys. Rev. D",
    volume = "35",
    pages = "2955",
    year = "1987"
}

@article{Spokoiny:1993kt,
    author = "Spokoiny, Boris",
    title = "{Deflationary universe scenario}",
    eprint = "gr-qc/9306008",
    archivePrefix = "arXiv",
    reportNumber = "KUNS-1201",
    doi = "10.1016/0370-2693(93)90155-B",
    journal = "Phys. Lett. B",
    volume = "315",
    pages = "40--45",
    year = "1993"
}

@article{schumann2019direct,
  title={Direct detection of WIMP dark matter: concepts and status},
  author={Schumann, Marc},
  journal={Journal of Physics G: Nuclear and Particle Physics},
  volume={46},
  number={10},
  pages={103003},
  year={2019},
  publisher={IOP Publishing}
}

@article{DEramo:2017gpl,
    author = "D'Eramo, Francesco and Fernandez, Nicolas and Profumo, Stefano",
    title = "{When the Universe Expands Too Fast: Relentless Dark Matter}",
    eprint = "1703.04793",
    archivePrefix = "arXiv",
    primaryClass = "hep-ph",
    reportNumber = "SCIPP-17-02",
    doi = "10.1088/1475-7516/2017/05/012",
    journal = "JCAP",
    volume = "05",
    pages = "012",
    year = "2017"
}
\appendix
\renewcommand\chaptername{Appendix}
\chapter{Analytical calculations} 

\section{Solutions for the $\phi-$radiation system}{\label{app:phi-r}}
Let us start by solving the evolution eqs.~(\ref{eq:cosmo2}) and (\ref{eq:cosmo3}) during the dominance of $\phi$, taking into account the decays. The Hubble parameter can be written as $H\approx \sqrt{\rho_\phi/(3M_P^2)}$. By changing variables, from time to scale factor $t\rightarrow R$, and defining $u(r)=\frac{\rho_\phi}{\rho_{\phi i}} r^\beta$, and $r=R/R_i$, the evolution for the $\phi$ field now reads
\be
\frac{du}{u^{1/2}}=-\frac{\Gamma_\phi}{H_i \sqrt{\kappa}}\,r^{\beta/2-1}\,dr,
\ee
whose solution to first order in $\Gamma_\phi/H_i$ is
\be
\rho_\phi(R)= \rRi \kappa \left[\frac{\Gamma_\phi}{\beta\, H_i}-\left(\frac{R_i}{R}\right)^{\beta/2}\left(1+\frac{\Gamma_{\phi}}{H_i\,\beta}\right)\right]^2.
\label{eq:phi_full}
\ee
Replacing into eq.~(\ref{eq:cosmo3}), using $s=(\rho+p)/T$, we find
\be
\rR(R)=\rRi\left[\left(\frac{R_i}{R}\right)^4+\frac{2\Gamma_\phi\, \sqrt{\kappa}}{(8-\beta)H_i}\left(\frac{R_i}{R}\right)^{\beta/2}-\frac{2\Gamma_\phi\, \sqrt{\kappa}}{(8-\beta)H_i}\left(\frac{R_i}{R}\right)^{4}\right].
\ee
When the decays of $\phi$ are unimportant, we can just retain the first terms in the above equations. If $\kappa<1$, we will start with radiation domination, until a point where the two energies become equal, and from then starts to dominate $\phi$ (if $\beta<4$). By equating the first terms we get an expression for the transition point between radiation to $\phi$ domination
\be
\Req= R_i \kappa^{1/(\beta-4)}.
\ee
It is also possible to get an educated guess for $\Rc$ and $\Rend$, by considering that when the decays start to influence the radiation in the universe, both terms in $\rho_R$ should be of the same order, meaning
\be
\left(\frac{R_i}\Rc\right)^4\approx \frac{2\sqrt{\kappa}\,\Gamma_\phi}{(8-\beta)H_i}\left(\frac{R_i}{\Rc}\right)^{\beta/2},
\ee
from here, we get
\be
\Rc= R_i \left( \frac{(8-\beta)}{2 \sqrt{\kappa}} \left( \frac{T_i}{\Tend}\right)^2 \right)^{\frac{2}{8-\beta}}.
\ee
Similarly, the decays of $\phi$ will stop, when the two terms inside the brackets of eq.~(\ref{eq:phi_full}) become comparable, thus
\be
\left(\frac{R_i}{\Rend}\right)^\beta\approx\kappa^{-1/2}\frac{2\Gamma_\phi}{\beta\, H_i}\left( \frac{R_i}{\Rend}\right)^{\beta/2},
\ee
from where we find 
\be
\Rend= R_i\left(\frac{\beta \sqrt{\kappa}}{2} \left( \frac{T_i}{\Tend}\right)^2\right)^{2/\beta}.
\ee
The case $\beta=0$ should be treated separately. In this case, we start with eq.~(\ref{eq:cosmo2}) and defining $u=\sqrt{\rho_\phi/\rho_{\phi i}}$ and $r=\ln(R/R_i)$. Thus, the equation to solve is
\be
du=-\frac{\Gamma_\phi}{2H_i\sqrt{\kappa}} dr,
\ee
whose solution is given by
\bea
\rho_\phi(R)&\eqsim&\rho_{\phi_i}\left(1+\frac{\Gamma_\phi}{H_i \sqrt{\kappa}}\ln\left(\frac{R_i}{R}\right)\right).
\label{app:phibeta0}
\eea
By replacing into the radiation equation, we get the evolution is 
\bea
\rho_R(R)&\eqsim&\rho_{R_i}\left(\left(\frac{R_i}{R}\right)^4 + \frac{\Gamma_\phi \sqrt{\kappa}}{4H_i}\right).
\label{app:radbeta0}
\eea
Following the same steps used before we can find the scale factor when the two energy densities are equals, this is given by
\be
\Req=R_i \kappa^{-1/4}.
\label{Reqbeta0}
\ee
Also we can find the moment when the radiation energy density start to be influenced by the $\phi$ field
\be
\Rc=R_i \left(\frac{4H_i}{\Gamma_\phi \sqrt{\kappa}}\right)^{1/4}.
\label{Rcbeta0}
\ee
Additionally the moment when the decays of the $\phi$ field are significant is represented by
\be
\Rend=R_i \times\exp\left(\frac{H_i \sqrt{\kappa}}{\Gamma_\phi}\right).
\label{Rendbeta0}
\ee
\section{Axion oscillations during fluid domination}
\subsection{Oscillations in Region I}{\label{app:reg1}}
To analyse this epoch, first we find the oscillation temperature, given by the condition $3H(\Tosc)=m_a(\Tosc)$. Since for this period it is assumed the universe is radiation dominated, the Hubble parameter is $H=\sqrt{\rho_R/(3M_P^2)}$, so the oscillation temperature is the usual found for a QCD axion, depicted in fig.~(\ref{fig:std_tosc}). There are two regimes for the oscillation temperature, depending on whether the effects of temperature on the axion mass, eq.~(\ref{eq:thermal_mass}) are important or not, and they are the same as the standard cosmological scenario, namely
\bea
\Tosc=\begin{dcases} 
\left(\alpha \, m_0 M_P \LambdaQCD^4\right)^{1/6}, &\text{for}\,\,m_0\gtrsim 10^{-5}\,\mu \mbox{eV},\\
\left(m_0\, M_P\right)^{1/2}, &\text{for}\,\, m_0\lesssim 10^{-5}\,\mu \mbox{eV}.
\end{dcases}
\eea
These two temperatures intersect at the QCD epoch, which can be expressed as a particular mass of the axion, which is the same as the standard cosmological scenario, but we re-write here, for completeness to be
\be
m_{R_1}=\sqrt{\frac{\alpha\LambdaQCD^4}{M_P^2}}.
\ee
The condition for $\Rosc$ to occur in this regime can be re-written to be 
\be
T_{\rm{osc}}\gg T_i\kappa^\frac{1}{4-\beta},
\ee
and by using the expressions for $\Tosc$ we can write it as a condition for the axion bare mass
\be
m_0\gg
\begin{dcases}\left(\frac{T_i^6\,\kappa^{6/(4-\beta)}}{M_P\, \alpha\, \LambdaQCD^4}\right), & \mbox{for}\,\,\, m_0\gtrsim m_{R_1}, \\
\left(\frac{T_i^2\, \kap^{2/(4-\beta)}}{M_P}\right), & \mbox{for}\,\,\, m_0\lesssim m_{R_1},
\end{dcases}
\label{eq:R1_boundInf}
\ee
where the first equation correspond to the limit of the axion mass in the case that the temperature has significant effects in the mass of the axion (before QCD transition) and the second is the limit after the QCD transition.

The relic abundance if the oscillation happens in this region, is given by $\Omega_a(T_0)=\Omega_a^{\rm std}(T_0)\gamma_{R_1} $, such that $\Omega_a^{\rm std}$ is the axion relic energy density today in the standard cosmology, $\rho_a(T_0)/\rho_{crit,0}$ in the standard scenario, and $\gamma_{R_1}=S_{\rm osc}/S_{\rm end}$ the entropy dilution factor. To find the latter, we take
\bea
\gamma_{R_1}&=&\frac{S_{\rm osc}}{S_{\rm end}}=\frac{S_{\rm c}}{S_{\rm end}}= \left(\frac{\Tc\, \Rc}{\Tend\, \Rend}\right)^3,\\
&=& \left(\frac{\kappa \beta^2}{4}\left(\frac{T_i}{\Tend }\right)^{4-\beta}\right)^{\frac{-3}{\beta}}.
\eea

For the case of $\beta=0$, following the same steps that we detailed before, we can find
\be
\gamma_{R_1,\beta=0}=\left(\frac{T_i}{\Tend}\right)^3 \times \exp\left(\sqrt{\kappa}H_i/\Gamma_\phi\right).
\ee

Now, let us replace the dilution factor into eq.~\eqref{eq:NSC_axion_density}, and we find
\bea
\Omega_{R_1}=
\begin{dcases}
\Omegastdo \left(\frac{\kappa \beta^2}{4}\right)^{-3/\beta}\tau_{\rm end}^{12/\beta-3} , & \mbox{for}\,\,\, m_0\lesssim m_{R_1},\\
\OmegastdT\left(\frac{\kappa \beta^2}{4}\right)^{-3/\beta}\tau_{\rm end}^{12/\beta-3} , & \mbox{for}\,\,\, m_0\gtrsim m_{R_1}.
\end{dcases}
\eea
Where we have defined the axion relic densities for a standard cosmology, when the temperature effects on the mass are unimportant and important, respectively, as
\bea
\Omegastdo&=&\frac{\chi_0 T_0^3}{2\rhocrit} (m_0 M_P)^{-3/2}\,\theta_i^2,\\
\OmegastdT&=&\frac{\chi_0 T_0^3}{2\rhocrit} (m_0 M_P)^{-7/6} (\alpha \LambdaQCD)^{-1/6}\,\theta_i^2.
\eea

\subsection{Oscillations in Region 2}{\label{app:reg2}}
We start by writing the Hubble parameter as
\be
H\approx\sqrt{\frac{\rho_\phi}{3M_P^2}}\approx\sqrt{\kappa} H_i \left(\frac{R_i}R\right)^{\beta/2},
\label{HR2}
\ee
since $T\propto R^{-1}$, eq.(\ref{HR2}) can be written in terms of the temperature as
\be H(\Tosc)=\sqrt{\kappa} H_i  \left(\frac{\Tosc}{T_i}\right)^{\beta/2},
\ee
and equating $3H(\Tosc)=m_a(\Tosc)$, we find the oscillation temperature for both cases, where the temperatures effects on the mass are and are not important, as
\bea
\Tosc^{R_2}=
\begin{dcases}
\left(\frac{T_i^{\frac{\beta-4}{2}}M_P\, m_0}{\sqrt{\kappa}}\right)^{2/\beta}& m_0\lesssim m_{R_2},
\\
\left(\frac{T_i^{\frac{\beta-4}{2}}M_P\, m_0}{\sqrt{\kappa}}\, \alpha\, \LambdaQCD^4 \right)^\frac{2}{\beta+8}& m_0\gtrsim m_{R_2}.
\end{dcases}
\label{app:eq:ToscR2}
\eea
These two temperatures intersect at a mass of the axion given by
\be
m_{R_2}= \frac{\sqrt{\kappa}\,(\alpha \LambdaQCD^4)^{\beta/8}\,T_i^{\frac{4-\beta}{2}}}{M_P}.
\label{app:eq:mR2}
\ee
By requiring $\Tc\ll \Tosc\ll \Teq$ we find the mass range of the axion in order to have the oscillation happening in this era. There are two possibilities: either we are interested in masses above $m_{R_2}$, so in that case the oscillation happens before the QCD transition and we only need to require $\Tc\ll T_{\rm osc,T}^{R_2}\ll \Teq$, or if we go to smaller masses than $m_{R_2}$, we have to require $\Tc\ll T_{\rm osc,0}^{R_2}$ and $T_{\rm osc,T}^{R_2}\ll \Teq$, where we have written the $0$ subscript for the oscillation temperature when temperature effects on the axion mass are unimportant and the subscript $T$ when they are important.  For each case, respectively we obtain
\bea
\frac{1}{M_P\,\alpha\LambdaQCD^4}\left(\frac{2 (\kappa\,T_i^{4-\beta})^{\frac{8}{8+\beta}}\,\Tend^2}{8-\beta}    \right)^{\frac{8+\beta}{8-\beta}}\ll m_0\ll \frac{\left(T_i\,\kappa^{\frac{1}{4-\beta}} \right)^{6}}{M_P\,\alpha \LambdaQCD^4},
\label{app:eq:NSC_R2_Tbound}\\
\frac{\sqrt{\kappa}\,T_i^2}{M_P}\left(\frac{2\,\Tend^2\sqrt{\kappa}}{T_i^2 (8-\beta)}\right)^{\frac{\beta}{8-\beta}}\ll m_0\ll \frac{\left(T_i\,\kappa^{\frac{1}{4-\beta}} \right)^{6}}{M_P\,\alpha \LambdaQCD^4}.
\label{app:eq:NSC_R2_0bound}
\eea

The axion abundance can be found by taking the expression for the energy density in the NSC, eq.~(\ref{eq:NSC_axion_density}), where $S_{\rm osc}/S_{\rm end}$ is the entropy dilution, which is the same found in eq.~(\ref{eq:gamma_R1}). By considering a temperature independent axion mass, we get:
\bea
\Omega_{\rm{R_2}}^{0}= \Omegastdo \left(\frac{\beta^2}{4} \right)^{-3/\beta} \left(\frac{\Tend^2}{m_0 M_P} \right)^{\frac{3}{2\beta} (4-\beta)} .
\label{eq:NSC_R2_relic0}
\eea

Note that the dependence on $\kappa$ and $T_i$ has cancelled (and therefore, the DM abundance is insensitive to it).
The expression of the axion abundance, in the case the mass of the axion receives important thermal contributions is

\bea
\Omega_{\rm{R_2}}^{T}= \OmegastdT \left(\frac{\beta^2 \kappa}{4} \right)^{-3/\beta} \left(\frac{\Tend}{T_i}\right)^{\frac{3}\beta (4-\beta)} \left[ \frac{\kappa^{7/(4-\beta)}\, T_i^7}{\left(\alpha \LambdaQCD^4 \, m_0 M_P\right)^{7/6}}\right]^{\frac{4-\beta}{\beta+8}}.
\label{eq:NSC_R2_relicT}
\eea

\subsection{Oscillations in Region 3}{\label{app:reg3}}

We will assume $\Rosc$ is still far from $\Rend$, such that the Hubble parameter, for our purposeses,  can still be considered as
\be
H\sim \sqrt{\frac{\rho_\phi}{3\, M_P^2}}=\sqrt{\kappa}H_i \left(\frac{R_i}R\right)^{\beta/2}.
\ee
Nonetheless, the decays of $\phi$ are affecting radiation, such that its energy density it is better described as
\be
\rho_R \sim \rho_{Ri}\,\frac{2\Gamma_\phi\, \sqrt{\kap}}{(8-\beta) H_i}\left(\frac{R_i}R\right)^{\beta/2}.
\ee
Now we can equate the above expression with the energy density as a function of the temperature, the well known $\rho_R=(\pi^2/30) g_*(T) T^4$, such that we can obtain the relation between scale factor and temperature in this region to be
\be
\left(\frac{R_i}R\right)^{\beta/2}=\left( \frac{8-\beta}{2\sqrt{\kappa}}\right) \frac{T}{ \Tend^2 T_i^2}.\label{eq:R_vs_T_3}
\ee
Replacing back into the Hubble parameter, we find\footnote{Dropping the degrees of freedom.}

\be
H(T)=H_i \left(\frac{8-\beta}2\right)\frac{T^4}{\Tend^2T_i^2}.
\ee
As in the previous regions, we can find the oscillation temperature for temperature independent masses of the axion, {\it i.e.} $3H(\Tosc)=m_0$, and when the thermal effects are important, meaning $3H(\Tosc)=m_a(\Tosc)$.  For the first case, we find
\bea
\Toscthree=
\begin{dcases}
\left(\frac{2 m_0 M_P\, \Tend^2}{8-\beta} \right)^{1/4} & \mbox{for}\,\,\, m_0\lesssim m_{R_3},\\
\left(\frac{2 m_0 M_P\, \alpha\, \Lambda_{\text{QCD}}^4\, \Tend^2}{8-\beta}\right)^{1/8}& \mbox{for}\,\,\, m_0\gtrsim m_{R_3}.
\end{dcases}
\label{eq:Tosc_R3}
\eea
So for this region the dependence of $\Tosc$ on $\beta$ is quite mild, and on $\kap$ nonexistent.  
The intersection between these temperatures gives the moment of the QCD phase transition, which can be used to obtain the mass of the axion at this point. We get
\be
m_{R_3}=\frac{(8-\beta)}2\frac{\alpha\, \LambdaQCD^4}{M_P \,\Tend^2}\approx 10^{-6}\, \mu\mbox{eV} \left(\frac{\mbox{GeV}}{\Tend}\right)^2.
\label{eq:int_mass_R3}
\ee
In this case it is easy to put them in a single expression for high and low temperatures, as
\bea
\Toscthree=
\begin{dcases}
6.3\, \text{GeV} \left(\left(\frac{m_a}{\mu\text{eV}} \right)\left(\frac{\Tend}{\text{GeV}}\right)^2\frac{1}{8-\beta}\right)^\frac{1}4, & \text{for}\, \,m_a\lesssim 10^{-6}\, \mu\text{eV}\,\left( \frac{\text{GeV}}{\Tend}\right)^2,\\
1\, \text{GeV}\left(\left(\frac{m_a}{\mu\text{eV}} \right)\left(\frac{\Tend}{\text{GeV}}\right)^2\frac{1}{8-\beta}\right)^\frac{1}8, & \text{for}\, \,m_a\gtrsim 10^{-6}\,\mu\text{eV}\,\left( \frac{\text{GeV}}{\Tend}\right)^2.
\end{dcases}
\eea
We now can write down the mass ranges expected for the oscillation of the axion field to happen in Region 3. Here it is better to work with the stronger requirement $\Rc\ll R_{\rm osc}^{R_3}\ll \Rend$, but it can be easily translated into a temperature range by using eq.~(\ref{eq:R_vs_T_3}). Firstly, let us assume the oscillation happens completely before QCD transition, where temperature effects on the axion mass are important. In that case 
\be
\frac{8\, \Tend^6}{(8-\beta)\,\beta^2 M_P\, \alpha\LambdaQCD^4}\ll m_0\ll \frac{T_i^6 \sqrt{\kappa}}{M_P\,\alpha\,\LambdaQCD^4}\left(\frac{2\,\Tend^2 \sqrt{\kappa}}{(8-\beta)T_i^2}\right)^{\frac{8+\beta}{8-\beta}},
\label{eq:mass_range_R3_T}
\ee
On the other hand, it is possible to extend to smaller masses by considering for the bounds $\Rc\ll R_{\rm osc,0}^{R_3}\ll \Rend$, such that the axion mass is temperature independent. The mass range is extended as
\be
\frac{2}{\beta} \frac{\Tend^2}{M_P}\ll m_0 \ll \frac{T_i^2 \sqrt{\kappa}}{M_P}\left(\frac{2\,\Tend^2 \sqrt{\kappa}}{(8-\beta)T_i^2}\right)^{\frac{\beta}{8-\beta}}.
\ee
We remind that the lower bound of $m_0$ is only valid till $m_{R_3}$ is reached. Further, has to be replaced by expression (\ref{eq:mass_range_R3_T}). Also note that the lower bound, in either case, is independent on $\kappa$.

As final step, we find the expression for the misalignment angles, in the low temperature regime {\it i.e.} $m_a(T)=m_0$, and high temperature one. To do so, we look into the energy density in this region, given by eq.~(\ref{eq:NSC_axion_density}).
The entropy dilution factor, $S_{\rm osc}/S_{\rm end}$, is different than eq.~(\ref{eq:gamma_R1}), because the axion is produced in an era where the SM entropy is increasing. We write
\be
\frac{S_{\rm osc}}{S_{\rm end}}=\left(\frac{\Tosc R_{\text{osc}}}{\Tend\Rend}\right)^3.
\ee
Inserting back into eq.~(\ref{eq:NSC_axion_density}) we have
\be
\rho_a(T_0)\approx\rho_a(\Tosc) \frac{m_0}{m_a(\Tosc)}\frac{T_0^3}{\Tend^3}\left(\frac{\Rosc}{\Rend}\right)^3,
\ee
where from eq.~(\ref{eq:R_vs_T_3}) and (\ref{eq:rend}) we can easily find the ratio $\Rosc/\Rend$ to be
\be
\frac{\Rosc}{\Rend}=\left[\frac{4}{\beta(8-\beta)}\left(\frac{\Tend}{\Tosc}\right)^4\right]^{2/\beta}.
\ee
Thus, the axion energy density today can be written as
\be
\rho_a(T_0)=\rho_a(\Tosc) \frac{m_0}{m_a(\Tosc)}\left(\frac{T_0}{\Tosc}\right)^3 \gamma_{R_3}.
\label{eq:energy_R3}
\ee
Where $\gamma_{R_3}$ is the entropy dilution factor, given by
\be
\gamma_{R_3}=\left(\frac{4}{\beta(8-\beta)}\right)^{6/\beta} \left(\frac{\Tend}{\Tosc}\right)^{24/\beta-3}.
\label{eq:gamma_R3_ap}
\ee
Now the axion abundance is easily found by $\rho_a(T_0)/\rhocrit$, with $\rho_a(T_0)$ from \cref{eq:energy_R3}
\bea
\Omega_{R_3}=\begin{dcases}
 \Omegastdo\left(\frac{2}{\beta}\right)^{6/\beta} \left(\frac{\Tend^2}{m_0 M_P}\right)^{\frac{3}\beta (4-\beta)} & \mbox{for}\,\, m_0\lesssim m_{R_3}\\
\OmegastdT\, (8-\beta)^{\frac{\beta+6}{2\beta}} \left(\frac{\Tend^6}{\alpha \LambdaQCD^4 m_0 M_P}\right)^{3/\beta-2/3} & \mbox{for}\,\, m_0\gtrsim m_{R_3}
\end{dcases}
\label{eq:R3_relic}
\eea
Both expressions are independent on $\kap$ and $T_i$, showing that for this region the main parameters are $\Tend$ and $\beta$.

\chapter{Numerical code} 
We develop our numeric work in Python. To solve the axion differential equation we have used integrator \textbf{dopri5}, corresponding to an explicit Runge-Kutta of order 4.

\begin{figure}[h]
\centering
\includegraphics[scale=0.6]{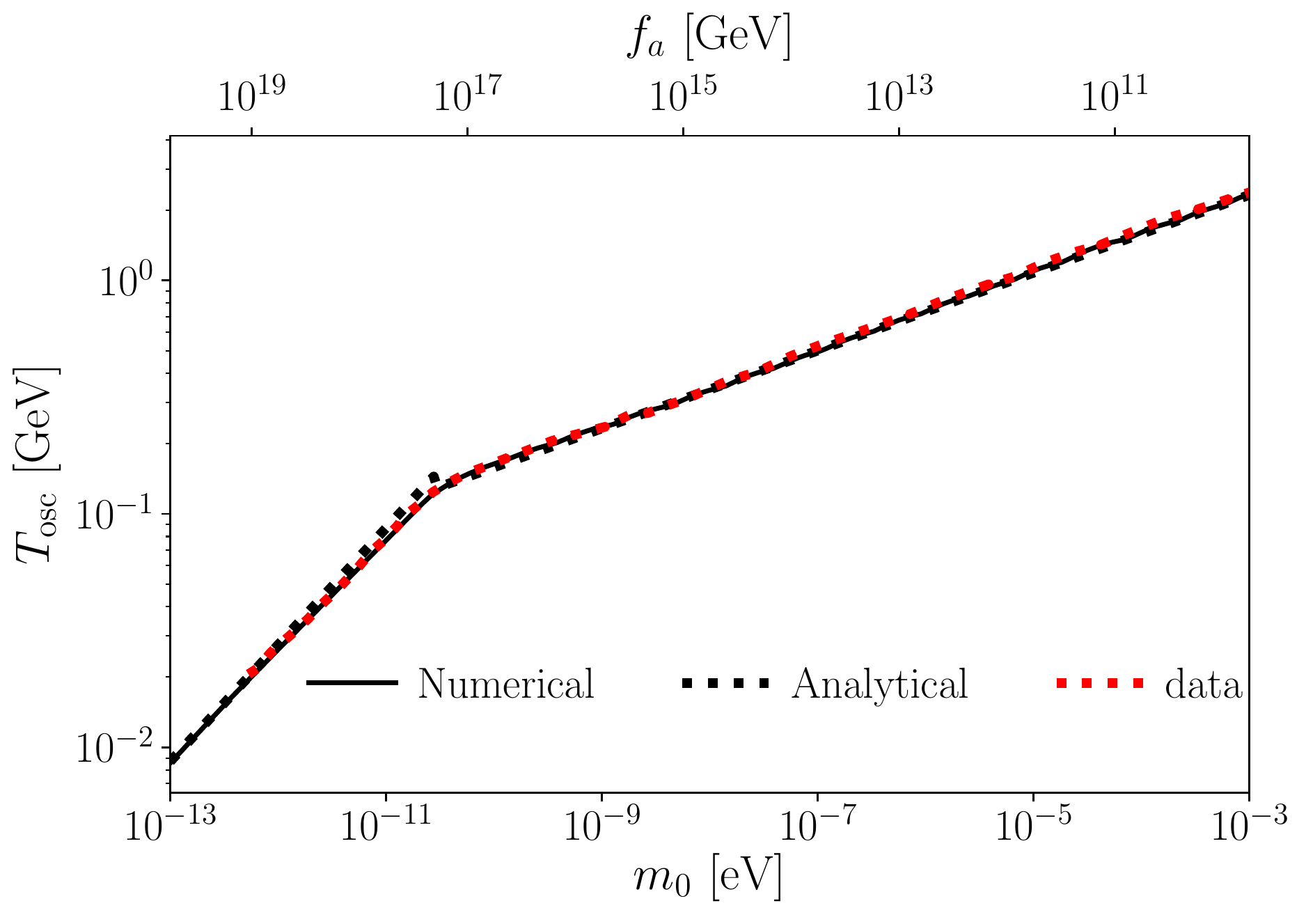}
\caption[Check to the oscillation temperature as function on the axion mass]{Check to the oscillation temperature as function on the axion mass. Black line is for the numerical solution, dotted black line is for analytical solution and red dots were obtained from \cite{Borsanyi:2016ksw}. The bend around $m_a\sim10^{-5}{\rm{\mu eV}}$ corresponds to the QCD transition.}
\label{fig:std_tosc_check}
\end{figure}

\begin{figure}[h]
\centering
\includegraphics[scale=0.6]{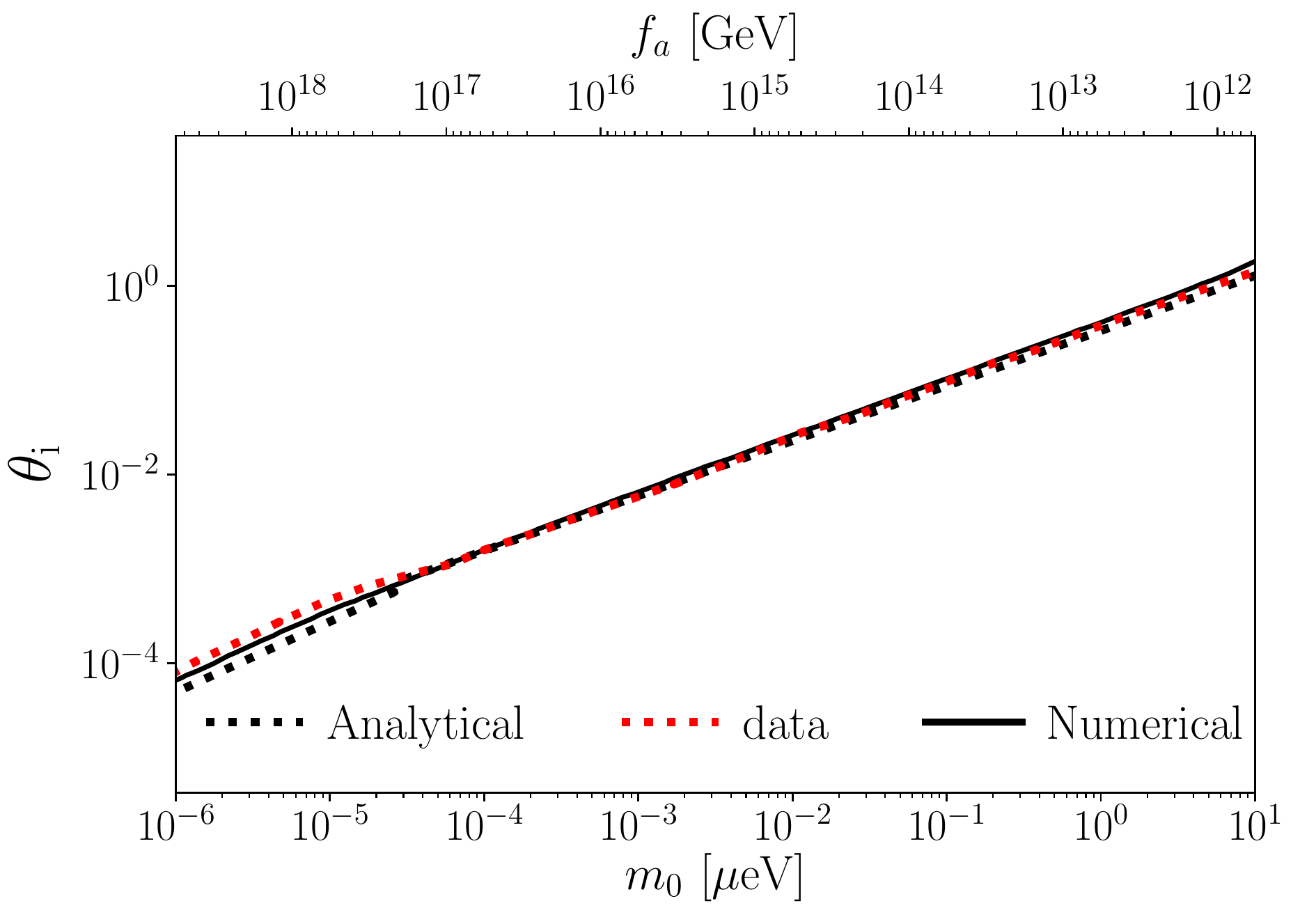}
\caption[Check to the initial misalignment angle as function on the axion mass]{Check to the initial misalignment angle as function on the axion mass. Black line is for the numerical solution and red dots were obtained from \cite{Borsanyi:2016ksw}. }
\label{fig:std_relic_check}
\end{figure}

Throughout this thesis we introduce the degrees of freedom by hand. For the case of standard cosmology we used that for $\Tosc<\Tosc(m_i)$, $g_\star(\Tosc)=10.75$ , while for  $\Tosc\geq\Tosc(m_i)$, $g_\star(\Tosc)=61.75$. We compared our results with bibliography, in \cref{fig:std_tosc_check} and \cref{fig:std_relic_check} we include data obtained in  \cite{Borsanyi:2016ksw}.

To solve the Boltzmann equations we have used the well$-$known numerical method \textbf{Radau} for solving stiff differential equations.

\end{document}